# Models of polymer solutions in electrified jets and solution blowing

Marco Lauricella

Istituto per le Applicazioni del Calcolo, Consiglio Nazionale delle Ricerche, Via dei Taurini 19, I-00185 Rome, Italy

Sauro Succi

Istituto per le Applicazioni del Calcolo, Consiglio Nazionale delle Ricerche, Via dei Taurini 19, I-00185 Rome, Italy

Center for Life Nanoscience at la Sapienza, Istituto Italiano di Tecnologia, Viale Regina Elena 295, I-00161, Rome, Italy

Harvard Institute for Applied Computational Science, Cambridge, Massachusetts 02138, United States

Eyal Zussman

Faculty of Mechanical Engineering, Technion-Israel Institute of Technology, Haifa 32000, Israel

Dario Pisignano*

Dipartimento di Fisica, Università di Pisa, Largo Bruno Pontecorvo 3, I-56127 Pisa, Italy

NEST, Istituto Nanoscienze–Consiglio Nazionale delle Ricerche, Piazza San Silvestro 12, I-56127 Pisa, Italy

Alexander L. Yarin

Department of Mechanical and Industrial Engineering, University of Illinois at Chicago, 842 W. Taylor St., Chicago, IL 60607-7022, United States

* dario.pisignano@unipi.it

**Abstract**

Fluid flows hosting electrical phenomena make the subject of a fascinating and highly interdisciplinary scientific field. In recent years, the extraordinary success of electrospinning and solution blowing technologies for the generation of polymer nanofibers has motivated vibrant research aiming at rationalizing the behavior of viscoelastic jets under applied electric fields or other stretching fields including gas streams. Theoretical models unveiled many original aspects in the underpinning physics of polymer solutions in jets, and provided useful information to improve experimental platforms. This article reviews advances in the theoretical description and numerical simulation of polymer solution jets in electrospinning and solution blowing. Instability phenomena of electrical and hydrodynamic origin are highlighted, which play a crucial role in the relevant flow physics. Specifications leading to accurate and computationally viable models are formulated. Electrohydrodynamic modeling, theories for the jet bending instability, recent advances in Lagrangian approaches to describe the jet flow, including strategies for dynamic refinement of simulations, and effects of strong elongational flow on polymer networks are reviewed. Finally, the current challenges and future perspectives of the field are outlined and discussed, including the task of correlating the physics of the jet flows with the properties of realized materials, as well as the development of multiscale techniques for modelling viscoelastic jets.

# CONTENTS

# I. INTRODUCTION

Rayleigh, and Sir Geoffrey Taylor (Taylor, 1964, 1966). For instance, Gilbert noted that a sessile droplet of water lying on a dry surface is deformed near the apex into a cone, when it is approached by a rubbed amber (Gilbert, 1600). Indeed, when rubbed with fur or wool, the amber acquires a net negative electric charge, through a kind of contact electrification which is known as triboelectric effect. A few simple experimental consequences of the triboelectric effect, such as the capability of rubbed amber to attract small wires or feathers, have been known for many centuries, since ancient Greece. The deformation of sessile droplets observed by W. Gilbert was also due to electrostatic attraction, namely to the interaction of the rubbed amber and droplets warped from their resting shape by an electric field. In the Gilbert's experiment, a water droplet is placed onto a glass surface, and the water-amber attraction is sustained by positive electric charges at the liquid surface, thus determining a conical shape as sketched in Figure 1a. A photograph of a conical surface made of canola oil, stretched by an electric field and captured by a high-speed camera (Collins *et al*., 2008, ), is shown in Figure 1b. The fact that conical liquid-air interfaces can be generated in this way is, at first sight, surprising. This stands in contrast with our everyday experience that a droplet of water tends to be pulled into a spherical shape, namely to reach a minimum surface area ( $A$ ) which is due to the imbalance of cohesive forces (i.e., the attraction of the liquid molecules to each other) at the surface layer. The surface tension ( $\alpha$ ) of the liquid is the physical quantity that is defined by such imbalance of cohesive forces and it is the Gibbs free energy ( $W_G$ ) per unit area at constant temperature and pressure, $\alpha = \partial W_G / \partial A$ , and the cohesive imbalance is expressed by the Laplace pressure which is $\alpha(r_{T1}^{-1} + r_{T2}^{-1})$ , where $r_{T1}$ and $r_{T2}$ are the radii of curvature in each of the axis parallel to the liquid-air interface. As a consequence, another force needs to be applied to the fluid to have the latter strained out of its unperturbed shape in which the surface area is minimum and the curvature is constant. The accumulation of an excess charge on the surface of a droplet, due to an electric field generated in the region of space surrounding the liquid-air interface resides, can provide such extra force (the Maxwell stresses) in an extremely effective way (Reznik *et al.*, 2004; Reznik *et al.*, 2006; Collins *et al.*, 2008), thus leading to the formation of an apex as displayed in Figure 1. As will be clear in this Review, producing conical shapes in liquid-air interfaces this way is highly important from a technological viewpoint, since it is at the base of the possibility of focusing and stretching polymer molecules dissolved in fluids into sub-micrometer features.

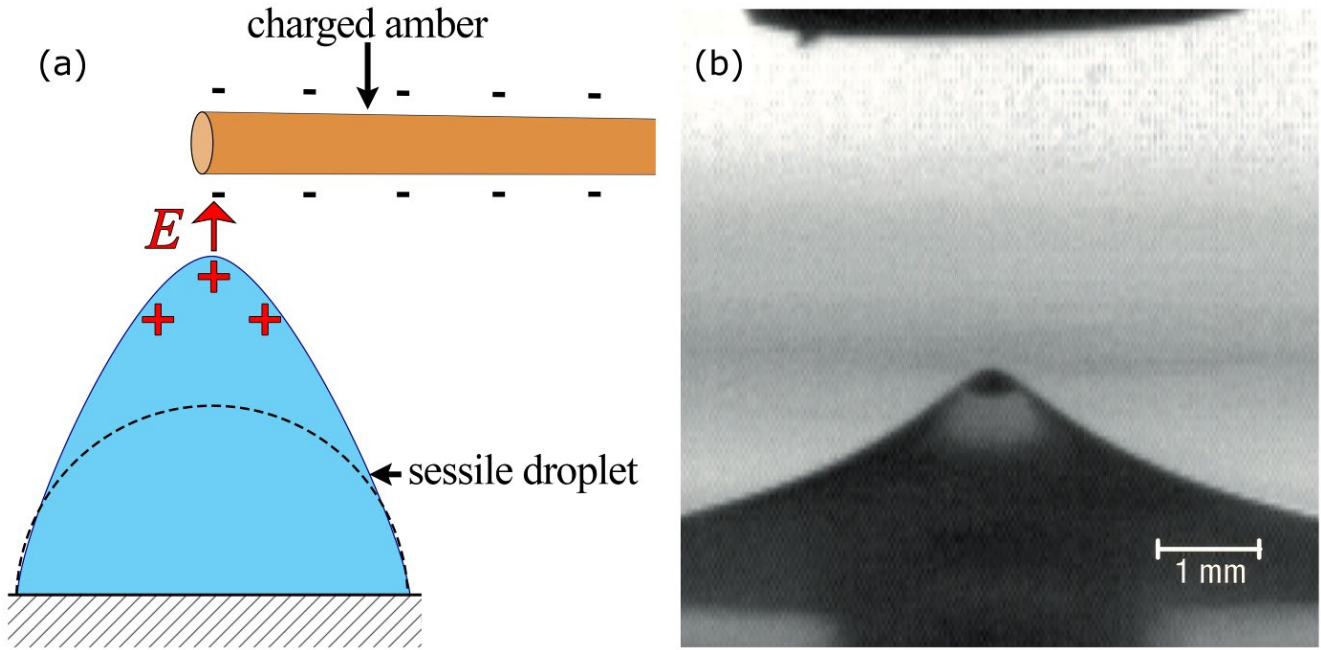

**Figure 1.** (a) Sketch of Gilbert's observation about the electrostatic deformation of sessile water droplets. A piece of negatively-charged amber is placed in proximity of a sessile droplet lying on a surface. The electric field ($E$) generated by the rubbed amber deforms the droplet, from the resting shape (dashed line) into a conical one (continuous line). (b) Conical liquid-air interface photographed in a region filled by an electric field. The liquid is canola oil. Adapted with permission (Collins *et al.*, 2008). Copyright © 2008, Nature Publishing Group.

In the 20th century, patents describing methods for dispersing fluids with the aid of electric fields were issued to Cooley (Cooley, 1902) and Morton (Morton, 1902). These early studies were followed by others by Zeleny (Zeleny, 1914, 1917), and finally by several patents by Formhals (Formhals, 1934, 1939, 1940, 1943, 1944), who focused on the formation of fibers by means of electrified jets of solutions of cellulose acetate. Later on, fibers with diameter smaller than 1 μm were realized from electrified jets of acrylic solutions in dimethylformamide (DMF), by DuPont researchers (Baumgarten, 1971). However, it was with two works from Reneker's group (Doshi and Reneker, 1995; Reneker and Chun, 1996) that the interest for the process named electrostatic spinning, i.e. electrospinning, started at our times. Such technology was largely developed during the years around 2000, concomitantly with a remarkable growth of the global interest to nanotechnology, that opened new routes for the headway of a wide variety of applications.

Depending on the process parameters, that involve the electric field applied on a polymer solution, the solution concentration, and various aspects of the used experimental set-up (all the details will be presented in the following Section III.B), the polymeric or ceramic (calcinated) fibers produced by electrospinning exhibit cross-sectional diameters ranging from a few nm (Huang *et al*., 2006) to a few microns. This property directly affects the surface area to material mass ratio. Indeed, polymer filaments with a cross-sectional radius about 100 nm and material density about 1.5 $g/cm^3$ have a surface area to mass ratio well above $10^5$ $cm^2 / g$ , which offers intriguing prospects for several practical applications. In addition, the electrospinning process was found to be chemically versatile, i.e., fibers could be formed from a wide variety of compounds, such as thermoplastic materials, conductive and light-emitting polymers, piezoelectric polymers, biomolecules, and blends. Many aspects of this technology could be engineered at a very advanced level. This aimed at fabricating nanofibers with desired shape, morphology, and composition (Sun *et al*., 2003; Loscertales *et al*., 2004; Li and Xia, 2004a; Ji *et al*., 2006; Zussman *et al*., 2006), assembled in specific architectures and networks (Theron *et al.*, 2001; Li *et al*., 2003; Zussman *et al*., 2003; Sun *et al*., 2006; Xie *et al*., 2010), or performing given functions. Finally, electrospinning technologies are operationally simple, they need a relatively low investment for equipment, and their throughput is generally much higher than those of other methods for producing nanostructures, such as high-resolution lithographies, nanoimprinting, molecular self-assembly, or colloidal synthesis (Xia *et al.*, 2003; Li and Xia, 2004a, Luo *et al.*, 2012). For these reasons, the applications of electrospun fibers span today the entire field of nano- and micro-technologies, including filtration, catalysis, energy harvesting and storage, photonics and optoelectronics, nanoelectronics, development of surface coatings with controlled wettability and thermal properties, new textile materials, chemical and biochemical sensors, systems for drug delivery, tissue engineering and regenerative medicine, as well as cancer research, as summarized in Figure 2. Electrospinning and related methods, as well as the applications of nanofibers and microfibers produced by them, have been the subject of several books (Ramakrishna *et al.*, 2005; Reneker and Fong, 2006; Wendorff *et al.*, 2012; Pisignano, 2013; Yarin *et al.*, 2014; Yarin *et al.*, 2017) and reviews (Dzenis 2004; Li and Xia, 2004b; Greiner *et al.* 2006; Reneker *et al.* 2007; Yarin *et al.* 2007; Greiner and Wendorff, 2007; Reneker and Yarin 2008; Yarin 2011; Choi *et al.*, 2017). Not equally covered is the major effort done to develop theoretical models of these processes, which is the subject of this Review.

Modelling approaches might offer a critical tool to investigate the underpinning physics of electrified polymer solutions, and provide valuable information for rationalizing observed phenomena, and then for substantially improving experiments. Here we report on the main modelling strategies pursued in aligned nanofibers deposited on a ring collector, and image of dura fibroblasts seeded on fibronectin-coated scaffolds made of the radially-aligned nanofibers. Reproduced with permission (Xie *et al*., 2010). Copyright © 2010, American Chemical Society. (l) Scheme of the fabrication of an implantable active-targeting micelle-in-nanofiber device (FM-Nanofiber) and of the delivery process of doxorubicin-loaded micelles (FM) from nanofibers to tumor tissues and cells. Reproduced with permission (Yang *et al*., 2015). Copyright © 2015, American Chemical Society.

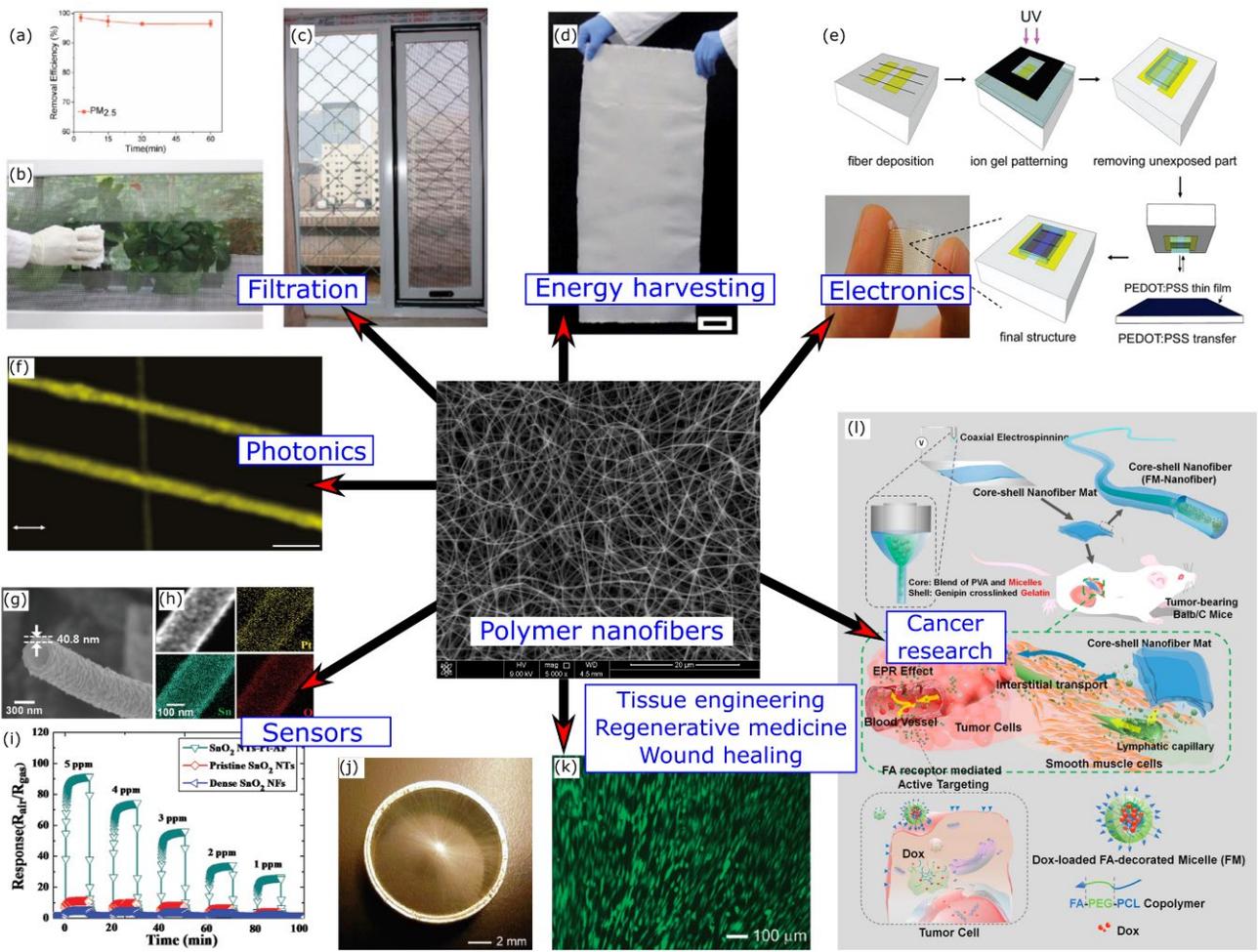

**Figure 2.** Applications of nanofibers realized through electrified jets or solution blowing from polymer solutions. (a-c) Filtration performance of a window screen coated by polyacrylonitrile nanofibers, used to remove $PM_{2.5}$ particles. The wiping of nanofibers from the window screen, performed by tissue paper, is also shown in (b). Adapted with permission (Khalid *et al*., 2017). Copyright © 2017, American Chemical Society. (d) Large-area mat of piezoelectric nanofibers. Scale bar: 5 cm. Reproduced with permission (Fang *et al*., 2013). Copyright © 2013, The Royal Society of Chemistry. (e) Process for the realization of an array of top-gate field-effect transistors, involving the deposition of electrospun poly(3-hexylthiophene) nanofibers across gaps between source and drain electrodes. The process include a transfer procedure of a poly(3,4-ethylenedioxythiophene) polystyrene sulfonate (PEDOT:PSS) layer. Reproduced with permission (Lee *et al*., 2010). Copyright © 2010, American Chemical Society. (f) Light-emitting polymer fibers for photonics, showing emission polarized along their longitudinal axis. The arrow displays the direction of the polarizer used to analyse the emitted light. Scale bar: 5 μm. Reproduced with permission (Pagliara *et al*., 2011). Copyright © 2011, American Chemical Society. (g-i) Chemoresistive sensors: Example of Pt-loaded $SnO_2$ nanotubes made through electrospinning, corresponding elemental distribution (Pt, Sn, and O), and response to acetone for $SnO_2$ nanotubes, $SnO_2$ nanofibers, and Pt-loaded $SnO_2$ nanotubes. Concentration range: 1-5 ppm at 350 °C. Adapted with permission (Jang *et al*., 2015). Copyright © 2015, The Royal Society of Chemistry. (j,k) Application in biomedical devices for wound closure or tissue engineering. Radially-

the last decades, introducing relevant physical characteristic lengths and highlighting instability phenomena of electrical and hydrodynamic nature, which play a crucial role in the physics of flow. The Review is organized as follows. The processes, their working principles and relevant dimensionless parameters describing the physics underneath are presented in Section II. Section III provides readers with an introductory yet comprehensive overview of the experimental methods and of the physical rationale behind them. Section IV presents the specifications that make modelling reliable in describing the dynamics of these fluids, namely what one needs in order to elaborate accurate and computationally viable models. These aspects include, for instance, required parameters from experiments/phenomenology, as well as specific computational platforms and resources. Section V.A discusses the modeling methods of the physical processes that affect electrified fluids, and that are based on a so-called electrohydrodynamic (EHD) description. Then the physical reasons at the base of the instabilities affecting electrospun fluids will be reviewed. A general and detailed model of the electrospinning process has been formulated in the works by the Reneker and Yarin groups since 2000 (Reneker *et al*., 2000; Reneker *et al.* 2007; Yarin *et al*., 2014). The basic building blocks on which these and other, more recent models are formulated are reported in Sections V.B and V.C. The main approaches developed for modelling the flow of electrified fluids under very high strain rates, namely under very fast deformation, and its effects on polymer networks, are described in Section V.D. Section VI reviews modelling methods for solution blowing, that is an air-jet spinning method and among the most recent and promising technologies to generate nanofibers (Yarin *et al*., 2014; Sinha-Ray *et al*., 2015). In this process, polymer solutions are delivered into a co-flowing, sub- or supersonic gas jet, which stretches them directly (i.e., without the application of an electric field). Exploiting the aerodynamic drag force results in a 100-fold increase in the production rate of nanofibers, and in higher compatibility with industrial equipment already designed for other uses such as meltblowing (Kolbasov *et al*., 2016). Finally, an outlook is provided in Section VII regarding currently open challenges, as well as possible future developments of theory and modelling methods for this class of technologies.

A few details on notation and units used in this paper deserve to be mentioned. Throughout the Review, boldfaced characters denote vectors. For equations containing terms depending on the electric field, we use Gaussian (CGS) units unless differently stated, since these are highly convenient where the highlighted physics encompasses both electrostatics and fluid mechanics. An advantage of the CGS system is in the compact dimensional description of physical quantities, which involves only three base quantities (length-mass-time). In particular, while in SI units the electric charge ($q$) needs to be defined as an independent quantity (in Coulomb units), in the CGS system setting the vacuum permittivity $\varepsilon_0 = 1$ and the Coulomb constant $k_C = 1$ (Sommerfeld, 1952), the Coulomb's law for the force, $F = q^2 / \ell^2$, between two charges at distance $\ell$, easily defines the charge as $q = \sqrt{\mathrm{dyn\,cm}^2} = \mathrm{cm}^{3/2} \mathrm{g}^{1/2} \mathrm{sec}^{-1} = \mathrm{statC}$. Hence, the electric field, i.e. the electrostatic force divided by the test charge, $E = F / q$, in CGS units is measured in $\mathrm{dyn/statC} = \mathrm{cm}^{-1/2} \mathrm{g}^{1/2} \mathrm{sec}^{-1} = \mathrm{statV/cm}$, and the electric potential, $\Phi = E \cdot \ell$, is measured in $\mathrm{statV} = \mathrm{cm}^{1/2} \mathrm{g}^{1/2} \mathrm{sec}^{-1}$. All the physical quantities used in this Review, and their units, are given in the List of Symbol.

**List of symbols**

| Symbol | | Units (Gaussian CGS) |
|---|---|---|
| $A$ | Surface area | $cm^2$ |
| Abs | Linear attenuation coefficient | $cm^{-1}$ |
| **B** | Unit binormal of jet axis | 1 |
| $Bo$ | Bond number | 1 |
| **C** | Unit vector pointing the jet curvature center | 1 |
| $Ca$ | Capillary number | 1 |
| $C_s$ | Solvent concentration | $mol\ cm^{-3}$ |
| $C_p$ | Polymer concentration | $mol\ cm^{-3}$ |
| $D$ | Diffusion coefficient | $cm^2\ s^{-1}$ |
| $De$ | Deborah number | 1 |
| $e_0$ | Electric charge per unit jet length | $statC\ cm^{-1}$ |
| $E$ | Electric field | $statV\ cm^{-1}$ |
| $g$ | Gravitational acceleration | $cm/s^2$ |
| $G$ | Elastic modulus | $g\ cm^{-1}\ s^{-2}$ |
| $h$ | Nozzle-collector distance | cm |
| $I$ | Current | $statC\ s^{-1}$ |
| $k$ | Jet curvature | $cm^{-1}$ |
| $\ell_{jet}$ | Jet arc length | cm |
| $M$ | Mass | g |
| **N** | Principal unit normal of jet axis | 1 |
| $N_s$ | Number of monomers | |
| $p$ | Pressure | $g\ cm^{-1}\ s^{-2}$ |
| $q$ | Charge | statC |
| $Q$ | Flow rate | $cm^3/s$ |
| $R$ | Cross-sectional radius | cm |
| $r_0$ | Cross-sectional radius (initial) | cm |
| $r_T$ | Droplet radius | cm |
| **R** | Position vector | cm |
| $t$ | Time | s |
| $T$ | Temperature | K |
| Tr | Radiation transmission | |
| $\alpha$ | Surface tension | $g/s^2$ |
| $\varepsilon$ | Dielectric constant | 1 |
| $\dot{\varepsilon}$ | Strain rate | $s^{-1}$ |
| $\Phi, \varphi, \varphi_0$ | Electric potential | statV |
| $\phi$ | Volume fraction | |
| $\dot{\gamma}$ | Shear rate | $s^{-1}$ |
| $\lambda$ | Stretching ratio | 1 |
| $\theta$ | Relaxation time (viscoelastic) | s |
| $\theta_C$ | Relaxation time (charge) | s |
| $\theta_H$ | Relaxation time  (hydrodynamic) | s |
| $\sigma$ | Stress | g cm$^{-1}$ s$^{-2}$ |
| $\sigma_e.$ | Electrical conductivity | s$^{-1}$ |
| $\sigma_q$ | Surface charge density | statC cm$^{-2}$ |
| $\rho$ | Mass density | g cm$^{-3}$ |
| $\rho_q$ | Volumetric charge density | statC cm$^{-3}$ |
| $\mu$ | Dynamic viscosity | g cm$^{-1}$ s$^{-1}$ |
| $\mu_e$ | Elongational viscosity | g cm$^{-1}$ s$^{-1}$ |
| $\nu$ | Kinematic viscosity | cm$^2$ s$^{-1}$ |
| $\mathbf{T}$ | Unit tangent of jet axis | 1 |
| $\boldsymbol{\upsilon}$ | Velocity | cm/s |
| $W_G$ | Gibbs free energy | g cm s$^{-2}$ |
| $\omega$ | Angular velocity | rad/s |
| $\xi$ | Mesh size of polymer network | cm |
| $\Sigma$ | Mass attenuation coefficient | cm$^2$ g$^{-1}$ |

## II. PROBLEM FORMULATION AND DIMENSIONLESS PARAMETERS IN ELECTROSPINNING

The electrospinning process (scheme in Figure 3) generates nanofibers starting from a polymer solution at a high enough concentration (for instance, 5-30% in polymer weight with respect to solvent) and applying an electric voltage bias ($\Delta\Phi_0 \approx 3 - 300$ statV, namely $1 - 100$ kV), between the solution and a metal surface onto which fibers are to be deposited. The motivation for having a quite high concentration of molecules in the polymer solution is that such molecules have to be entangled, in order to provide the fluid with remarkable viscoelastic properties, as better described below. In a typical process, the solution and the metal surface of the target (a counter-electrode) are initially at a typical distance ( $h$ ) ranging from a few cm to a few tens of cm thus leading to electric fields, $E$, from

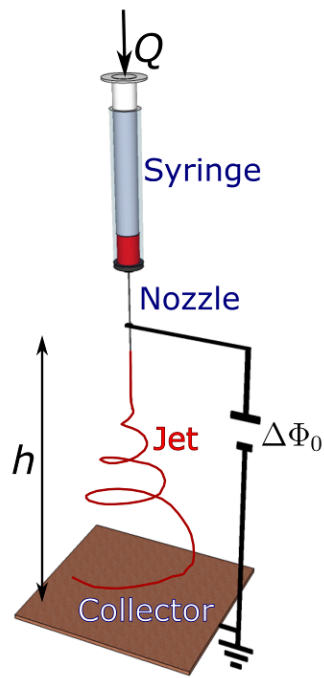

**Figure 3.** Scheme of electrospinning. The various process parameters are indicated: nozzle-collector distance ($h$), nozzle-collector voltage bias ($\Delta\Phi_0$), polymer solution flow rate ($Q$), together with main experimental equipment (syringe, metal needle, collecting surface).

a few statV/cm to a few tens of statV/cm (i.e., $\sim 2-10$ kV/cm). The accumulation of an excess charge on the solution surface (i.e., a droplet, or a free liquid surface) then leads to the formation of an apex with a well-defined shape and local radius ($r_T$) at the same surface. The determination of the exact shape of this apex has been the subject of intense research, which will be reported in Section V.A.1. Upon increasing the applied voltage, electric forces overcome the elastic forces characteristic to viscoelastic polymer solutions and surface tension ($\alpha$) and destabilize the pristine fluid body. Limiting for simplicity to the contributions from electric forces and surface tension, this effect can be conveniently expressed by a dimensionless parameter, which is the electric Bond number,

$Bo = r_T E^2 / \alpha$, where $E$ is the electric field at the solution apex. The Bond number directly compares the electrostatic pressure at the liquid surface ($\varepsilon_0 E^2 / 2$ for a liquid acting as a perfect electric conductor in the vacuum with $\varepsilon_0 = 1$ in CGS units) with the Laplace pressure ($2\alpha / r_T$), which is related to surface energy. Hence, when $Bo$ approaches unity (Reznik $et\ al.$, 2004), a significant deformation of the solution surface can be achieved. Considering typical values of $\alpha = 50$ g/s$^2$ and

$E = 10$ statV cm$^{-1}$ which corresponds to $3$ kV cm$^{-1}$, one would have a $Bo$ value up to 0.2 for an apex radius, $\leq 0.1$ cm. As a result, strain occurs in the fluid body, and an electrically-charged solution jet is issued in electrospinning.

Viscoelasticity plays a crucial role in the jet onset. Since the used solution concentration is quite high, a significant amount of polymer entanglements are formed in the fluid. Too low values of the polymer concentration in the solution might instead lead to electrospraying (or the so-called dripping regime), thus forming mono- or poly-dispersed, charged sprayed droplets instead of a jet and nanofibers (Collins $et\ al.$, 2007; Gañán-Calvo and Montanero, 2009; Herrada $et\ al.$, 2012; Collins $et\ al.$, 2013; Gañán-Calvo $et\ al.$, 2018). The dimensionless quantity allowing one to distinguish the two cases of jetting (electrospinning) and dripping (jet breaking) is the capillary number, $Ca = (\mu_e \upsilon / \alpha)$, where $\upsilon$ is the jet velocity and the characteristic dynamic viscosity involved is the elongational viscosity, $\mu_e = \dfrac{\sigma}{\dot{\varepsilon}}$ ($\sigma$: longitudinal stress and $\dot{\varepsilon}$: strain rate). The capillary number is the ratio between the (elongational) viscous forces and capillary forces (Anna and McKinley, 2001; Montessori *et al.*, 2019), and it also allows 'high-viscosity' and 'low-viscosity' fluids to be distinguished in electrospinning experiments, depending on the obtained $Ca$ values. In the jetting regime, the electric field is strong enough to imprint a suited velocity, or the solution viscosity is high enough so that the viscous forces overcome the surface tension ($Ca > 1$). For instance, in aqueous solutions of poly(ethylene oxide) (PEO) at different concentrations, with elongational viscosity spanning from $500 \text{ g cm}^{-1}\text{s}^{-1}$ to $5000 \text{ g cm}^{-1}\text{s}^{-1}$ (Xu *et al.*, 2003; Reneker *et al.*, 2007), and typical values, $\upsilon = 10 \text{ cm s}^{-1}$ and $\alpha = 50 \text{ g/s}^2$, the corresponding capillary number is in the range $10^2$-$10^3$. On the other hand, if the capillary force is dominant ($Ca < 1$), the jet rapidly necks down pinching off into droplets due to the Rayleigh-Tomotika instability (Tomotika *et al.*, 1935), and tiny droplets can be easily emitted from the solution surface (dripping regime). In other words, in this regime the polymer jets cannot be stabilized against the capillary instability (Entov and Yarin, 1984; Yarin, 1993).

During the fast (≤0.1 s) path to the counter-electrode, the electrospun jet is dramatically stretched, initially as a short almost straight section, and then, in the course of various other instabilities that *bend* the trajectory of the fluid, generating spiraling loops as shown in Figure 4 (Reneker and Fong, 2006; Reneker and Yarin, 2008). The term, *bending instability*, in electrospinning (Reneker *et al.*, 2000) is motivated by the similar term applied to the kindred aerodynamically-driven bending instability (Weber, 1931), and a basic similarity was also recognized between bending jets and the elastic bar bending in the classical Euler-Bernoulli theory (Landau and Lifshitz, 1970). Bending instabilities occur because small perturbations rapidly trigger their growth, which is driven by the fact that lateral electric forces, namely electrostatic repulsion of charges along the fluid filament, appear at any curved section of the jet. These instabilities influence the electrospinning outcome in many ways, increasing the overall length of the jet trajectory to a practically fractal-like one and thus leaving longer time for fluid stretching and diameter reduction, and also leading to a highly disordered configuration of nanofibers (non-woven mat, central panel of Figure 2) deposited on flat counter-electrodes. Due to instabilities, an electrospun jet is a highly complex system, namely a continuous fluid object whose cross-sectional radius, $r$, and curvature, $k$ (defined as the inverse of the curvature radius and measured in $cm^{-1}$) both are strongly dependent on the position along the fluid filament. In addition, the solvent simultaneously evaporates while the jet is moving, which allows solid polymer nanofibers to be deposited. Fortunately, some physical estimates allow the description of electrospun solutions to be simplified in the different regions of the jet. Indeed, lot of the jet dynamics depends on the magnitude of the characteristic hydrodynamic time, $\theta_H$, vs. the relaxation time, $\theta_C$, of ionic charges present in the fluid. $\theta_H$ can be estimated as $\mu\, r / \alpha$ (Reznik et al., 2004). With $\mu$ of the order of of 10 g cm$^{-1}$s$^{-1}$ (1 g cm$^{-1}$s$^{-1}$ = 1 Poise = 0.1 Pa s ) which is reasonable for electrospun solutions, and $\alpha = 50$ g/s$^2$, one would obtain a $\theta_H$ value of the order of 10 ms for $r = 0.1$ cm and well below the ms for $r \leq 10$ μm, respectively. The characteristic charge relaxation time, $\theta_C$, can be estimated as $\varepsilon / \sigma_e$, where $\varepsilon$ is the dielectric permittivity, and $\sigma_e$ is the electric conductivity (expressed in s$^{-1}$ in CGS units). With plausible values of $\varepsilon = 40$ and $\sigma_e$ in a range $10^3 - 10^4$ s$^{-1}$ (corresponding to $10^{-7} - 10^{-6}$ S m$^{-1}$ in SI units), $\theta_C$ is up to a few ms (~4-45 ms) for polymer solutions used in electrospinning (Reneker et al., 2007; Yarin et al., 2014). These solutions are in fact leaky dielectrics, i.e. quite poor ionic conductors (Melcher and Taylor, 1969; Saville, 1997). However, whenever $\theta_H$ is significantly higher than $\theta_C$, the electric behavior of an ionic conductor reduces to that of a perfect conductor, even though it is actually a poor conductor compared to such truly good conductors as metals. On the contrary, when $\theta_H \ll \theta_C$, the fluid body can be considered as a perfect dielectric, with 'frozen' charges (Reneker et al., 2000). This means that, while the electrospun jet continuously reduces its diameter when moving from the initial fluid apex to the counter-electrode, its physical description might actually change from that typical of a conductor to that typical of a dielectric system. We will see in the continuation of this Review how this feature is exploited in models of the electrospinning dynamics.

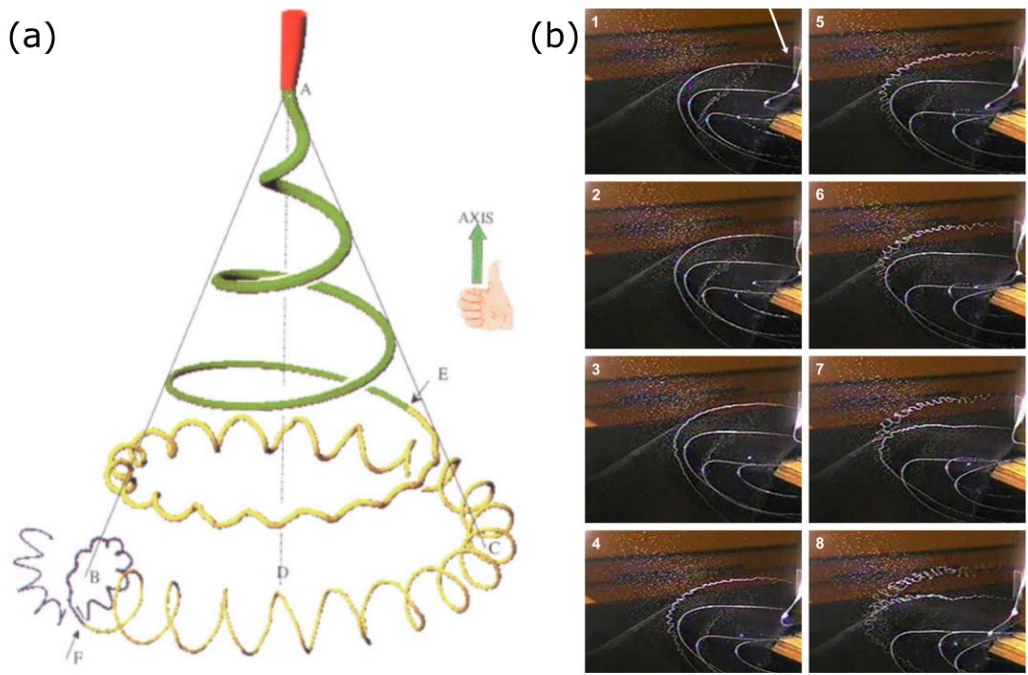

**Figure 4.** (a) Schematics of an electrified polymer solution jet. The various regimes of jet flow are highlighted, with a straight region (A) followed by various bending instability stages featuring spiraling loops with different characteristic frequency. D indicates the projection of the straight region along the vertical direction. AB and AC: jet envelope cone. E and F indicate points of onset of the second and third bending instability, respectively. Reproduced with permission (Reneker and Fong, 2006). Copyright © 2006, American Chemical Society. (b) Photographs of a jet of polyisobutylene in a viscous mixture of acetone and paraffin oil, highlighting bending instability. The frames are captured by a video camera at a rate of 30 per second. The diameter of the largest coils is about 20 cm. Reproduced with permission (Reneker and Yarin, 2008). Copyright © 2008, Elsevier Ltd..

The strong elongational flow in the jet also causes a substantial stretching of involved macromolecules and polymer networks. Indeed, strain rates exerted by fluid bodies during elongational processes, that are comparable to or higher than the reciprocal relaxation time ($\theta^{-1}$) of the involved polymer molecules, favour the transition of random coils into stretched and relatively aligned molecular assemblies (de Gennes, 1974). This can be expressed in terms of the dimensionless Deborah number, $De = \dot{\varepsilon}\theta$, that reaches very quickly a value above unity along the jet path (Bellan *et al*., 2007). Such effect can be observed experimentally by various methods, including Raman, luminescence, birefringence, infrared spectroscopy, mechanical measurements, and X-ray inspection (Fong and Reneker, 1999; Reneker *et al.*, 2007; Kakade *et al*., 2007; Arinstein *et al*., 2007; Pagliara *et al*., 2011; Pai *et al*., 2011; Camposeo *et al*., 2013; Richard-Lacroix and Pellerin, 2013, 2015; Yarin *et al*., 2014), and it was also discussed theoretically (Greenfeld and Zussman, 2013; Deng *et al*., 2017). For instance, below a certain crossover diameter of electrospun nanofibers, that is dependent on the polymer molar mass, the elastic moduli of the fibers begin to rise sharply (Arinstein *et al.*, 2007; Burman *et al.*, 2008; Ji *et al.*, 2008; Burman *et al.*, 2011; Liu *et al.*, 2011). Similarly, in optically-active polymers, the effective conjugation length of chromophores might increase as a result of electrospinning, and the optical absorption and emission become polarized along the fiber axis (Camposeo *et al.*, 2013; 2014). These findings are highly important for applications of polymer nanofibers. They also make further clear that rationalizing the dynamics of electrospun jets by means of proper models is essential to understand how the jet properties affect, or are inherited by, the obtained nanofibers. The main phenomena to be caught by models are described in the next Section.

## III. PHENOMENOLOGY OF ELECTRIFIED JETS

### A. Formation and characteristics of electrified jets

In electrospinning, polymer solutions are usually delivered in a continuous way through a syringe (Figure 3), terminated by a metallic needle with a diameter of a few hundreds of μm. This leads to the formation of a pendant droplet at the tip of the needle. However, most of the mechanisms producing electrified jets are quite general, and they hold not only for pendant droplets, but also for any other free surface (Yarin and Zussman, 2004; Lukas *et al*., 2008). For instance, upon inserting an electrode into a charged (sessile or pendant) droplet of polymer solution and applying an electric voltage bias with respect to a counter-electrode, the droplet can acquire a stable shape whenever the potential difference is not too high (Reznik *et al*., 2004). As introduced in Section II, this can be achieved by connecting electrodes to a high-voltage generator, and the counter-electrode works as collecting surface onto which polymer nanofibers are to be deposited. The corresponding electric field imposes the electric Maxwell stresses pulling and stretching the droplet toward the counter-electrode. The surface tension of the fluid would tend to minimize the droplet surface and to shape it as a spherical volume (thus, minimizing the surface energy) through the Laplace pressure. The elastic effect in the viscoelastic polymer fluid, which might be much higher than the surface tension, also plays a restraining role. A steady-state droplet shape then arises as a result of the interplay between the Maxwell stresses and the restoring forces. The shape at the apex of the droplet was originally described by the so-called Taylor cone (Taylor, 1964, 1966). This shape, at the transition region between the stressed droplet and the formed jet, is important. Indeed, correlations were recently found between observable features of the droplet-jet shape and the diameter of the obtained electrospun nanofibers (Liu and Reneker, 2019).

Clearly, the condition on the threshold electric field for jet activation corresponds to a condition in terms of the minimum surface charge density accumulated at the fluid-air interface nearby the apex (so-called Rayleigh condition). In addition, due to the high solution concentration, in electrospinning also significant elastic stresses should be overcome by the electric Maxwell stresses.

Once formed, the jet proceeds very quickly towards the counter-electrode. The jet velocity ranges between the order of ten to a few hundreds of cm/s, which can be measured by various experimental techniques such as particle imaging velocimetry (Bellan *et al*., 2007; Reneker *et al*., 2007) and direct high-speed imaging (Yarin *et al*., 2001a; Han *et al*., 2008; Montinaro *et al*., 2015). Corresponding accelerations reach the order of $10^4$ cm/s$^2$. The very high strain rate, $\dot{\varepsilon}$, so reached, exceeding the reciprocal relaxation time ($\theta^{-1}$) of the solution (de Gennes, 1974; Thompson *et al*., 2007), might stretch the polymer matrix into a non-equilibrium conformational state (Greenfeld *et al*., 2011). This effect is at the origin of the significant orientational anisotropy inherited by the produced nanofibers. In this way, polymer chains might become prevalently oriented along the longitudinal axis of electrospun fibers, which in turn affects several optical, electronic, and mechanical properties, although some relaxation of the anisotropic structure can still occur after formation.

The jet continuously delivers an electric charge from the spinneret to the counter-electrode, which results in a current, $I = Q\rho_q$, where $Q$ is the jet flow rate and $\rho_q$ is the volumetric charge density. The overall charge carried by the jet rapidly transforms into surface charges, leading to bulk and surface advection components that contribute to the current (Fridrikh *et al.*, 2003; Reznik *et al*., 2006). During the jet flight and the development of the bending instability, the electric charges are basically at rest with respect to the jet, because the characteristic charge relaxation time, $\theta_C$, is larger than the characteristic hydrodynamic time, $\theta_H$ (see Section II). In other words, in this regime transport processes associated with viscous relaxation and hydrodynamics are faster than electric transport phenomena (Saville, 1997), leading to the conclusion that during its path toward the counter-electrode (collector) the jet may be assumed as a perfect dielectric (Yarin *et al*., 2014). Some of these aspects will be further detailed in Section V.A. Overall, a quantity denoted by $\rho_q$ can still be used to indicate carried charge density, meaning an effective charge density that would account for both bulk and surface advection components contributing to the current. The current delivered by electrospun jets was investigated in dedicated experiments (Theron *et al*., 2004; Bhattacharjee *et al*., 2010). Depending on the used polymer solution and on other process parameters, values ranging from the order of nA to hundreds of µA ($\sim 3$ to $3 \cdot 10^5$ statC/s ) were measured (Reneker and Chun, 1996; Deitzel *et al*., 2001a; Hohman *et al*., 2001b; Theron *et al*., 2004; Kalayci *et al*., 2005; Bhattacharjee *et al*., 2010).

The evaporation of the solvent from the jet is another phenomenon very important for the process outcome, contributing to diameter reduction and determining whether still wet or fully dried nanofibers are deposited on the collector (Yarin *et al*., 2001a). In addition, if the evaporation rate is sufficiently high, the solvent evaporation being faster from the external layers of the jet might lead to the formation of a polymer skin along the fluid body (Koombhongse *et al*., 2001; Guenthner *et al*., 2006). Such skin can then collapse and cause the formation of electrospun fibers with different cross-sectional shapes, including belts, ribbons, hollow filaments, fibers with elliptical cross-section, etc (Figure 5). Solvent evaporation effects were modelled by nonlinear mass diffusion transfer to estimate the transient solvent concentration profiles in the jets (Wu *et al*., 2011). Furthermore, solvent evaporation strongly affects the morphology and porosity of the surface of electrospun nanofibers (Srikar *et al*., 2008). It was shown that the characteristic times of (i) polymer-solvent mutual diffusion, (ii) solvent evaporation, and (iii) phase separation of immiscible components influence the ultimately achieve nanofibers, which can be smooth, exhibit corrugations as displayed in Figure 5b (Pai *et al*., 2009), or pores as displayed in Figure 5c,d (Bognitzki *et al*., 2001).

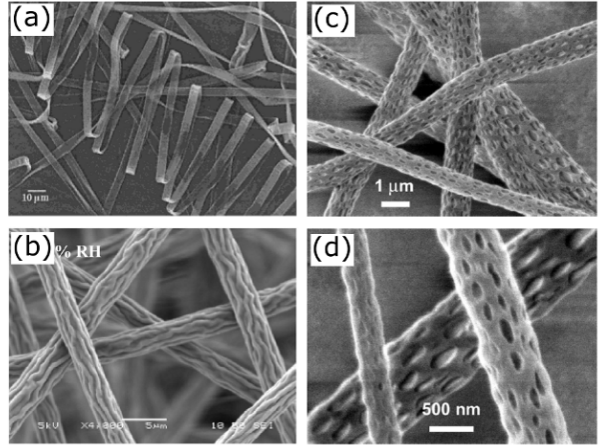

**Figure 5.** (a) Electrospun ribbons made of poly(ether imide). Scale bar: 10 μm. Reproduced with permission (Koombhongse *et al.*, 2001). Copyright © 2001, John Wiley & Sons, Inc. (b) Corrugated fibers electrospun from a 30 wt% polystyrene/DMF solution under 22% relative humidity. Scale bar: 5 μm. Adapted with permission (Pai *et al.*, 2009). Copyright © 2009, American Chemical Society. (c,d) Porous fibers made of poly-L-lactide. Scale bar: 1 μm (c), 500 nm (d). Reproduced with permission (Bognitzki *et al.*, 2001). Copyright © 2001, WILEY-VCH Verlag GmbH & Co. KGaA.

Since early studies in late 1990s, it was found that during its path from the spinneret to the collector the electrified jet does not follow a straight trajectory, but it is affected instead by a rich variety of bending instabilities (Reneker *et al.*, 2000). An analogous effect is rapid whipping, that is also non-axisymmetric and involves a deformation of the centerline of the electrified jet (Hohman *et al.*, 2001a; 2001b; Shin *et al.*, 2001). Other mechanisms observed are jet branching, consisting in secondary filaments separating from the main jet and possibly leading to fibers with complex shape (Yarin et al., 2005), and buckling at the collector, which produces many different coiled geometries as displayed in Figure 6 (Han *et al.*, 2007). Buckling effects are superimposed to those from bending instabilities, and strikingly resemble those previously observed with larger non-electrified jets (Chiu-Webster and Lister, 2006). The buckled coils found in electrospinning experiments show typical diameters in the range of μm up to tens of μm, and patterns built at MHz characteristic frequencies, depending on the jet velocity and viscoelastic properties. While the origin of bending instability is in the existence of lateral forces of electric origin, which appear at any curved section of a jet and lead it in further bending, buckling arises from compressive forces acting along the jet from an obstacle. In this sense, buckling is a close counterpart of the classical elastic buckling of compressed columns discovered and explained by Euler (Landau and Lifshitz, 1970). Readers are referred to dedicated papers for the theory of buckling of free liquid jets (Tchavdarov *et al.*, 1993; Yarin, 1993).

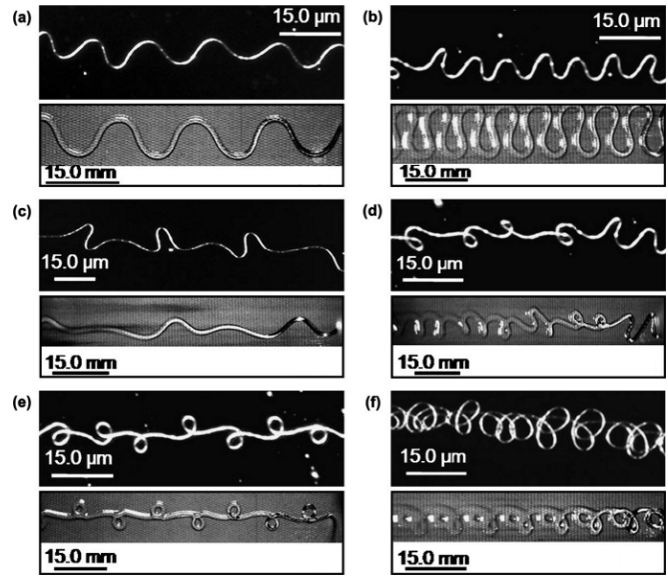

**Figure 6.** Buckled patterns with various geometries, formed by electrified jets of PEO in water (upper panel in each pair), collected on glass slides, and similar patterns produced by buckling of uncharged, gravity-driven syrup jets (corresponding bottom panel). The fall of viscous threads leads to buckling patterns ~$10^3$ times larger than those from electrified jets. The obtained patterns include sinusoidal features (a), compressed meanders (b), double meanders (c), irregular features (d), elongated figures of eight (e), and coils (f). Top panels: Reproduced with permission (Han *et al.*, 2007). Copyright © 2007, Elsevier Ltd.. Bottom panels: Adapted with permission (Chiu-Webster and Lister, 2006). Copyright © 2006, Cambridge University Press.

## B. Experimental parameters

The electrospinning is a process depending on a number of governing parameters (Theron *et al*., 2004). Ideally, by optimizing the set of these variables, an experimentalist can obtain nanofibers with desired composition and morphology. For this reason, also theory and models of electrified jets carefully consider these parameters and generally start from them in order to properly describe the process (Reneker *et al*., 2000; Fridrikh *et al*., 2003). The values of many of these parameters (so-called primary parameters) can be directly controlled by experimentalists (such as chemicals, solution concentration, the set-up parameters). However other quantities are not chosen directly, but depend in turn on the values of the primary parameters. These aspects make optimizing the overall process quite complex, and also highlight how important modelling can be to better rationalize observed phenomena, to identify most relevant variables that affect the formation of nanofibers, and to provide suitable ranges or starting values of relevant parameters to guide the experiments. Already in pioneering papers (Doshi and Reneker, 1995), the electrospinning parameters were grouped in three main classes:

(i) variables related to *solution properties*, that depend on the used polymer species, their molecular weight, concentration, and elastic relaxation time, as well as on the solvent used; these parameters include the solution density, viscosity, electrical conductivity, dielectric constant, and surface tension. In addition, the solvent properties include its volatility, determining the evaporation rate;

(ii) variables related to the used *set-up*, that include the applied electric potential difference ($\Delta\Phi_0$), the solution flow rate ($Q$), the distance between the spinneret and the collector ($h$), and the internal diameter of the spinning needle;

(iii) variables related to *ambient*, such as relative humidity, temperature, and pressure.

The various classes of parameters are summarized in Table 1.

| | Primary parameters (directly chosen) | Secondary parameters (depend on the primary ones) | |
|---|---|---|---|
| Solution properties | Polymer species<br>Solvent species<br>Solution concentration ($C_p$, mol cm$^{-3}$)<br>Polymer molecular weight | Solution density ($\rho$, g cm$^{-3}$)<br>Dynamic viscosity ($\mu$, g cm$^{-1}$ s$^{-1}$)<br>Conductivity ($\sigma_e$, s$^{-1}$)<br>Surface tension ($\alpha$, g/s$^2$)<br>Relaxation time ($\theta$, s)<br>Dielectric constant ($\varepsilon$) | Deitzel *et al.*, 2001a; Megelski *et al.*, 2002; Theron *et al.*, 2004; Jarusuwannapoom *et al.*, 2005; Shenoy *et al.*, 2005; Thompson *et al.*, 2007; Montinaro *et al.*, 2015 |
| Set-up settings | Applied voltage bias ($\Delta\Phi_0$, statV)<br>Solution flow rate ($Q$, cm$^3$/s)<br>Inter-electrode distance ($h$, cm)<br>Needle internal diameter | $E$, Electric field ($\text{statV cm}^{-1}$) | Deitzel *et al.*, 2001a; Megelski *et al.*, 2002; Theron *et al.*, 2004; Thompson *et al.*, 2007; Montinaro *et al.*, 2015 |
| Ambient properties | Atmosphere composition<br>Atmosphere pressure<br>Relative humidity<br>Temperature ($T$, K) | | Megelski *et al.*, 2002; Casper *et al.*, 2004; Thompson *et al.*, 2007; Wang *et al.*, 2007; Pai *et al.*, 2009; Fasano *et al.*, 2015 |

**Table 1.** Different classes of process parameters in electrospinning. Quantities used in models are reported with their symbol and Gaussian (CGS) units. Relevant references are also highlighted for each class of parameters.

The influence of the solution and set-up parameters or of combinations of them on the dynamics of electrified jets and on the morphology of electrospun nanofibers was investigated in several studies (Deitzel *et al.*, 2001a; Megelski *et al.*, 2002; Theron *et al.*, 2004; Jarusuwannapoom *et al.*, 2005; Shenoy *et al.*, 2005; Thompson *et al.*, 2007; Montinaro *et al.*, 2015). Other works focused on the effects of the ambient variables, particularly on humidity that is relevant to control the surface morphology and porosity of nanofibers (Casper *et al.*, 2004). Recent experiments showed that performing electrospinning in controlled nitrogen atmosphere (Figure 7a) might be useful to reduce the surface roughness of nanofibers, as well as to improve light emission properties of fibers made of conjugated polymers, because of the reduced incorporation of oxygen in the jets (Fasano *et al.*, 2015).

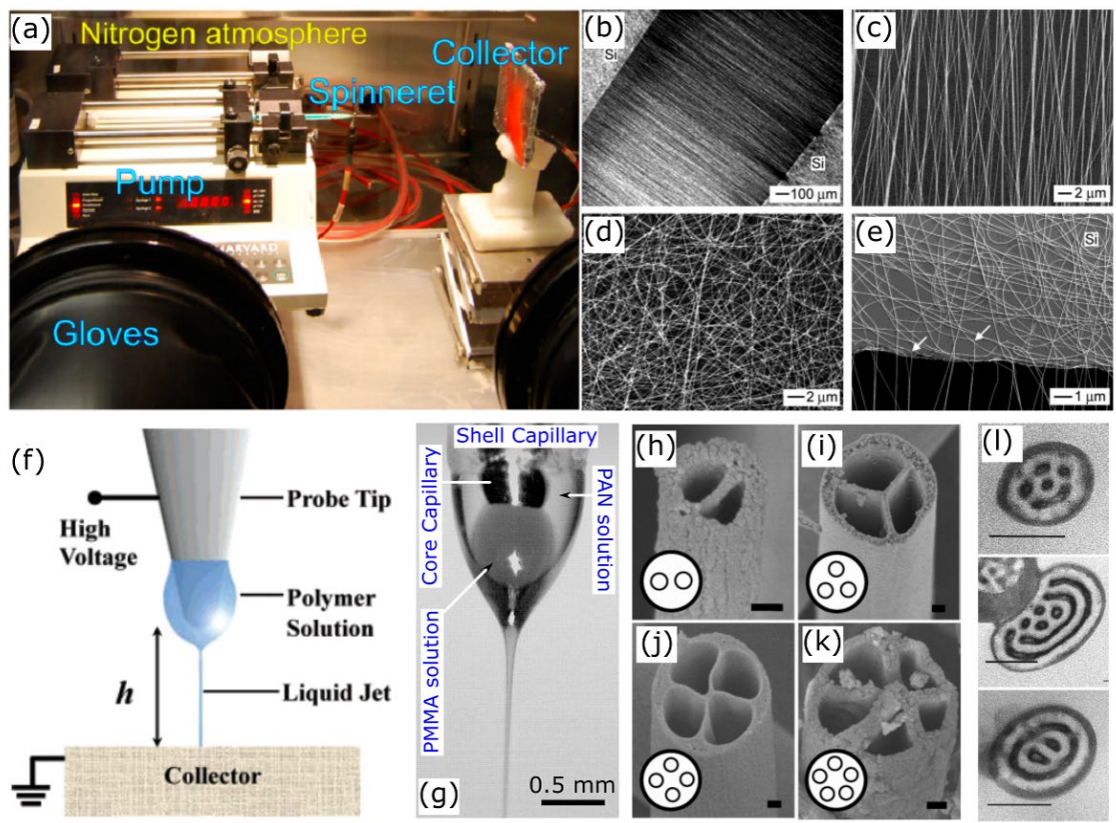

**Figure 7.** Examples of process engineering. (a) Set-up for producing electrified polymer solution jets in nitrogen atmosphere. Reproduced with permission (Fasano *et al*., 2015). Copyright © 2015, American Chemical Society. Doi: 10.1021/acs.macromol.5b01377. Further permissions related to the material excerpted should be directed to the American Chemical Society. (b-e) Images of uniaxially aligned arrays of electrospun polymer nanofibers produced with a gap collector: dark-field optical micrograph (b), and scanning electron microscopy (SEM) images (c,d) comparing aligned (c), and random fibers deposited onto a flat surface (d). A SEM micrograph of the gap edge is also shown, displaying a transition from aligned to random nanofibers (e). Reproduced with permission (Li *et al*., 2003). Copyright © 2003, American Chemical Society. (f) Scheme of the near-field electrospinning process. Here, *h* is from a few hundreds of μm to a few mm. Reproduced with permission (Sun *et al*., 2006). Copyright © 2006, American Chemical Society. (g) Compound droplet and bicomponent Taylor cone encompassing two different polymer solutions in co-axial electrospinning. Used polymers: poly(methyl methacrylate) (PMMA) for the core, and polyacrylonitrile (PAN) for the shell, respectively. Adapted with permission (Zussman *et al*., 2006). Copyright © 2006, WILEY-VCH Verlag GmbH & Co. KGaA. (h-k) SEM micrographs of multichannel tubes with variable diameter and channel number, obtained by a multi-channel spinneret. Scale bars: 100 nm. Reproduced with permission (Zhao *et al*., 2007). Copyright © 2006, American Chemical Society. (l) Transmission electron micrographs of cross-sections of annealed block copolymer fibers. Scale bars: 100 nm. Reproduced with permission (Kalra *et al*., 2009). Copyright © 2009, WILEY-VCH Verlag GmbH & Co. KGaA.

Finally, it should be mentioned that the polymer solution can be doped in various ways, particularly by adding nanoparticles with the aim of realizing nanocomposite fibers. Classes of quantum dots and of other nanoparticles used in electrified jets included those made of metals (Au, Ag), semiconductors (CdS, CdSe, ZnSe, ZnS), oxides (ZnO, $Fe_3O_4$, silica, titania), minerals such as hydroxyapatite and tricalcium phosphate, two-dimensional (2D) materials, carbon nanotubes, etc. (Dror et al., 2003; Salalha et al., 2004; Sui *et al*., 2005; Schlecht *et al*., 2005; Liu *et al*., 2006; Lu *et al*., 2009; Zhang *et al.*, 2016; Resta *et al*., 2017). Because nanoparticles are dispersed in the solution, the polymer component works as a three-dimensional (3D) topological network, with the particles constituting distributed solid domains. These can significantly affect the jet rheology and dynamics, as highlighted by dedicated models (Lauricella *et al*., 2017a).

## C. Electrified jet engineering

Using a planar collecting surface, electrified jets produce a random distribution of nanofibers on the plane connected to the counter-electrode as shown in the central panel of Figure 2, with a layer-by-layer stacked deposition along the direction perpendicular to that surface. This is fine for some applications, such as for building filters, or nonwoven mats for catalysis or textiles, but for other applications parallel nanofibers, or nanofibers arranged in different architectures are much more convenient. To this aim, collector geometries were engineered in various ways, to achieve contraction of the pattern of the electric field lines on sharp edges and, consequently, desired assemblies of nanofibers (Theron *et al*., 2001; Zussman *et al*., 2003; Sundaray *et al*., 2004; Teo and Ramakrishna, 2006). For instance, it was found that nanofibers can be obtained in almost uniaxially-oriented arrays (Figure 7b-e) between electrodes with parallel conductive regions separated by a gap (Li *et al*., 2003; Li *et al*., 2004; Xie *et al*., 2010). Upon approaching such regions, the electrified jet experiences two electrostatic forces. One is due to the electrospinning electric field, while the other is the attractive interaction between advection charges in the jet and the corresponding image charges in the collector electrodes. The combination of the two forces make fibers to stretch across the gap between the conductive regions. The effects of residual charges and of the gap size on fiber alignment were studied by numerical simulations, and the alignment of fibers was found to improve upon increasing the gap distance from 3 to 8 mm (Liu and Dzenis, 2008). Another successful method consists in collecting nanofibers on rotating cylinders (Sundaray *et al*., 2004) or disks (Theron *et al*., 2001). This approach might also lead to extra pulling and stretching of the electrified jets and possibly to enhanced order of polymer macromolecules in nanofibers. The magnetic-field assisted electrospinning has also gained importance as method to align nanofibers that incorporate magnetic particles (Rahmani *et al*., 2014; Mei *et al*., 2015; Huang *et al*., 2016; Guarino *et al*., 2019). Pioneering experiments showed that magnetic fields might be more efficient than electric fields in aligning fibers into parallel arrays (Yang *et al*., 2007). A conceptually different approach for controlling fiber positions (Figure 7f) consists in reducing the distance between the spinneret and the collecting surface to the sub-cm or even sub-mm scale, thus exploiting the initial, straight part of the jet (i.e., before the onset of bending instabilities), in so-called near-field electrospinning (Sun *et al*., 2006; Chang *et al*., 2008).

Electrified jets were also engineered to deliver more than one polymer compound in co-flows, in order to obtain nanofibers with core-shell or even multicomponent architecture, or hollow nanofibers (following removal of an internal sacrificial material). Core-shell structures in fibers are important for a wide range of technologies, including waveguides with layers showing refractive index contrast, electrical nanowires with insulating sheath, polymer structures for drug delivery, nanofluidics, and biological scaffolds. Co-axial electrospinning was implemented by means of concentric needles delivering different polymer solutions (Sun *et al*., 2003; Loscertales *et al*., 2004; Li and Xia, 2004a; Yu *et al*., 2004; Zhang et al., 2004; Zussman *et al*., 2006). A bicomponent Taylor cone can be obtained in this way (Figure 7g). Interestingly, in coaxial jets with two immiscible solutions, instabilities might be reduced, and the flow of the internal solution is supported by the external one thus making possible to use as core fluid a solution that has not, by its own, a sufficiently viscoelastic behavior to form electrospun fibers. This is the case of many solutions of light-emitting or conductive polymers, low-molar-mass molecules, biomolecules, and drugs to be encapsulated for controlled and sustained delivery. The method can be scaled up to more fluids, and nanofibers with several, either concentric or parallel layers or cavities (Figure 7h-k) formed by means of spinnerets featuring three or more fluidic channels (Zhao *et al*., 2007; Chen *et al*., 2010). An especially intriguing application of multi-fluid jets was shown in confining block-copolymers within the small volumes of individual nanofibers, to study the resulting microphase configuration (Ma *et al*., 2006; Kalra *et al*., 2006; 2009). For instance, nanofibers with a block-copolymer core and a shell made of silica or of a polymer with high glass transition temperature can be annealed to induce ordered domains in the internal region (Figure 7l). Core-shell nanofibers can also be electrospun from a single nozzle, by using emulsions of two polymers in a single solvent as a working fluid (Bazilevsky *et al*., 2008; Yarin *et al*., 2014). In the case of emulsion spinning, the dispersed phase forms the core, whereas the continuous phase forms the shell. Recently, emulsion spinning of core-shell nanofibers found application in generating self-healing vascular nano-textured composite materials (Yarin *et al*., 2019).

## D. From electrospinning to solution blowing

Methods to enhance the throughput in producing nanofibers could make use of forces other than electrostatic. For instance, a gas stream delivered by a nearby distributer would support polymer solution jets through additional shearing and stretching. Such gas-assisted electrospinning or electro-blowing methods (Um *et al*., 2004; Wang *et al*., 2005; Hsiao *et al*., 2012) might be implemented through gas distributers surrounding the spinneret, and allow thinner nanofibers to be obtained. In fact, co-flowing subsonic gas jets have been used for a long time to assist the formation of microscopic fibers in meltblowing (Pinchuk *et al*., 2002; Fedorova and Pourdeyhimi, 2007; Yarin *et al*., 2014). These technologies start from polymer melts, and at variance with electrospinning, the bending instability in this case arises purely aerodynamically, due to a distributed lateral force acting on curved sections of polymer jets, driving the instability process. To obtain nanofibers, air-jet spinning i.e. solution blowing (Sinha-Ray *et al*., 2010a) rather then meltblowing is used. This process is schematized in Figure 8. The needle is concentric with the nozzle issuing the air jet at a given delivery pressure. Solution blowing arise an increasing interest due to their superior throughput and capability to generate very thin (20-50 nm) nanofibers (Sinha-Ray *et al*., 2013; Yarin *et al*., 2014; Sinha-Ray *et al*., 2015; Daristotle *et al*., 2016; Polat *et al*., 2016). The method has already been scaled up to industrial equipment (Kolbasov *et al*., 2016). As shown in the sketch in Figure 9a, in this technique polymer solutions are employed similarly to electrospinning, however the jet is issued into a co-flowing sub- or supersonic gas jet without the application of an external electric field, similarly to meltblowing.

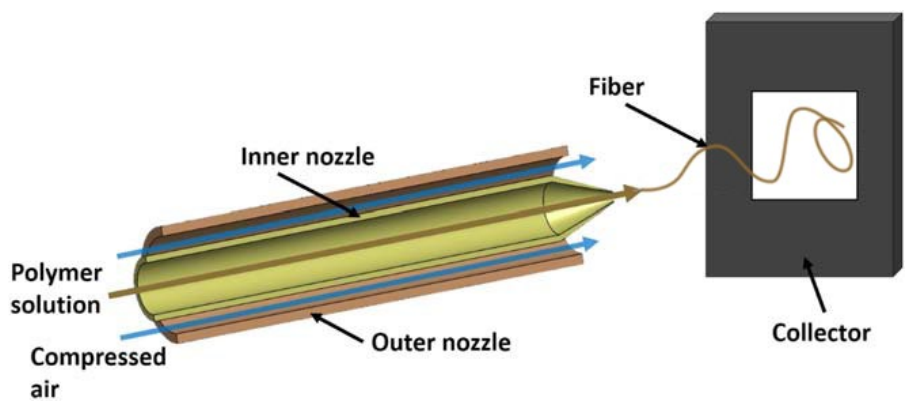

**Figure 8.** Scheme of a solution-blowing set-up and process. Adapted with permission (Polat *et al*., 2016). Copyright © 2015, WILEY-VCH Verlag GmbH & Co. KGaA.

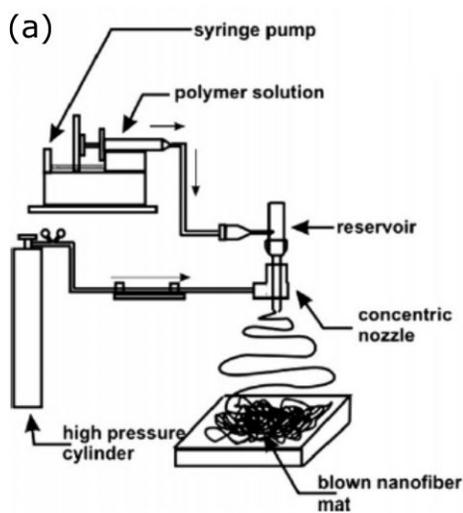 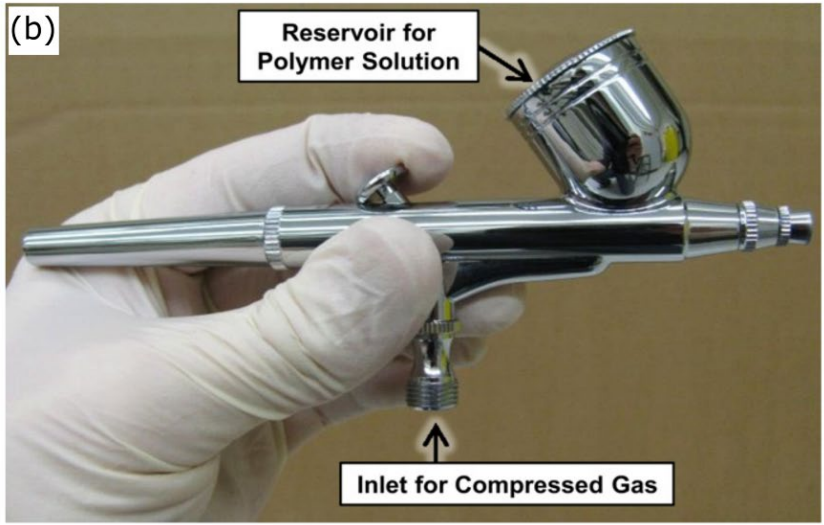

**Figure 9.** Scheme of a solution blowing bench setup. Reproduced with permission (Sinha-Ray *et al*., 2010b). Copyright © 2010, Elsevier Ltd. (b) A commercially available airbrush for portable solution blowing of nanofibers. Reproduced with permission (Tutak *et al*., 2013). Copyright © 2013, Elsevier Ltd.

Solution blowing shares many of its process parameters with electrospinning, including solution properties (polymer and solvent species, concentration, zero-shear viscosity and elastic relaxation time). Also of importance are several variables related to the used set-up (particularly, solution flow rate, needle gauge), and such ambient parameters as humidity and temperature. Here aerodynamically-driven bending perturbations can be triggered by turbulence of the surrounding gas flow, whereas polymer viscoelasticity, similarly to electrospinning, plays a restraining role. In some cases, two streams of polymer solution and pressurized gas can be also generated by means of a commercially available airbrush as that displayed in Figure 9b (Tutak *et al*., 2013). Solution-blown nanofibers were used to obtain carbon nanotubes (Sinha-Ray *et al*., 2010b), filter media (Zhuang *et al*., 2013; Liu *et al*., 2019), for biomedical engineering (Tutak *et al*., 2013; Khansari *et al*., 2013; Behrens *et al*., 2014;

2015; Magaz *et al.*, 2018), as well as for the formation of self-healing vascular nanocomposites (Yarin *et al.*, 2019).

Another fundamental feature of liquid jets concentric with gas flows is the resulting hydrodynamic flow-focusing effect driven by the external fluid. For example, this can be implemented in a configuration with the external gas stream surrounding a manifold jet in turn composed by two immiscible solutions, and with the internal capillary, delivering the core fluid, protruding by about one diameter length from the outer capillary, delivering the sheath fluid. Such arrangement was analysed in terms of working conditions (solution viscosities, capillary diameters, gas pressure, etc.), and found to overcome surface tension effectively and to lead to the generation of continuous, steady capillary jets down to the sub-μm cross-sectional size (Gañán-Calvo *et al.*, 2007).

## IV. REQUIREMENTS OF THEORETICAL METHODS FOR MODELLING ELECTRIFIED AND BLOWING FLUIDS

Modelling of electrified and solution-blown fluids is a practical tool to achieve reliable predictions of the geometries of nanofiber sizes, assembly, or internal molecular structure, guiding the choice of specific process parameters. For instance, this could aim at providing a likely starting point for the fabrication process of fibers with desirable geometry or architecture, thus significantly saving experimental resources and time. Modeling would be even more important in order to rationalize the fundamental properties of polymer solutions, in particular, their spinnability in fiber-forming processes, which are predominantly uniaxial elongation processes. The description is based on a system of highly non-linear partial differential equations (PDEs) for the electrified or blowing jet dynamics. In principle, this could be tackled by standard grid methods, such as finite-differences or finite volumes. However, the huge material stretching in space and the characteristic times of the competing physical phenomena in electrospinning and solution blowing are such that make a standard grid approach (the Eulerian approach) extremely challenging on computational grounds. The main problem is the major stretching of polymer jets accompanied by reduction in its cross-sectional size from, say, ~100 microns at the exit of the nozzle, down to 1 micron or less at the collector. Even modelling the shrinkage of the polymer jet via a constitutive law, the jet representation would require at least 10 grid points, hence to describe correctly a jet diameter of about 1 micron at the collector the length scale would be 0.1 μm. With a static Eulerian grid, covering a 3D cubic domain of about 10 cm in size would then require $10^{18}$ grid points. The simple storage of a corresponding array of a scalar variable in double floating numbers would require eight millions of terabytes which is totally out of reach for any foreseeable computer in many years to come.

Of course, the naive estimate above can be considerably softened by resorting to moving-grid (Lagrangian) techniques (Section V.B), whereby high resolution would also be put in place where needed, i.e. in the vicinity of the jet, by meshing adaptively 'on-the-fly' in the course of the simulation (sub-Section V.C.4). The corresponding procedure is computationally less demanding. For instance, the discretization of a jet length equal to 300 cm with a typical length step of 0.02 cm would be stored in only 15,000 grid points (120 kilobytes in double floating numbers to storage an array of a scalar variable). The Lagrangian approach is one of the two options traditionally used in fluid mechanics, and it implies grid following the individual fluid particles, i.e., the grid 'frozen' in the fluid, rather than an arbitrary surrounding space as in the case of the Eulerian grid (Lamb, 1959; Loitsyanskii, 1966; Batchelor, 2002). Moreover, this strategy provides low data traffic, low input/output operations and lean communication between threads on parallel machines. The latter point is of extreme importance, since it is increasingly apparent that, as performance ramps up, accessing data could become more expensive than perform floating-point operations (Succi *et al.*, 2019).

The discrete-particle representation is obviously an approximation of the fluid in a quasi-one dimensional (1D) object, but in problems related to electrospinning, melt- and solution blowing and hydroentanglement (Reneker *et al.*, 2000; Yarin *et al.*, 2001a; Yarin *et al.*, 2011; Sinha-Ray *et al.*, 2015; Yarin, *et al.*, 2014; Li *et al.*, 2019a; Li *et al.*, 2019b), it has the substantial advantage that particles naturally flow in the spatial regions where the relevant physics of the jet takes place.

In fact, denoting with $r$ the jet cross-sectional radius and with $z$ the longitudinal coordinate along which the jet is delivered, jets can be frequently considered as slender bodies with slow changes of $r$ along $z$, i.e. with $|\mathrm{d}r/\mathrm{d}z| << 1$. This is also referred to as slenderness assumption (Eggers, 1997). These long and thin fluid threads, whose velocity fields are directed mostly along one axis, can be well-described by 1D fluid flows. In this framework, the velocity field in the radial direction is usually expanded to the lowest-order terms, which should be sufficient to describe the radial motion (Lee, 1974; Pimbley, 1976; Entov and Yarin, 1984; Yarin, 1993; Eggers and Dupont, 1994). Hence, equations of motion (EOM) are written only for the expansion coefficients depending on the axial variable, $z$. Nonetheless, observables related to the remaining dimensions such as radial velocity, curvature radius, and jet cross-section, to name a few, survive, but they are now dependent on the 1D expansion coefficients. This is the core of the quasi-1D description, which can be implemented in jets, waves, drop dynamics, and thin-film flows (Middleman, 1995). In the jet context, several other assumptions are usually included in the quasi-1D description (Yarin, 1993; Eggers, 1997; Eggers and Villermaux, 2008;), such as incompressible velocity field, fluid volume-preservation, axisymmetric flow, isotropic expansion or contraction in the jet cross-section, and absence of shearing forces in the lateral jet surface. However, some experimentally relevant cases may go beyond one or more of these assumptions (e.g., considering an elliptical cross-section for the thread), so that a critical evaluation of all the approximations adopted in each specific model should always be addressed. Quasi-1D equations for electrified jets are presented in the following sub-Section V.A.3 in their EHD form [see Eqs. (5)-(8)], and in sub-Section V.A.4 in their fundamental Lagrangian descriptions [see Eqs. (12)]. The limitations of this approach, leading to the need of a fully 3D representation for comprehensively catching bending jets, will be discussed in sub-Section V.B.3.

In fact, in Lagrangian descriptions of a jet in $n$ (individual) particles (Sections V.B and V.C) as schematized in Figure 10, the computational complexity of solving the associated system of equations scales like $n^2$, mostly on account of unscreened electrostatic interactions (all-to-all, long-range coupling). This is the so-called 'direct' summation technique, where each particle interacts non-hierarchically with all other particles. With the simpler paradigm of direct summation, on a standard personal computer the number of viable particles would be limited to about 1,000, which, based on the existing literature, appears to be largely sufficient to deliver informative insights on the jet behavior. Nonetheless, more sophisticated approaches imply a hierarchical organization of the long-range Coulomb interactions, whose complexity scales like $n \log(n)$. In this framework, the charge elements of the discretised jet are recursively clustered and the monopole coefficients of the clusters are computed (Kowalewski *et al*., 2009). The scheme can be augmented by including higher orders of the multipole expansion, usually referred to as the fast multipole method.

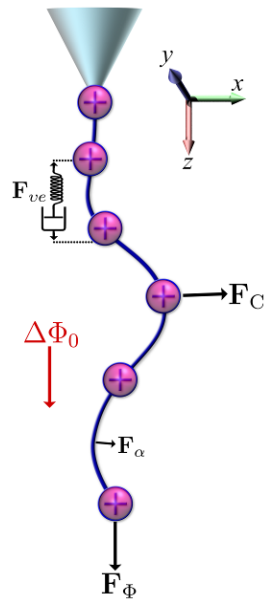

**Figure 10.** Sketch of the Lagrangian representation for electrospun jets. Each discrete element representing a jet segment is drawn by a bead, here with a plus sign denoting the positive charge of the segment. The scheme shows the viscoelastic force, $\mathbf{F}_{ve}$, the surface tension force, $\mathbf{F}_{\alpha}$, pointing to the center of curvature to restore a rectilinear shape, and the Coulomb repulsive term, $\mathbf{F}_C$, which is in opposite direction with respect to the local center of curvature, thus destabilizing laterally the filament. The cone at the top of the figure represents the nozzle, whereas the external electric potential bias, $\Delta\Phi$, provides a field indicated by the red arrow, and the electric force term, $\mathbf{F}_{\Phi}$, on the jet beads.

The considerations above are also valid within the framework of parallel computing. Two main paradigms are usually pursued for exploiting the parallelism. The first is Replicated Data (RD) strategy, where fundamental data of the simulated system are reproduced on all processing nodes. The second paradigm is based on the Domain Decomposition (DD) strategy, which exploits the decomposition of the space in sub-domains distributed over several processors, so that each process deals with only its sub-set of particles. The RD strategy provides satisfactory results of parallel efficiency in simulations on up to 100 processors, involving up to 30,000 particles (Smith *et al*., 1996). Given the small number of particles used to represent the jet, practical cases (Lauricella *et al*., 2015a; Lauricella *et al*., 2016b) have shown that such volume of data is by no means prohibitive on current parallel codes exploiting the RD strategy. However, all floating-point operations should be distributed in equal portion (as much as possible) for each processor in order to balance the load over participated processors. The load balancing is crucial since it guarantees that the computational work is performed cooperatively and simultaneously on all processing nodes.

In the parallel framework, all the *n*-body force terms, such as Coulomb intra-jet forces, must be obtained as a global sum of the contributing terms calculated over all nodes. As a consequence, a communication overhead is paid whenever the forces should be updated. Both DD and RD strategies pay a time lag, which depends on the information size to be communicated. Nonetheless, the exchanged information could be collected in smaller pieces in order to mitigate the latency time in data communication. For instance, the aforementioned fast multipole method may be used to pass cluster information of the repulsive Coulomb interaction, so decreasing the communication data size. Although this is just a flavour of the type of problems encountered in parallel coding, an efficient numerical implementation should always consider a fair trade between the largest distribution of computational work and the smallest data communication between processors.

To describe electrified and solution-blown jet, a number of parameters from experiments are generally needed. In addition, different models might have specific region of validity. For instance, the validity of polymer network modeling (Section V.D) is restricted to the initial stage of the jet, nearby the Taylor cone, where elastic elongation is still possible. Assuming mass conservation in this region of the jet, details are required relating to jet radius, velocity, and strain rate. Also, data on the polymer solution is required such as the shear viscosity, concentration, and type of solvent (e.g, $\theta$-solvent). The voltage applied in electrospinning, needle-collector distance, needle internal diameter, bulk velocity of the polymeric fluid in the needle, as well as the mass and charge density, zero-shear viscosity, elastic modulus, and surface tension of the used polymer solutions are generally the parameters needed as input of simulation tools based on Lagrangian models of electrospinning (Lauricella *et al*., 2015a). Nonetheless, simulating advanced electrospinning experiments may require further input parameters. As an example, magnetic and gas-assisted electrospinning simulations (Section V.C) need the magnetic field and gas velocity flow field as an extra parameter, respectively. Whenever the case, additional input parameters will be explicitly introduced in the following Sections.

# V. MODELLING METHODS FOR ELECTROSPINNING

In the following we present the current status of modelling methods developed for electrified polymer solution jets, not just as a list of sequential findings but with a logical flow in which the historical background will be accompanied by a critical review of research progress.

## A. Electrohydrodynamic models

### 1. Taylor cone

As explained in Section II, the liquid in a static drop attached to an electrode located at some distance from a counter-electrode can be considered as a perfect ionic conductor, even though its conductivity might be very low (Taylor, 1964; Ramos and Castellanos, 1994a). This means that the excess anions or cations have enough time to escape to the droplet surface, on which they are distributed non-uniformly and such to maintain zero electric field inside the liquid. This means that that droplet surface is equipotential. The shape evolution of small droplets attached to a conducting surface and subjected to electric fields was studied both experimentally and numerically (Reznik *et al*., 2004, Collins *et al*., 2008). The following scenarios can be distinguished for either perfectly conducting (Reznik *et al*., 2004) or moderately conducting (Collins *et al*., 2008) drops:

(i) in sufficiently weak electric fields, droplets are stretched by the electric Maxwell stresses in air, and take steady-state shapes, where equilibrium is achieved by means of surface tension;

(ii) in stronger electric fields, the Maxwell stresses overcome surface tension, and jetting can be initiated.

Conditions corresponding to the situation (i) ad (ii) above can be named as 'subcritical' and 'supercritical', respectively. The ultimate equilibrium droplet shape (i.e., the 'critical' condition) resulting from the competition of electric and surface forces reveals the Taylor cone configuration (Saville, 1997; Yarin *et al*., 2001b; Yarin *et al*., 2014). This corresponds to a solution in power-law form of the Laplace equation for the electric potential ($\Phi$) in the surrounding air. Such solution is scaling-invariant over different length scales of its variables, a property consisting in reproducing itself at different time and space scales (Barenblatt, 1994), which is also called self-similarity.

An axisymmetric liquid body kept at a potential ($\varphi_0$+const) with its tip at a distance, $d_0$, from an equipotential plane (that might be the counter-electrode in electrospinning experiments) is considered in Figure 11a, where the distribution of the electric potential, $\Phi = \varphi$+constant, is described in terms of the spherical coordinates, $R$ and $\theta$ (Yarin *et al*., 2001b). The potential, $\varphi_0$, can always be expressed in terms of the surface tension coefficient, $\alpha$, and of $d_0$, i.e. $\varphi_0 = C\left(d_0\alpha\right)^{1/2}$, where $C$ is a dimensionless factor, which follows from the dimensional analysis. In addition, also due to the dimensional arguments, the general representation of the electric potential, $\varphi$, is $\varphi = \varphi_0 F_1\left(R/d_0, \theta\right)$, where $F_1$ is a dimensionless function of the $R/d_0$ ratio and of the $\theta$ angular coordinate (Yarin, 2012). Hence, the value of the potential, $\Phi$, throughout the space that surrounds the liquid droplet is given by:

$$\Phi = \left(d_0\alpha\right)^{1/2} F\left(R/d_0, \theta\right) + \text{constant} , \tag{1}$$

where $F = CF_1$ is a dimensionless function. At distances $R \gg d_0$, equivalent to the mathematical assumption $R \to \infty$, one can assume that the influence of the gap, $d_0$, is small. Under this condition, the function $F$ should approach a power-law scaling form (Yarin, 2007):

$$F\left(R/d_0, \theta\right) \approx \left(R/d_0\right)^{1/2} \Psi\left(\theta\right), \tag{2}$$

where $\Psi\left(\theta\right)$ is a dimensionless function. Finally, Eq. (1) takes the asymptotic self-similar form, independent of $d_0$:

$$\Phi = \left(\alpha R\right)^{1/2} \Psi\left(\theta\right) + \text{constant} . \tag{3}$$

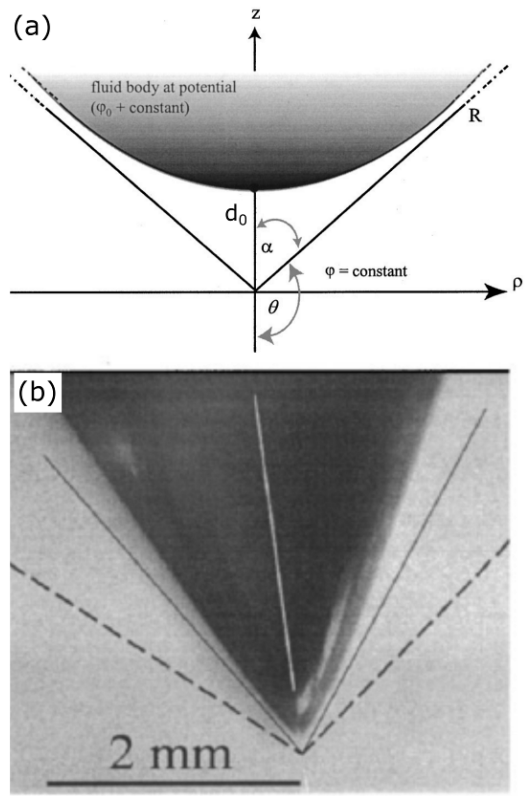

**Figure 11.** (a) Axisymmetric, 'infinite' fluid body kept at potential, $\Phi_0 = \varphi_0$+const at a distance $d_0$ from an equipotential plane sustained at $\Phi$ = const. (b) Critical shape observed for a pendant droplet in this configuration. The half-angle associated with the self-similar solution for the Taylor cone is indicated by the dashed lines. The tangent to the experimentally observed droplet of polymer solution has half-angle 31° (shown by solid line). Adapted with permission (Yarin *et al*., 2001b). Copyright © 2001, American Institute of Physics.

The solution in Eq. (3) should also satisfy the Laplace equation (Landau *et al*., 1984; Smythe, 1989, Feynman *et al*., 2006). Thus, one finds the function $\Psi$ (Taylor, 1964):

$$\Psi\left(\theta\right) = P_{1/2}\left(\cos\theta\right), \tag{4}$$

where $P_{1/2}$ is the Legendre function of order 1/2, which takes values $P_{1/2}\left(\cos\theta\right) = 0$ whenever $\theta = \theta_0$, with $\theta_0$ being the angle value matching the equipotential condition $\Phi = \text{constant}$. In other words, the free surface becomes equipotential only when $\theta$ corresponds to the single zero of $P_{1/2}(\cos\theta)$ in the range $0 \leq \theta \leq \pi$, which Taylor found to be equal to $\theta_0 = 130.7°$ (Taylor, 1964) using the tabulated values (Gray, 1953) of the function $P_{1/2}(\cos\theta)$. According to the self-similar equilibrium solution, the fluid body should so be enveloped by a cone with the half-angle at its tip, $\alpha_T = \pi - \theta = 49.3°$, i.e. the half-angle value associated with the Taylor cone (Figure 8b). Other studies on self-similar solutions analyzed dielectric liquids with varied permittivity (Ramos and Castellanos, 1994a; 1994b), finding that stationary cones can be formed only for $\varepsilon > 17.6$ (in CGS). By including the effect of the deviation of the surface shape from the conic one, such inferior limit for the permittivity was later increased to 22.2, corresponding to a half-angle value 39.25° (Zubarev, 2002). Also, the local behavior of various physical quantities (e.g. fluid velocity, surface curvature, electric current) could be determined, analyzing the nonlinear dynamics of the liquid cone for inviscid incompressible fluids (Zubarev, 2002; 2006; Belyaev *et al.*, 2019).

Power-law scalings resulting in self-similar solutions as that shown in Eq. (3) are common in the boundary-layer theory (Schlichting, 1979; Zel'dovich, 1937, collected in Zel'dovich, 1992; Yarin, 2007). In particular, such self-similar solutions for jets and plumes, considered as issuing from a pointwise origin, in reality correspond to the non-self-similar solutions of the boundary-layer equations (the Prandtl equations) for jets and plumes that are issued from finite-size needles, at distances much larger than the needle size (i.e., such solutions constitute remote asymptotics). For instance, the self-similar solution for capillary waves generated by a weak impact of a droplet onto a thin liquid layer emerges at distances from the center of impact much greater than the droplet diameter (Yarin and Weiss, 1995; Yarin *et al.*, 2017). The self-similar solution in Eq. (3), motivated by the same idea, was expected by Taylor to correspond to the limiting behavior of all non-self-similar solutions at $R \gg d_0$. In fluid dynamics, there are several contexts where self-similar asymptotics can be experimentally realised. For example, fluid flows involving boundary layers near a solid wall or free flows show the self-similar asymptotics, even though the experimental setup contains some details that are non-compliant with the assumptions of self-similarity (Yarin, 2007). In this situation, one can say that the flow pattern evolves to the self-similar solution, i.e. that a dynamic system evolves towards its attractors. Otherwise speaking, the non-self-similar solutions (e.g., the initial flow pattern) 'are attracted' by the self-similar solution. Indeed, the fact that the self-similar behavior can be experimentally realized directly evidences that it attracts the initially non-self-similar fluid flows, a behavior that is usually consistent with physical phenomena governed by parabolic PDEs for submerged jets (Schlichting, 1979; Yarin, 2007) and plumes (Zel'dovich 1992). On the other hand, the self-similar Taylor's cone solution, stemming from the elliptic Laplace equation, significantly disagrees with the experimental data (Yarin *et al.*, 2001b; Yarin *et al.*, 2014) for electrified polymer jets, leading to the conclusion that realizing the self-similar solution could be experimentally impracticable. It was also shown, by means of numerical simulations, that it does not attract the transient evolution at the tip of the fluid cone (Reznik *et al.*, 2004). An analogous observation on a self-similar solution that does not attract the corresponding non-self-similar one was found in the problem described by the biharmonic (elliptic operator squared) equation, namely, in the case of a wedge subjected to a concentrated couple of forces at its tip. This is known as the Sternberg-Koiter paradox (Sternberg and Koiter, 1958; Barenblatt, 1996) in the theory of elasticity. In electrospinning, an approximate non-self-similar solution was found instead in the form of a prolate hyperboloid of revolution (Yarin *et al.*, 2001b; Yarin *et al.*, 2014), which has finite curvature at the tip. Readers are referred elsewhere for the detailed explanation of the procedure used to find this solution (Yarin *et al.*, 2001b). Here we recall that the half-angle at the tip of the cone to which the hyperboloid leans is 33.5°, which is significantly smaller than the angle originally obtained for the Taylor cone (49.3°), and quite close to experimental data (Figure 11b).

In experiments with water (Taylor, 1964), it would be very difficult to approach the critical drop shape, because perturbations disrupt the equilibrium much earlier thus making impossible to measure the critical half-angle at the tip accurately. On the other hand, in experiments with polymer solutions, perturbations are suppressed by the viscoelastic behaviour of the fluid jet (Yu *et al.*, 2006), and equilibrium can be approached closely (Figure 11b). Figure 12 shows the predicted and measured shapes of a polycaprolactone (PCL) droplet at different moments. The numerical predictions slightly underestimate the stretching rate, but the overall agreement is fairly good. The early supercritical regime leads to jets generated from the cones with half-angles of 25° to 30°, which would support the assumption that the critical droplet shapes are close to those predicted with half-angle of 33.5° rather than to 49.3° in this specific regime (Reznik *et al*., 2004). However, this statement should be made with caution, because in early supercritical dynamical cases half-angles can be smaller because of the presence of the jet protrusion.

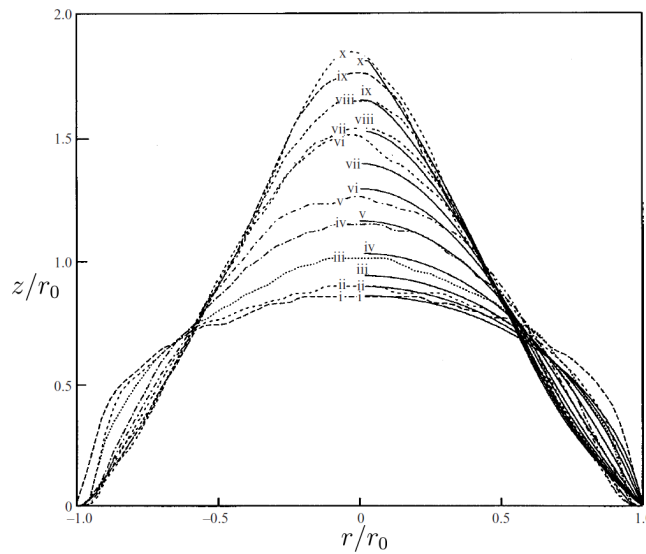

**Figure 12.** Measured and predicted shapes (vertical coordinate $z$, *vs.* radial coordinate) of a PCL droplet at different times, during deformation under an applied electric field: (i) $t$=0, (ii) 101.5, (iii) 201.5, (iv) 351.5, (v) 501.5, (vi) 601.5, (vii) 651.5, (viii) 701.5, (ix) 731.5, (x) 756.5 ms. Both $r$ and $z$ are rescaled by the initial base radius, $r_0$, of the droplet. Results from calculations are shown by solid lines in the right side of the drop (numerals located at their tip points). The experimental shapes are plotted as dotted lines. On the left side of the drop, the values of the radial coordinate (here, $r$) are made negative. Adapted with permission (Reznik *et al.*, 2004). Copyright © 2004, Cambridge University Press.

Furthermore, this dynamics strongly depends on the viscoelastic properties of the used fluid. For instance, experiments on drop evolution in a high-voltage electric field were also reported with water (Zhang and Basaran, 1996), showing a behavior quite different from that of highly viscous and elastic fluids used in electrospinning. For low-viscosity liquids, i.e. with $Ca$ in the range $10^{-5}$-$10^{-3}$ (Zhang and Basaran, 1996), electrospraying can occur i.e. the dripping regime is easily reached as described in Section II. Sometimes tiny droplet emission from the cone tip begins at half-angles close to 45° (Michelson, 1990), sometimes close to 49° (Fernandez de la Mora, 1992; 2007). The dynamical evolution of the fluid cone in presence of an electric field was theoretically investigated in the context of liquid-metal ion sources, and found to reach an angle equal to the Taylor value (Zubarev, 2001; Suvorov and Zubarev, 2004; Boltachev *et al*., 2008). The cone dynamics was also studied for a perfectly conducting liquid, in order to describe the mode of drop formation changing from simple dripping to so-called microdripping (Notz and Basaran, 1999). Furthermore, the critical droplet shapes and the dripping regime from a liquid film of finite conductivity were carefully studied and numerically simulated in the framework of electrospray ionization, showing that EHD tip streaming phenomena do not occur if the liquid is perfectly conducting or perfectly insulating and highlighting a universal scaling law for the size and for the charge carried by the droplets that are emitted (Collins *et al*., 2007; Collins *et al*., 2013). Recent works then aimed at including also the effects of charge relaxation in order to catch the transient EHD response in low-conductivity fluids (Pillai *et al*., 2016; Gañán-Calvo *et al*., 2016).

In general, dripping from low-viscosity liquids can lead to significant space charge from these tiny droplets. The backward electric effect of the charged droplets on the tip of the cone was shown to lead to the half-angle in the 32°-46° range (Fernandez de la Mora, 1992). On the contrary, since breakup of tiny threads and filaments is generally prevented by viscoelastic effects in electrospinning (Yarin, 1993; Reneker *et al*. 2000; Yarin *et al*., 2001a; Yarin *et al*., 2014), it is highly unlikely that the half-angle values found in the experiments with electrospun polymeric liquids can be attributed to space charge effects.

## 2. Onset of electrified jets

Electrospun jets are straight in their initial path (Figure 13), where the growing bending perturbations are still very small (Reneker *et al*., 2000; Yarin *et al*., 2014). Indeed, the longitudinal stress ($\sigma$) due to the external electric field (acting on the charge carried by the jet) stabilizes the fluid filament for some distance from the nozzle (Reneker *et al*., 2007). A similar trend was also noted in uncharged fluid threads of cylindrical shape, where the time of growth of the hydrodynamic instability was reported to scale logarithmically with the strain rate in hyperbolic extensional flow (Khakhar and Ottino, 1987). Furthermore, until the jet is sufficiently thick, it has a high bending stiffness, since this stiffness scales with the jet cross-section radius as $\sim r^4$ (Reneker *et al*., 2000; Yarin *et al*., 2014). The combined effects of the bending stiffness and of the longitudinal stress so lead to the initial straight path of the jet. A detailed investigation, including measurements of longitudinal stresses and thinning rates on a 6 wt% aqueous solution of PEO, found that longitudinal stress rescaled to the capillary pressure, ($\sigma \cdot r / 2\alpha$) span approximately from 10 up to 100 in the straight path zone (Han *et al*., 2008).

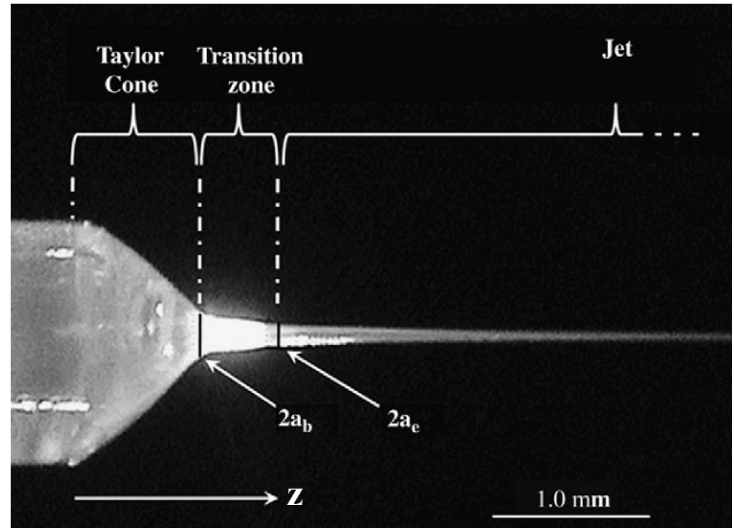

**Figure 13.** Modified Taylor cone, with the transition zone, and the beginning of a thin, electrified polymer solution jet. Here the cross-sectional radii at the beginning and at the end of the transition zone are here denoted $a_b$ and $a_e$, respectively. These quantities are measured by analizing a 6 wt% aqueous solution of PEO (solution density $\cong 1$ g/cm$^3$, zero-shear viscosity 50 g cm$^{-1}$ s$^{-1}$, surface tension = 61 g/s$^2$. Adapted with permission (Han *et al*., 2008). Copyright © 2008, Elsevier Ltd.

This straight region is relatively easier to be investigated experimentally, compared to that of the droplet-jet transition. Therefore, the theoretical description of the straight jet profile attracted significant attention. For describing the flow near the origin of the electrified jet, it is natural to use quasi-1D equations (Yarin, 1993; Yarin *et al*., 2014) and to consider jets as slender bodies with slow changes of the cross-sectional radius ($r$) in the longitudinal direction, as illustrated in Section IV. This approach was adopted in a number of works for Newtonian fluids (Taylor, 1966; Melcher and Warren, 1971; Kirichenko *et al*., 1986; Li *et al*., 1994; Gañan-Calvo, 1997a; 1997b; 1999; Cherney,1999a; 1999b; Stone *et al*., 1999; Hohman *et al*., 2001a; Feng, 2002; 2003; Fridrikh *et al*., 2003; Barrero and Loscertales, 2007; Gañan-Calvo, 2007, Gañan-Calvo *et al.* 2007). The regime of steady, straight stretching, and particularly the asymptotic behavior at sufficiently large distance from the spinneret, was also studied for non-Newtonian fluids with rheology described by the Oswald-deWaele power law (Spivak and Dzenis, 1999; Spivak *et al*., 2000). Other approaches, incorporating empirical models for the elongational viscosity (Feng, 2002) or the Giesekus constitutive equation (Feng, 2003) in the slender body theory, were also developed.

In solving the quasi-1D equations, solutions for the jet flow should also be matched with the droplet or meniscus region. For example, this could be achieved by the standard method of matched asymptotic expansions (van Dyke, 1964). In this way, one could match the jet flow with a conical semi-infinite meniscus (Cherney,1999a; 1999b). A drop shape with Taylor cone of 49.3° was chosen, which could be rather questionable, as described in the previous Section. Moreover, complete asymptotic matching was not achieved, as the solutions for the velocity, the potential and the field strength, and the free-surface configuration are all discontinuous (Cherney, 1999a; 1999b). A formal inconsistency of Cherney's analysis was pointed out in a later study (Higuera, 2003). Approximate approaches were largely tested in order to tackle these difficulties, particularly by extending the quasi-1D jet equations through the entire droplet up to its attachment to the needle (Gañan-Calvo, 1997a; 1997b; 1999; Hohman *et al*., 2001a; Feng, 2002; 2003). This is quite reasonable as a first approximation, however one should keep in mind that the flow in the drop region might have 2D character. Another complication arises from the electric part of the problem, where the image effects (Hohman *et al*., 2001a) at the solid wall should be taken into account. In fact, the electrical pre-history effects, namely, the image effects responsible for a detailed electrode shape etc., were found to be important only in a very thin boundary layer (Feng, 2002), adjacent to the cross-section where the initial conditions are imposed (i.e., at the needle exit). Accordingly, the quasi-1D equations (Feng, 2002) could be applied moving the jet origin to a cross-section in the droplet (to the distance of the order of the apparent height of the droplet tip). Based on this idea, the flow in the jet region was matched to the one in the droplet (Reznik *et al*., 2004). The electric current-voltage characteristic, $I=I(\Delta\Phi_0)$, was predicted in this way, as well as the flow rate, $Q$, in electrospun viscous jets. The predicted $I(\Delta\Phi_0)$ dependence is nonlinear due to the convective mechanism of charge redistribution superimposed on the conductive (ohmic) one. Several other 2D calculations of the transition zone between a droplet and the electrically pulled Newtonian jet were published (Hayati, 1992; Higuera, 2003; Yan *et al*., 2003; Reneker *et al*., 2000; Han *et al*., 2008; Yarin *et al*., 2014), studying the straight part of electrified, viscoelastic jets, and highlighting huge elastic stresses, and thus, elongational viscosity, which is an additional stabilizing factor disfavouring the early onset of bending instabilities, as well as suppressing the capillary instability.

## 3. Electro-hydrodynamic behavior

Models highlighting EHD effects in jets benefitted from studies in the fundamental physics of uncharged, fluid cylindrical threads moving in air or other fluids (Tomotika, 1936; Kase and Matsuo, 1965; Matovich and Pearson, 1969; Khakhar and Ottino, 1987). Results developed in these contexts were supplemented by elements of electro-hydrodynamics to analyze charged fluids stressed by electric fields (Saville, 1970; 1971a; 1971b). Both axisymmetric (Bassett, 1894) and non-axisymmetric (Saville, 1971b) modes were predicted for charged viscous cylindrical threads, and viscous effects were found to damp axisymmetric deformations (responsible for the capillary instability) more than non-axisymmetric ones (Saville, 1971b). In particular, following experimental insights (Magarvey and Outhouse, 1962; Huebner, 1969; 1970), Saville theoretically observed as an increasing amount of electrical charge carried by a liquid jet causes the amplification of non-axisymmetric disturbances, today commonly known as bending/whipping instabilities, which imprint a sinusoidal shape on the cylindrical thread (Saville, 1971b). Later on, several groups investigated the stability of non-axisymmetric modes in electrified cylindrical jets under an external electric field (Reneker *et al*., 2000; Yarin *et al*., 2001a; Hohman *et al*., 2001a; 2001b; Shin *et al*., 2001; Fridrikh *et al*., 2003; Li *et al*., 2013). Extending the overview to the dripping regime, Collins and co-workers (Collins *et al.*, 2007) numerically probed the breakup of electrified jets in the range of slightly viscous and moderate viscous jets, corresponding to capillary number values ∼ $10^{-3}$ and $10^{-1}$. Further, López-Herrera and co-workers investigated the capillary jet breakup of conducting liquids with different viscosities, under external electrostatic fields (López-Herrera *et al.*, 1999; 2004). In the framework of electrospinning, starting from previous 1D approximation of the Navier–Stokes equation (Eggers and Dupont, 1994; Eggers, 1997), Hohman and co-workers (Hohman *et al.*, 2001a; 2001b) extended the Saville's model in order to account for the presence of surface charge on the jet with a finite conductivity between two plane metallic electrodes, leading to a $E_z$ axial component of the electric field. The jet was schematized as shown in Figure 14, namely as a continuous slender body with surface charge density, $\sigma_q$, and the fluid assumed to have permittivity, $\varepsilon$, density, $\rho$, and kinematic viscosity, $\nu$. The dynamics was described in a gravitation field with acceleration, $g$, directed along the jet axis, and in a surrounding medium with dielectric constant, $\varepsilon_{air}$ (free space outside the jet). The following quasi-1D equations account for the mass conservation (the continuity equation), the charge conservation, the momentum balance (following from Navier-Stokes) of a fluid element along a cylindrical jet of radius $r$, and the electrostatic potential, $\Phi$, between the parallel electrodes, respectively (Hohman *et al.*, 2001a):

$$\frac{\partial}{\partial t}(\pi r^2) + \frac{\partial}{\partial z}(\pi r^2 \upsilon) = 0 \; , \tag{5}$$

$$\frac{\partial}{\partial t}(2\pi r \sigma_q) + \frac{\partial}{\partial z}(\pi r^2 \sigma_e E_z + 2\pi r \sigma_q \upsilon) = 0 \; . \tag{6}$$

$$\frac{\partial \upsilon}{\partial t} + \frac{\partial}{\partial z}\left(\frac{\upsilon^2}{2}\right) = -\frac{1}{\rho}\frac{\partial p}{\partial z} + g + \frac{2\sigma_e E_z}{\rho r} + \frac{3\nu}{r^2}\frac{\partial}{\partial z}\left(r^2 \frac{\partial \upsilon}{\partial z}\right), \tag{7}$$

$$\Phi(z,r') = \Phi_\infty - \left[\frac{1}{\varepsilon}\sigma_q \, r - \frac{(\varepsilon - \varepsilon_{air})}{2\varepsilon_{air}}\frac{d(E_z r^2)}{dz}\right]\ln\frac{r'}{L}, \tag{8}$$

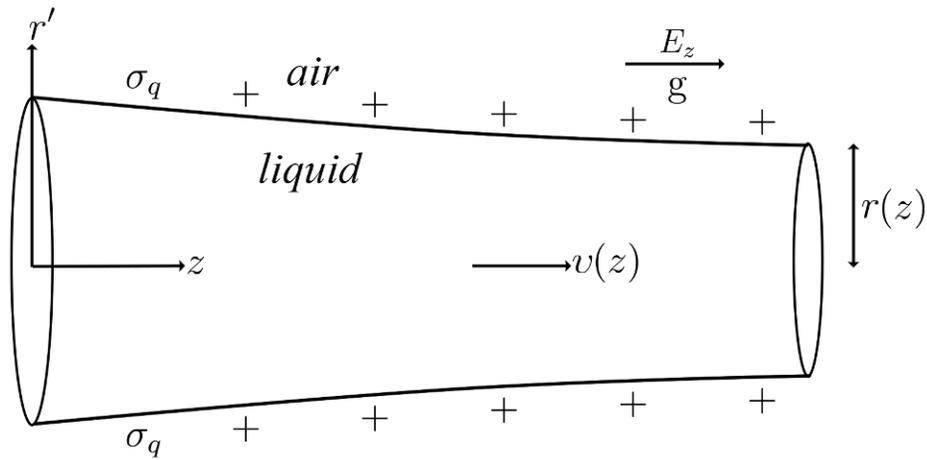

**Figure 14.** Scheme of the continuous slender body representing the jet in Eqs. (5)-(8). The coordinates $(z, r')$ used in Eqs. (5)-(8) are reported alongside the main parameters employed in the model.

Here, $p$ is the internal pressure in the fluid, estimated by Hohman and co-workers as

$p = k\alpha - E^2[(\varepsilon - \varepsilon_{air})/2] - \sigma_q^2[(1/(2\varepsilon_{air})]$, with $k$ local curvature and $\alpha$ surface tension (since the pressure term also includes the Laplace contribution due to the surface tension). $\upsilon$ is the axial velocity of the jet (to the leading order constant in the jet cross-section). In Eq. (8), $L$ indicates the characteristic axial lengthscale (determined by the shape of a jet as it thins away from the nozzle), whereas $\Phi_\infty$ indicates the electrostatic potential at a very large radial distance ($r$'). Asymptotic descriptions of the EHD equations were developed for a quantitative comparison with experiments (Hohman *et al.*, 2001a). It was also noted that the Laplace contribution is itself depending on the jet radius, which could be explicitly taken into account, i.e.

$k\alpha = \alpha /[1 + (\partial r / \partial z)^2]^{1/2} \cdot \{(1/r) - (\partial^2 r / \partial z^2)/[1 + (\partial r / \partial z)^2]\}$, where $r(z,t)$ is the radius at the jet position $z$ at time $t$ (Lee, 1974). The electric current ($I$) carried by the jet is given by the sum of a bulk ohmic component and a surface advection component associated with the surface charge density, $I = \pi r^2 \sigma_e E_z + 2\pi r \sigma_q \upsilon$. Finally, the solvent evaporation and the jet viscoelasticity are neglected.

Using the linear stability analysis, three modes of the instability could be identified (Reneker *et al.*, 2000; Yarin *et al.*, 2001a; Hohman *et al.*, 2001a; Hohman *et al.*, 2001b; Shin *et al.*, 2001; Yarin et al., 2005): (i) an axisymmetric instability that extends the classical phenomenon of the Rayleigh capillary instability (Rayleigh, 1878) to the case of electrified jets; (ii) another instability mode that is also axisymmetric, and named 'conducting' since it is only found for fluids with finite, non-zero conductivity; (iii) a non-axisymmetric, bending/whipping instability, that is also electrically-driven, in which the jet axis bends but the cross-section stays circular. Branching instability is also found, which is the electrically-driven instability developing on the background of the bending instability, but with the jet cross-section acquiring multi-lobe shapes rather than staying circular.

In other works, the boundary conditions at the nozzle were analyzed in detail, and non-Newtonian rheology was considered (Feng, 2002, 2003), which is extremely important for polymer solutions. Furthermore, the effects of the electrical conductivity and viscoelasticity on the jet profile during the initial stage of electrospinning were examined in depth (Carroll and Joo, 2006). Viscoelasticity could be incorporated in the EHD equations by modelling the fluid rheology with the Oldroyd-B constitutive equation (Prilutski *et al.*, 1983; Mackay and Boger, 1987), or the Upper-Convected Maxwell (UCM) model (Reneker *et al.*, 2000; Yarin *et al.*, 2001a; Yarin *et al.*, 2014). Increasing the conductivity (which can be attempted, for instance, by adding a salt in the solution) or the fluid viscoelasticity resulted in delayed and more rapid jet thinning, respectively. Indeed, how fast the jet thins is complex and it also depends on other parameters, such as the applied potential difference, namely the electric field in the region of space where electrospinning takes place (Carroll and Joo, 2006). Axisymmetric instabilities, and particularly axisymmetric conducting modes, were re-analyzed for viscoelastic polymer solutions (Carroll and Joo, 2008, 2009). The EHD equations for viscoelastic fluids were also studied by using the PDE module in the COMSOL® Multiphysics software, that was also applied to analyze the electric field in a multi-jet configuration (Angammana *et al*., 2011a, 2011c).

## 4. Electrically-driven bending instability

Significant stretching and thinning of electrically-driven polymer jets diminishes bending stiffness (which, as anticipated in sub-Section V.A.2, is proportional to $r^4$, where $r$ is the local cross-sectional radius of the jet). Then, at some distance from the needle, non-axisymmetric perturbations begin to grow and the electrically-driven bending instability sets in (Reneker *et al*., 2000). A typical bending path is shown in Figure 4b. As explained in Section II, in this regime the characteristic hydrodynamic time is significantly shorter than the charge relaxation time. Under such condition the same liquid, which behaved as a perfect conductor in the Taylor cone, behaves as a perfect dielectric in the jet. Then, the conduction component of the electric current along the jet can be neglected, and the charge transport is attributed entirely to the jet flow (i.e., the charge is 'frozen' in the liquid, leading to a purely advection current).

The reason for the bending instability observed in experiments may be understood as follows (Reneker *et al*., 2000). In the frame of reference moving with a rectilinear electrified jet, the electrical charges can be regarded as static and mutually interacting according to Coulomb's law (without the external field). Such systems are known to be unstable according to Earnshaw's theorem (Jeans, 1958). Indeed, an off-axis misalignment rapidly triggers bending instability driven by electrostatics, with an exponentially increasing amplitude. To illustrate this mechanism, one can consider three point-like charges, each with a value of $q$ , and originally located on a straight line at *A*, *B*, and *C*, along the longitudinal axis of the jet, as shown in Figure 15. Two Coulomb forces with magnitudes $F = q^2 / \ell_1^2$ push against charge $B$ from opposite directions. Here, $\ell_1$ indicates the distance between two charges and the Coulomb constant is left out because $k_C = 1$ in CGS units (Sommerfeld, 1952). If a perturbation causes the charge $B$ to move off the line by a distance ($\delta$) to point $B$', a net force $F_1 = 2F \cos \beta = 2q^2 / d_C^3 \cdot \delta$, acts on that charge, where $\beta$ is the angle between the perturbed $AB$ and $CB$ directions and the perpendicular to the jet axis. This net force is acting in the direction perpendicular to the original jet line, and leads the charge initially at point $B$ to move further in the direction of the perturbation, namely away from the line between the two fixed charges, $A$ and $C$. The growth of the small bending perturbation in the linear approximation is then:

$$m \frac{d^2\delta}{dt^2} = \frac{2q^2}{\ell_1^3} \delta ,$$ (10)

where $m$ is the mass of the particle and $\ell_1$ is shown in Figure 15. The growing solution of Eq. (10), $\delta = \delta_0 \exp[(2q^2 / m\ell_1^3)^{1/2} t]$, highlights that small bending perturbations increase exponentially, sustained by the corresponding decrease of the electrostatic potential energy. If the charges, $A$, $B$, and $C$ are attached to a liquid jet, other forces that are associated with the liquid tend to counteract the electrically-driven instability. For very thin liquid jets, the influence of the shearing force related to bending stiffness ($\sim r^4$) can be neglected compared with the stabilizing effect of the longitudinal forces, $f_\ell$, that are of the order of $r^2$ (Yarin, 1993). In fact, the longitudinal force at the cross-section at the onset of the bending instability is determined by the end of the straight section of the jet. The forces, $f_\ell$, are directed along $BC$ or $BA$ according to the scheme in Figure 15, and are opposite to the local Coulomb force. If such Coulomb force, $F$, is larger than the viscoelastic resistance, the bending perturbation continues to increase, but at a rate diminished by $f_\ell$. To describe the viscoelastic response, the Maxwell model (Maxwell, 1867; Ferry, 1980; Morozov and Spagnolie, 2015) assumes that the dumbbell defined by each pair of particles supports a stress, $\sigma$, that changes as that of system with a spring (Hooke's law) and a damper having a relaxation time $\theta = \mu / G$, where $G$ is the elastic modulus and $\mu$ is the viscosity. In addition, the surface tension also counteracts the bending instability, because bending always leads to an increase in the jet surface area. Hence, surface tension resisting the reaching of significant curvature tends to limit the smallest possible perturbation wavelengths, although surface tension effects might be negligibly small compared to the electric and viscoelastic forces in electrospinning.

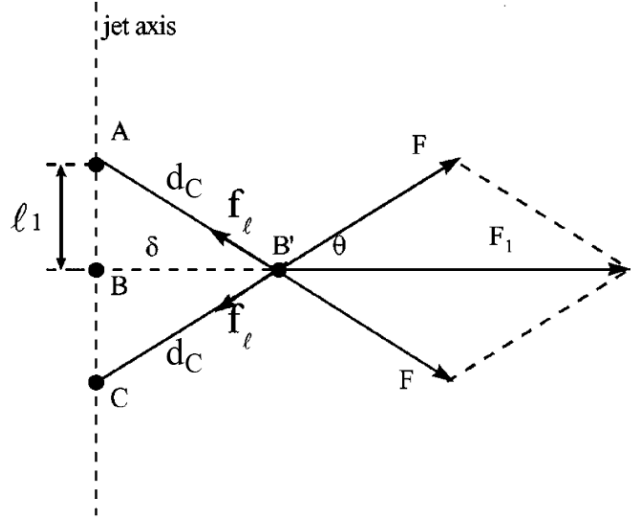

**Figure 15.** Sketch of the Earnshaw instability, leading to bending of electrified jets. Adapted with permission (Reneker *et al*., 2000). Copyright © 2000, American Institute of Physics.

The linear stability theory of bending in electrified polymer solution jets yielded the following characteristic equation for the growth rate ($\gamma$) of such perturbations (Yarin *et al*., 2001b):

$$\gamma^2 + \frac{3}{4} \frac{\mu \chi^4}{\rho r_0^2} \gamma + \left[ \frac{\alpha}{\rho r_0^3} - \frac{e_0^2 \ln(L / r_0)}{\pi \rho r_o^4} \right] \chi^2 = 0 , \tag{11}$$

where $\chi = 2\pi r_0 / \ell_P$ is the dimensionless wavenumber ($\ell_P$ is the wavelength of the fastest-growing perturbation), $\mu$ is the dynamic viscosity, $r_0$ and $e_0$ are the unperturbed values of the cross-sectional jet radius and of the electric charge per unit jet length, respectively, and $L$ is the cut-off jet length. This equation accounts for the shearing force and moment of forces in the jet cross-section, namely for the bending stiffness. It also shows that the destabilizing electric force overcomes the stabilizing effect of the surface tension (the only stabilizing effect here) if $e_0^2 \ln(L/r_0) > \alpha \pi r_0$. If this condition is not fulfilled, the capillary and bending instabilities are concurrent, as indeed observed in experiments with viscous organic oils (Malkawi *et al.*, 2010) and shown in Figure 16. More details about these aspects are reviewed in sub-Section V.B.2.

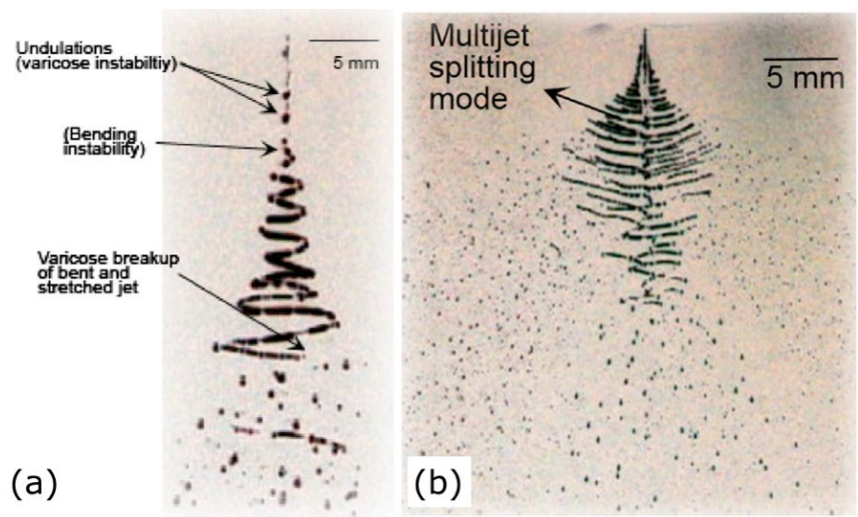

**Figure 16.** Breakup of electrified corn oil jets, generated with a fixed flow rate of 30 mL/min (corresponding to a velocity of 10 m/s at the injection orifice) and applied voltages -8 kV (a) and -10 kV (b) applied between the needle orifice and the nozzle of an atomizer. Adapted with permission (Malkawi *et al.*, 2010). Copyright © 2010, American Institute of Physics.

The nonlinear theory of the bending instability of viscoelastic polymeric jets in electrospinning was developed by the Reneker and Yarin groups (Reneker *et al*., 2000; Yarin *et al*., 2001b; Yarin *et al*., 2014). As explained above, for very thin jets the effect of the shearing force, as well as the bending stiffness, can be neglected (Yarin, 1993). Then, in this momentless approximation, a curvilinear parameter, *s*, can be introduced. It might be considered as a Lagrangian coordinate reckoned along the jet, and it takes the values $s = 0$ and $s = \ell_0$ at the nozzle and at the free jet termination, respectively. Here, $\ell_0$ is the initial arc length of the jet, and any value of $s \in (0, \ell_0)$ represents a jet element in an univocal way (the *s* value is 'frozen' in the jet elements). Thus, the quasi-1D continuity equation and the momentum balance equation take the form:

$$\lambda f = \lambda_0 f_0 \tag{12a}$$

$$\rho \lambda_0 f_0 \frac{\partial \mathbf{v}}{\partial t} = \mathbf{\tau} \frac{\partial P}{\partial s} + \lambda |k| P \mathbf{n} - \rho g \lambda_0 f_0 \mathbf{k} + \lambda |k| \left[ \alpha \pi r_0 - e_0^2 \ln \left( \frac{L}{r_0} \right) \right] \mathbf{n} - \lambda e_0 \frac{\Delta \Phi}{h} \mathbf{k} \tag{12b}$$

Eq. (12a) is the continuity equation, where $\lambda$ is the geometrical stretching ratio, i.e. $\lambda = | \partial \xi / \partial s |$, with $\xi(s,t) \in [0, \ell]$ being the arc length reckoned along the (bent) jet axis, and *f(s,t)*=$\pi r^2$ the cross-sectional area. The cross-section can be assumed to stay circular even in bending jets, which is a plausible approximation (Yarin, 1993). The fluid is incompressible (fluid volume preserved). The subscript zero denotes the parameter values at time *t*=0, namely the unperturbed values. Eq. (12b) expresses the momentum balance, with $\mathbf{v}(s,t)$ being the liquid velocity, $P(s,t) = (\lambda_0 f_0 / \lambda) \sigma(s,t)$ the longitudinal force in the jet cross-section (of viscoelastic origin in the case of electrospun polymer jets, or solution-blown jets), *g***k** gravity acceleration (the unperturbed jet is, in general, implied to be in the vertical direction), *t* time, **n** the local principal normal unit vector of the jet axis. $\Delta \Phi_0 / h$ is the external electric field intensity (the external field is assumed to be parallel to the unit vector **k**, with $\Delta \Phi_0$ and *h* being

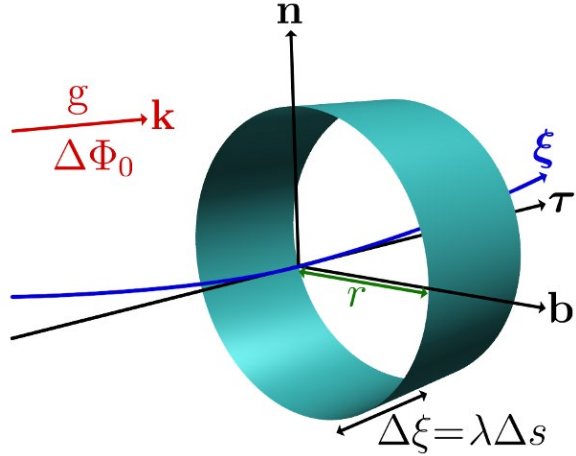

**Figure 17.** Sketch of an element of a curved jet of length $\Delta\xi$ and the associated internal frame of reference, with normal ($\mathbf{n}$), binormal ($\mathbf{b}$), and tangent vector ($\boldsymbol{\tau}$), used in Eq. (12).

the values of the electrical potential bias and the distance between the jet origin and a collector, respectively). The overall configuration is displayed in the scheme of Figure 17. On the right-hand side of the momentum Eq. (12b), the following forces are accounted for: the longitudinal internal force of rheological origin acting in the jet (the first two terms), gravity (the third term), the bending electric force *vs*. the stabilizing effect of the surface tension (the fourth term), and the force due to the external electric field (the fifth term).

## B. Lagrangian models

This sub-Section reviews the general features of the Lagrangian formulation for modelling of electrospinning. The advantages of the Lagrangian formulation are summarized at first, contextualizing some aspects anticipated in Sections IV and V.A, through some electrospinning models based on it. Then, several application scenarios are reviewed, paying specific attention to physical insight as obtained in the theoretical framework.

# 1. Why the Lagrangian formulation?

It is possible to write the continuity equation, quasi-1D in Eqs. (12), in the 3D framework, allowing for the simulation of the full electrospinning process by numerical solvers. At a first inspection of the force terms, it is easy to realize that, although the external electric field provides a global force, the other force terms driving the jet dynamics are mostly local [Eq. (12b)]. The locality would seem to suggest to numerically solve the continuity equation at specific positions in space as a function of time, i.e. with an Eulerian frame of reference. Nonetheless, the presence of constitutive relations between stress and strain, describing the deformation of the jet, makes the Eulerian description not optimal for the numerical solution of actual models. Indeed, jets flowing at high speed undergo stress-induced deformations which depend on rheology. As a consequence, a memory of such deformations is locally stored on the travelling jet as a 'fingerprint'. If If viscous forces do not dissipate the local signatures, these are transported by the flow field. The electrospinning process presents exactly these features, with a deformable interface moving rapidly over the space towards the collector counter-electrode.

The strength of the Lagrangian formulation is the retention of fluid element identity which is stored in memory with a set of observables (e.g. position, velocity, etc.) describing its state at a given instant. Instead, the Eulerian description of the jet over a grid of fixed points would suffer memory loss, because the jet elements are not tracked over the fluid dynamics.

Summarizing, the choice of the optimal description is usually dependent on how far a piece of local information is transported by the dynamics. As a consequence, it is not surprising that the Lagrangian formulation was mostly adopted in simulations of turbulent flows, where the ratio of inertial forces to viscous forces (Reynolds number) is of the order of several thousand (Meneveau *et al*.,1996; Bennett, 2006). Moreover, a number of Lagrangian methods have been introduced to treat several problems of fluid dynamics where a local state of a system is moved over the space. Among them, we recall molecular dynamics, dissipative particle dynamics, smoothed particle hydrodynamics (Gui-rong, 2003; Liu and Liu, 2016), to name a few.

In 1994, a first notable example of a Lagrangian particle method for the simulation of electrically driven fluid dynamics was presented in the context of electrospray modelling (Gañán-Calvo *et al*., 1994). This effort highlighted the efficiency of the Lagrangian formulation in modelling highly dispersed charged droplets, emitted from an electrified conical meniscus towards a collector. In this description, each particle represented a charged droplet with appropriate mass and charge.

An electrospun jet is a more complex system, which, however, lies in a 1D space over its arc length $\ell_{jet}$ (namely, the length of the curve drawn by the axis of the jet). As mentioned in Section IV, in 2000, the Reneker and Yarin groups introduced the first Lagrangian model for electrospinning (Reneker *et al.*, 2000), where the jet was treated as a 1D chain of particles, connected pair by pair through viscoelastic springs extending over the jet curve. In this framework, each particle represents a volumetric portion of the jet with given mass and charge. Each pair of two consecutive connected particles along the chain acts as a viscoelastic dumbbell, with the distance between the two extreme points modelling the stretch ratio of the jet (as well as the surface to volume ratio).

## 2. Quasi one-dimensional Lagrangian models

An electrospun fluid undergoes an increase in the tensile stress as it passes from the Taylor cone to the straight jetting region through a transient zone. The experimental data show that the rate of strain $\dot{\varepsilon}$ is of the order of 100-1000 s$^{-1}$ in this region, which provides an extremely high longitudinal viscoelastic stress (Han *et al*., 2008). This widely affects both the shape and charge distribution of the jet already at the initial stage of the dynamics, hence developing a reliable modelling for this zone is critically important. The charge distribution here formed has a substantial effect on the further evolution of the jet, considering that the bending instabilities are mainly related to the intra-jet repulsive Coulomb force. Thus, the Lagrangian discretization of the quasi-1D set of Equations (12a) and (12b), reported in sub-Section V.A.4 has been employed in several theoretical investigations (Reneker *et al.*, 2000; Kowalewski *et al.*, 2005; Carroll and Joo, 2011; Rafiei *et al.*, 2013; Lauricella *et al.*, 2015). In a 1D framework, given the $z$ axis and approximating the stretching ratio $\lambda = |\partial\xi/\partial s| \approx \ell_{jet}(t)/\ell_0$, the position of a jet element under stretching is traced by the relation $z(t) = \lambda(t)\,s$ in an univocal way. Thus, a discrete set of positions values $\{z_i\}_{i=1,\dots,n}$ with $z_i = \lambda\,s_i$ can be inserted in Eq. (12b), obtaining a Lagrangian discretization of the continuous object.

The simplest model exploits the jet discretization in two particle-like beads, (u,d) with the same mass $m$ and charge $q$ describing a charged drop (Figure 18). The upper bead (u) is held fixed to the nozzle at $z_u = 0$, while the lower one (d) is free to move and initially placed at distance $z_d = \ell_0$. Neglecting the gravitational force, the quasi-1D momentum balance equation along the unit tangent jet vector $\vec{\tau}$ for the free (d) bead reads:

$$m\frac{\partial \upsilon}{\partial t} = -\pi r^2 \sigma + q\frac{\Delta\Phi_0}{h} + \frac{q^2}{\ell^2}, \tag{13}$$

where $\ell = (z_d - z_u)$ is the mutual distance between the two beads equal to the entire jet length, $\ell = \ell_{jet}$, by construction, $r$ is the cross-sectional radius and $\Delta\Phi_0$ is, as usual, the difference in electric potential between the nozzle and the collector placed at distance $h$. Here, the stress force, $\pi r^2 \sigma$, is modelled by a constitutive equation. The UCM model is a plausible rheological representation for semi-dilute and concentrated polymeric liquids under uniaxial, strong elongational flows (Chang and Lodge, 1972; Stelter *et al*., 2000; Yarin *et al*., 2014). Denoting $G$ the elastic modulus and $\mu$ the viscosity of the fluid jet, the simplest UCM in its linear version, namely the Maxwell model (Maxwell, 1867; Ferry, 1980; Morozov and Spagnolie, 2015), reads:

$$\frac{\partial \sigma}{\partial t} = G\dot{\varepsilon} - \frac{G}{\mu}\sigma, \tag{14}$$

where $\dot{\varepsilon} = \frac{\partial \ell}{\ell \partial t}$ is the strain rate. Applying the condition of volume conservation $\pi r_0^2 \ell_0 = \pi r^2 \ell$ and the kinematic relation:

$$\frac{\partial z}{\partial t} = \frac{\partial \ell}{\partial t} = \upsilon \tag{15}$$

this simple model allows one to probe the initial dynamics of the electrified jet. In order to adopt a dimensionless form of the EOM, the dimensionless quantities reported in Table 2 can be conveniently used, so that, for instance, the dimensionless form of Eq. (13) reads:

$$\frac{\partial \overline{\upsilon}}{\partial \overline{t}} = F_{\upsilon e} \frac{\overline{\sigma}}{\ell} + V + \frac{Q_e^2}{\ell^2}. \qquad (16)$$

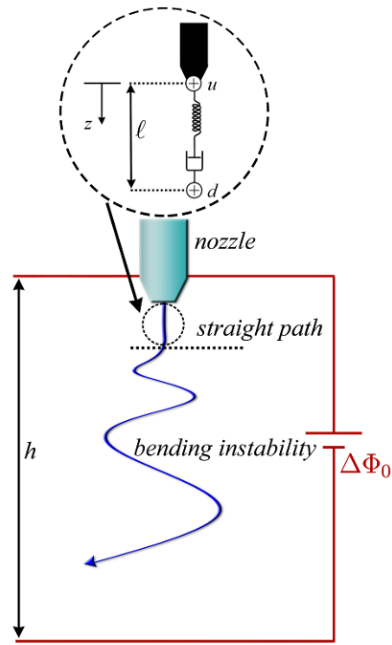

**Figure 18.** Schematic drawing of the electrospinning process (not in scale), highlighting the two particle-like beads used in the quasi-1D model. Adapted with permission (Lauricella *et al.*, 2015c). Copyright © 2015, Elsevier Ltd. The two beads are connected by a linear, viscoelastic dumbbell. $h$: distance between the collector plate and the nozzle. $\Delta\Phi_0$: applied voltage difference between these two elements. $z$: reference axis coordinate, whose origin is fixed at the injection point.

Using this Equation, Reneker and collaborators inspected the time evolution of the viscoelastic force term for the typical values of experimental relevance, $Q_e = 12$, $V = 2$ and $F_{\upsilon e} = 12$ (Reneker *et al*., 2000). As shown in Figure 19, the external electric field acts along all the dynamics, stretching the initial jet with magnitude proportional to the distributed charge (Lauricella *et al*., 2015c). While the dynamics proceeds, the stress force increases up to a peak value within the dimensionless time $\overline{t} \leq 1$, remaining the dominant force term up to $\overline{t} \sim 2$. As the stress reaches its peak value (Figure 19a), lasting about up to $\overline{t} \sim 1$, the velocity comes to nearly constant value, then reaching a linear regime in the jet length evolution in time, $\overline{\ell} \propto \overline{t}$ (Figure 19b). After this point, the stress force starts to decay due to viscoelastic relaxation with relaxation time $\theta$ and is no longer able of sustain the expanding 'pressure' of the electrostatic interactions (repulsive Coulomb and external potential forces). Thus, the linear regime cannot last long and the jet expansion is described by a free fall regime with the jet length evolving in time as $\bar{\ell} \propto \bar{t}^2$. In this portrait of regimes, the quasi-straight path is experimentally observed as the initial dynamics of the process (the linear regime). Here, the viscoelastic force not only provides stiffness necessary to maintain the jet straight, but it plays the dominant role in stabilizing it. Given the importance of the viscoelastic term, the model was extended in order to account for different non-Newtonian behaviors in the initial jet dynamics (Pontrelli *et al.*, 2014), particularly by adding a Herschel–Bulkley stress term in Eq. (14) and including the yield stress for the description of Bingham fluids (Bird *et al.*, 1987). In this way, Eq. (14) takes the form $\dot{\sigma} = (G / \mu)\left[\sigma - \sigma_Y + K(\dot{\ell} / \ell)^n\right]$, so that the effective viscosity is $\mu = K\left|\dot{\ell} / \ell\right|^{n-1}$, with $K$ a prefactor having dimension g s$^{n-2}$cm$^{-1}$ and $n$ a power-law exponent. In particular, whenever the ratio $\sigma_Y / G$ is set equal to 0.8, a halved value was found in the jet linear extension during the initial stage of electrospinning, due to an increase by nearly two times in the value of the longitudinal viscoelastic stress, $\sigma$ (Pontrelli *et al.*, 2014). As a consequence, the jet would soon start to show bending oscillations.

| Characteristic scales | |
|---|---|
| $L_0 = \ell_{step} \sqrt{\dfrac{\pi a_0^2 \rho_V^2}{G}}$ | $t_0 = \dfrac{\mu}{G}$ |
| $\sigma_0 = G$ | |
| **Dimensionless derived variables** | |
| $\overline{\ell}_i = \dfrac{\ell_i}{L_0}$ | $\overline{R}_{ij} = \dfrac{R_{ij}}{L_0}$ |
| $\overline{k} = kL_0$ | . |
| **Dimensionless groups** | |
| $V_i = \dfrac{q_i V_0 \mu^2}{m_i h L_0 G^2}$ | $Q_{e,ij} = \dfrac{q_i q_j \mu^2}{L_0^3 m_i G^2}$ |
| $F_{ve,i} = \dfrac{\pi a_0^2 \mu^2}{m_i L_0 G}$ | $A_i = \dfrac{\alpha \pi a_0^2 \mu^2}{m_i L_0^2 G^2}$ |
| $F_g = \dfrac{g \mu^2}{L_0 G^2}$ | $K_s = \omega \dfrac{\mu}{G}$ |
| $H = \dfrac{h}{L_0}$ | $L_{step} = \dfrac{\ell_{step}}{L_0}.$ |

**Table 2.** Definitions of the characteristic scales, dimensionless derived variables, and groups employed in the sub-Section V.B.2

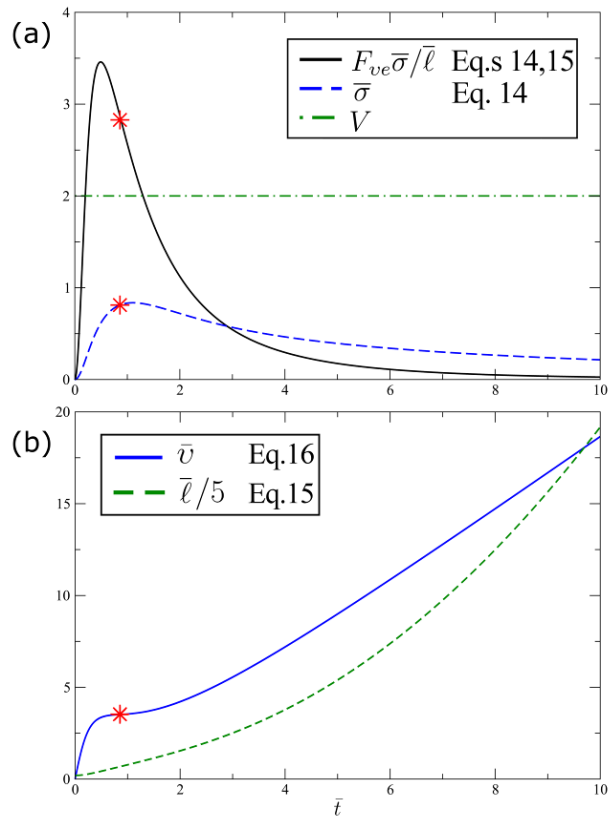

**Figure 19.** (a) The longitudinal force, $F_{ve}\bar{\sigma}/\bar{\ell}$ (continuous line), and the longitudinal stress, $\bar{\sigma}$ (dashed line), in the rectilinear part of the jet for the case $Q_e = 12$, $V = 2$, $F_{ve} = 12$. (b) Time evolution of the velocity $\bar{\upsilon}$ (continuous line) and the length $\bar{\ell}$ (dashed line), here rescaled by a factor 1/5 (dotted line) to have both quantities conveniently plotted with the same vertical axis. Two stages of the elongation process are observed. The first stage comes to a quasi-stationary point (star symbol in both the panels). In the second stage, the velocity comes to a near linearly increasing regime. Adapted with permission (Lauricella *et al.*, 2015c). Copyright © 2015, Elsevier Ltd.

In order to simulate the electrospinning process in its whole dynamics, one also needs to properly model the fluid injection at the nozzle. Several algorithms addressed this issue (Reneker *et al.*, 2000; Kowalewski *et al.*, 2009; Carroll and Joo, 2011; Lauricella *et al.*, 2015a). Let us consider a simulation starting again with two beads: a single mass-less point labelled $i = 0$ and fixed at $z_0 = 0$ representing the spinneret nozzle, and a second bead (labelled $i = 1$) modelling a pendant element of fluid of mass $m_1$ and charge $q_1$. The second bead is placed at the initial distance $z_1 = \ell_0 = \ell_{step}$ from the nozzle along the $z$ axis, with initial velocity $\upsilon_s$ equal to the bulk fluid velocity in the syringe needle. $\ell_{step}$ represents the length step used to discretise the jet in a sequence of beads. Once the travelling jet bead reaches the distance $z_1 = 2\,\ell_{step}$ away from the nozzle, a new jet bead ($i = 2$) is placed at a distance $z_2 = \ell_{step}$ from the nozzle with the initial velocity $\upsilon_2 = \upsilon_s + \upsilon_d$, where $\upsilon_d$ denotes the dragging velocity computed as $\upsilon_d = (\upsilon_1 - \upsilon_s) / 2$. The dragging velocity acts as an additional term accounting for the drag effect of the electrospun jet on the last inserted segment, which preserves the strain rate value at the jet point, $z_2$, before and after the bead insertion. Hence, the procedure is repeated until a chain of $n$ beads representing the jet is obtained (Figure 20). Note that the injection bead does not alter the curvilinear parameter, $s$, which preserves its domain, $s \in [0, \ell_0]$, but it only adds an extra element in the set of values used to discretize the jet path. On the other hand, the added bead represents a new jet parcel of volume $\ell_{step} \pi r_0^2$, mass $m_i$ and charge $q_i$, so that the extensive properties of the jet (total volume, mass and charge) increase after the insertion.

Adding the injection algorithm to Eq. (13), the momentum balance equation along the unit tangent jet vector $\boldsymbol{\tau}$ for the $i$−th bead reads:

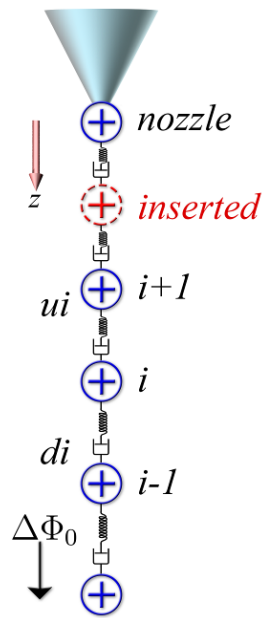

**Figure 20.** Diagram of the jet modelled as a chain of 1D discrete elements. Each element representing a jet segment is drawn as a circle with a plus sign denoting its positive charge. The $i$−th element is along the chain between the upper and downer dumbbell elements with extremes $(i+1, i)$ for the $ui$ and $(i, i-1)$ for the $di$ dumbbell, respectively, and it is stretched under the external electric potential difference, $\Delta\Phi_0$. The cone at the top of the figure represents the nozzle. Adapted with permission (Lauricella *et al.*, 2015b). Copyright © 2015, American Physical Society.

$$m_i \frac{\partial \upsilon_i}{\partial t} = -\pi r_{ui}^2 \sigma_{ui} + \pi r_{di}^2 \sigma_{di} + q_i \frac{\Delta \Phi_0}{h} + q_i \sum_{\substack{j=1,n \\ j \neq i}} \frac{q_j}{|z_i - z_j|^3}(z_i - z_j),$$ (17)

where the Coulomb repulsive force takes into account all the interactions with the other $n-1$ jet segments, and the tensile force is computed as the stress difference between the upper and downer dumbbell elements with extremes $(i+1, i)$ for the $ui$ and $(i, i-1)$ for the $di$ dumbbell, respectively, along the bead chain (Figure 20). In fact, the finite difference, $(-\pi r_{ui}^2 \sigma_{ui} + \pi r_{di}^2 \sigma_{di})$, approximates the derivative of the longitudinal force along the fiber, that is the term $\partial P / \partial s$ in the continuum description of Eq. (12b).

Results from this quasi-1D model were compared to experimental data (Carroll and Joo, 2011), finding some discrepancy in stable jet profiles, which was subsequently recovered by an amended version of the model. As a first point, the polymer fluid was described as an Oldroyd-B fluid, whose rheology is described using two distinct contributions. These include both a $\sigma_p$ stress term, from the viscoelastic dumbbell term (using the Maxwell model) due to the polymeric component, and $\sigma_s = \mu_s \dot{\gamma}$, from a Newtonian solvent in which the viscoelastic elements are immersed. Consequently, the total tensile stress, $\sigma = \sigma_p + \sigma_s$, is computed for each $i$−th bead and inserted in Eq. (17). Second, the liquid jet was considered as a leaky dielectric rather than a perfect conductor, thereby including the effect of finite conductivity in the fluid. Thus, the charge $q_i$ in Eq. (17) is replaced with an effective charge $q_{eff,i} = c_{eff,i} \, q_i$, where $c_{eff,i} = 1 - (I_{\text{conduction}} / I_{\text{total}})$ is the charge fraction on the jet surface which actually interacts with the electric field and the other charges. In particular, $c_{eff,i}$ is computed by setting the total current $I_{\text{total}} = I_{\text{convection}} + I_{\text{conduction}}$ as an input parameter and evaluating step by step at each $i$−th bead the quantity, $I_{i,\text{conduction}} = \pi r_i^2 \sigma_e \Delta \Phi_0 / h$, with $\sigma_e$ the electrical conductivity. With these amendments, the jet cross-section profiles $r(x)$ obtained by the Lagrangian model were found to be much closer to the experimental data (Carroll and Joo, 2011). This is illustrated in Figure 21, comparing the jet profiles of polyisobutylene Boger solutions at concentration of 4000 ppm in low molecular weight polybutene solvent. Moreover, a trend of the tensile stress was provided (Carroll and Joo, 2011), consisting in a rapid increase during the initial jet stretching and subsequent decrease due to the stress relaxation, which is in agreement with theoretical findings (Reneker *et al*., 2000). Thus, the comparison shows that the simplest two-body version of the Lagrangian model was already able to qualitatively catch the stretching process of the quasi-straight path in the jet. In a recent variant, the convection current term, $I_{i,\text{convection}} = 2\pi r_i^2 \upsilon_i \sigma_q$, is included in an extra ordinary differential equation of the charge, which is solved to compute the conduction charge fraction along the time evolution (Divvela and Joo, 2017).

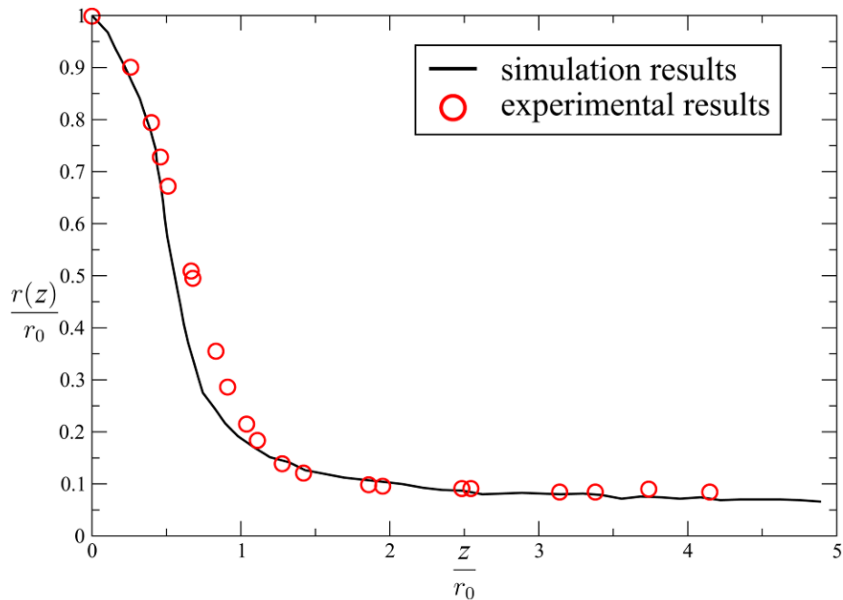

**Figure 21.** Radius profiles obtained from bead-spring simulations and from experiments, for the electrospinning of 4000 ppm polyisobutylene Boger fluid. The parameters used in the simulation are estimated from experiments (Carroll and Joo, 2006): jet initial radius $r_0 = 0.05\,\text{cm}$, surface tension $\alpha = 14.6\,\text{g s}^{-2}$, elastic modulus $G = 328\,\text{g cm}^{-1}\text{s}^{-2}$, electrical conductivity $\sigma_e = 270\,\text{s}^{-1}$, zero-shear viscosity $\mu_0 = 1.02\,\text{g cm}^{-1}\text{s}^{-1}$, solvent viscosity ratio $\mu_s / \mu = 0.471$, and total current carried by the jet $I_{\text{total}} = 5.4\,\text{StatC s}^{-1}$. Adapted with permission (Carroll and Joo, 2011). Copyright © 2011, American Physical Society.

Clearly, the fluid representation as a 1D chain of beads is valid only if the dynamics lies entirely in the jetting regime. In this respect, the magnitude of the tensile force plays a crucial role. In fact, thin rectilinear jets may endure varicose perturbations under the action of capillary forces, whose magnitude is driven by the surface tension, $\alpha$ (Khakhar and Ottino, 1987). In addition to the dimensionless capillary number described in Section II, another way to measure the competition between the instability growth and the viscoelastic response is to compare their respective time scales. The viscoelastic response takes place within the relaxation time, $\theta = \mu_e / G$, while the time scale of the capillary instability growth is the inverse of the dimensionless instability growth rate, $\omega_c$. The fastest-growing capillary instability occurs with dimensionless frequency, $\omega_c^{max} = 1 / [\sqrt{2}\left(1 + 3S\sqrt{Ca/2}\right)^{1/2}]$, where $S = 1/\left(1 + G\theta / \mu_s\right)$ accounts for non-Newtonian behavior of the jet, being $\mu_s$ the zero shear viscosity of the solvent (Chang $et$ $al$., 1999). The ratio of the viscoelastic relaxation time and instability growth time leads to another, convenient expression of the Deborah number (Section II),

$De = \theta\, \omega_c^{max} / t^*$, given the characteristic time $t^* = r_0^2 S \rho / \mu_e$ (Jian $et$ $al$., 2006). For the typical case of the polyisobutylene Boger fluid at concentration of 4000 ppm in polybutene (Carroll and Joo, 2011), taking a jet radius $r_0 = 0.05\,\text{cm}$, jet velocity $\upsilon = 200\,\text{cm s}^{-1}$ (Montinaro $et$ $al$., 2015), surface tension $\alpha = 14.6\,\text{g s}^{-2}$, relaxation time $\theta = 3.11 \cdot 10^{-3}\,\text{s}$, zero-shear viscosity $\mu_0 = 1.02\,\text{g cm}^{-1}\text{s}^{-1}$ with $\mu_s / \mu_0 = 0.471$ (Carroll and Joo, 2006), and the extensional viscosity $\mu_e \sim 20\mu_0$ (Ng $et$ $al$., 1996), one would estimate the dimensionless numbers $Ca \sim 280$ and $De \sim 680$, indicating that the capillary instability is arrested by viscoelastic forces at the very early stage of instability growth. In this framework, the dimensionless critical stress $\bar{\sigma}^*$ in the jet, necessary for a complete suppression of the Rayleigh-Tomotika instability, was also assessed and found to be equal to $\bar{\sigma}^* = r(z^*)/(2r_0 Ca)$, where $r(z^*)$ is the cross-sectional radius measured at the distance $z^*$ from the needle, at which the instability is observed to begin (Jian *et al.*, 2006). Taking the ratio $r(z^*)/r_0 = 0.1$, the critical value of the tensile stress is thousands of times smaller than values obtained in simulations (Lauricella *et al.*, 2015a; Lauricella *et al.*, 2016a), $\sigma(z^*) \sim 4000\sigma^*$. Hence, one can conclude that the quasi-1D representation of the polymeric fluid as a chain of beads is fully justified given the stability of the early jetting regime.

## 3. Three-dimensional Lagrangian models

At a certain point during stretching of electrified polymer jets, the destabilizing electric force prevails. As a consequence, the bending instability starts to grow along the jet. Although, the slenderness assumption is still valid, a complete description of the spiraling and looping path requires a 3D model of the electrospun jet. Thus, the quasi-1D model can be extended to describe the jet evolution in the 3D framework, keeping however the fundamental original assumptions (e.g., isotropic expansion-reduction of the jet cross-section, absence of shearing forces in the lateral surface, and dependence of the jet cross-section on the longitudinal expansion). To this purpose, one can exploit the usual curvilinear parameter $s \in [0, \ell_0]$ (see Section V.A) to describe the jet in the 3D framework (Reneker *et al.*, 2000), by introducing the vector $\vec{R}(s)$ of coordinates $(x(s), y(s), z(s))$ in Eq. (17) for a discrete set of values $\{s_i\}_{i=1,\dots,n}$. Exploiting the approximated form of the stretching ratio, $\lambda(t)$, introduced in sub-Section V.B.2, the total arc length of the jet path at time $t$ is $\ell_{jet}(t) = \lambda(t)\ell_0$. The momentum balance for $i-$th bead provides:

$$m_i \frac{\partial \boldsymbol{v}_i}{\partial t} = -\pi r_{ui}^2 \sigma_{ui} \boldsymbol{\tau}_{ui} + \pi r_{di}^2 \sigma_{di} \boldsymbol{\tau}_{di} + + k\pi \left( \frac{r_{ui} + r_{di}}{2} \right)^2 \alpha \mathbf{n}_i + q_i \frac{\Delta \Phi_0}{h} \mathbf{k} + q_i \sum_{\substack{j=1,n \\ j \neq i}} \frac{q_j}{|\mathbf{R}_i - \mathbf{R}_j|^2} \mathbf{u}_{ij} +$$

$$+ m_i g \, \mathbf{k} \qquad (18)$$

where the subscripts $ui$ and $di$ denote, as in sub-Section V.B.2, the upper and downer dumbbell elements in the bead chain, $\boldsymbol{\tau}_{ui} = (\mathbf{R}_i - \mathbf{R}_{i+1})/|\mathbf{R}_i - \mathbf{R}_{i+1}|$ and $\boldsymbol{\tau}_{di} = (\mathbf{R}_{i-1} - \mathbf{R}_i)/|\mathbf{R}_{i-1} - \mathbf{R}_i|$ are the unit tangent jet vector of the upper and downer dumbbells, respectively, $\mathbf{u}_{ij}$ is the unit vector pointing $i-$th bead from $j-$th bead, and $\mathbf{n}_i$ is the principal unit normal vector pointing to the centre of the local curvature from $i-$th bead. Here, $\mathbf{n}_i$ is multiplying the force term associated with surface tension, i.e. $k\alpha\pi\left[\left(r_{ui} + r_{di}\right)/2\right]^2$, which acts to restore the rectilinear shape of the bent portion of the jet with curvature $k$. For each $i-$th bead, the set of EOM is completed with three kinematic relations (in vector notation $\partial \mathbf{R}_i / \partial t = \mathbf{v}_i$) and two constitutive equations of the type reported, for instance, in Eq. (14) for the independent variables $\sigma_{ui}$ and $\sigma_{di}$ and schematized in Figure 20, obtaining a total of eight ordinary differential equations. Finally, the model exploits the injection bead strategy (sub-Section V.B.2), with the discretization step length $\ell_{step}$ as input parameter, and it is accompanied by a specific set of two EOM of the nozzle coordinates describing possible mechanical perturbations at the spinneret. In particular, the spinneret nozzle is represented by a single mass-less point (labelled 0) of charge $q_0$ fixed at $z_0 = 0$. The charge, $q_0$, is taken equal to the mean charge value of the jet beads and it can be interpreted as a small portion of jet, which is glued at the nozzle. The set of EOM describing the evolution in time of the nozzle reads $\partial x_0(t) / \partial t = \omega y_0(t)$ and $\partial y_0(t) / \partial t = -\omega x_0(t)$, where the initial position at time $t = 0$ is defined by an input phase $\varphi$, so that $x_0 = A\sin(\varphi)$ and $y_0 = A\cos(\varphi)$, being $\omega$ and $A$ the frequency and amplitude of the perturbation, respectively. Alternatively, the nozzle perturbation can be modelled by adding a simple random displacement in the position of the inserted bead (labelled 1) in the injection step, so that $x_1 = x_1 + A_{rand} \sin \varphi_{rand}$ and $y_1 = y_1 + A_{rand} \cos \varphi_{rand}$ with $A_{rand}$ and $\varphi_{rand}$ the random amplitude and phase, respectively (Kowalewski *et al.*, 2005).

The dimensionless form of the EOM, exploiting the quantities reported in Table 2, is:

$$\frac{\partial \overline{\mathbf{v}}_i}{\partial \overline{t}} = L_{step} \left( -F_{ve,ui} \frac{\overline{\sigma}_{ui}}{\overline{\ell}_{ui}} \boldsymbol{\tau}_{ui} + F_{ve,ui} \frac{\overline{\sigma}_{di}}{\overline{\ell}_{di}} \boldsymbol{\tau}_{di} \right) + L_{step} \frac{\overline{k}}{4} A_i \left( \frac{1}{\sqrt{\overline{\ell}_{ui}}} + \frac{1}{\sqrt{\overline{\ell}_{di}}} \right)^2 \mathbf{n}_i + V_i \mathbf{k} +$$

$$+ \sum_{\substack{j=1,n \\ j \neq i}}^{n} \frac{Q_{e,ij}}{\overline{\ell}_{ij}^2} \mathbf{u}_{ij} + F_g \mathbf{k} \qquad (19)$$

where $\overline{\ell}_{ui} = |\mathbf{R}_i - \mathbf{R}_{i+1}| / L_0$ and $\overline{\ell}_{di} = |\mathbf{R}_{i-1} - \mathbf{R}_i| / L_0$ denote the dimensionless distance module between the upper and lower bead with respect to $i-$th bead, respectively, while $\overline{\ell}_{ij} = |\mathbf{R}_i - \mathbf{R}_j| / L_0$.

In order to compare the 3D model of the electrospun jet with the quasi-1D representation [Eqs. (13)-(16)], one can consider the time evolution of the jet termination, $\ell_z$, and its velocity, $\upsilon_z$, along the $z$-axis, which is shown in Figure 22 (Lauricella *et al.*, 2015a). Here, the same, typical input values are used for the dimensionless parameter, $Q_e = 12$, $V = 2$ and $F_{ve} = 12$, as those employed in the quasi-1D case analyzed in Figure 19 (Reneker *et al.*, 2000). Also in the 3D case, two sequential stages can be identified in the elongation process: the first stage is mainly biased by the sum of the viscoelastic and Coulomb forces, and the second one is dominated by the external electric field. Hence, a free fall regime is observed in the second part of the dynamics, with the jet length evolving in time as $\ell_z \propto t^2$ and the velocity nearly linearly increasing in time. While all these results are consistent with findings for the quasi-1D case, the 3D model provides higher values of the achieved velocity (Figure 22). In fact, this is related to the injection algorithm used to add new particles at the nozzle, which leads to extra charge i.e. to further repulsive Coulomb force (~1.7 times the corresponding force value obtained in the 1D simulation).

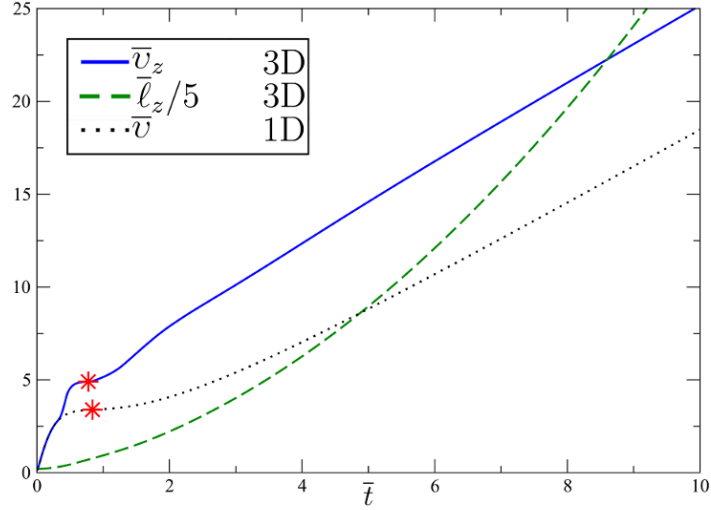

**Figure 22.** Time evolution of the velocity, $\overline{\upsilon}_z$ (continuous line), and the position, $\overline{\ell}_z$ (here rescaled by a factor 5, dashed line) of the jet termination, along the $z$-axis and in dimensionless units (see Table 2), from a 3D model [Eqs. (13)-(16)]. Input parameters: $Q_e = 12$, $V = 2$ and $F_{\upsilon e} = 12$. The corresponding velocity from from quasi-1D simulations [Eqs. (18), (19)], run with identical input parameters, is also shown for comparison (dotted line). A quasi-stationary point is found during the initial elongation (star symbol).

A comparison between 3D simulations and experimental data was performed considering an electrospun solution of polyvinylpyrrolidone (PVP) prepared by a mixture of ethanol and water (17:3 v:v), at a concentration about 2.5 wt% (Lauricella *et al*., 2015a). The applied voltage was around 30 statV (~ 10 kV), and the collector placed at 16 cm from the nozzle. Other parameters of the test case were: jet radius $r_0 = 0.05$ cm, jet velocity $\upsilon = 200$ cms$^{-1}$ (Montinaro *et al*., 2015), surface tension $\alpha = 21.1\,\mathrm{g\,s^{-2}}$ (Yuya *et al.*, 2010), elastic modulus $G = 5 \cdot 10^4\,\mathrm{g\,cm^{-1}s^{-2}}$ (Morozov and Mikheev, 2012), charge density $\rho_q = 44000\,\mathrm{statC\,cm^{-3}}$, zero-shear viscosity $\mu_0 = 0.2\,\mathrm{g\,cm^{-1}s^{-1}}$ (Yuya *et al.*, 2010;

Bühler, 2005), and extensional viscosity $\mu_e \sim 100\mu_0$ (Jian *et al.*, 2006). The simulations show that the model is able to reproduce all the slight perturbation from the linear path to bending instability, along with fully 3D, out-of-axis motion. This can be appreciated from Figure 23, comparing the bending oscillations between two snapshots of the simulated jet (a,b) and high-frame-rate experimental micrographs (c,d, Montinaro *et al.*, 2015) collected at the early stage and later regime of instability.

Further, the bending instability can be described in term of angular aperture ($\Theta$) of the instability cone.

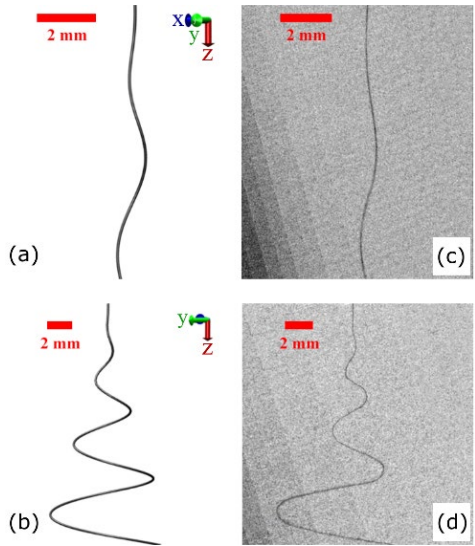

**Figure 23.** Snapshots of the simulated (a,b) and experimental (b,d) jets, taken close to the nozzle in the early stage (a,c), and in the bending regime (b,d) of their dynamics. The experiment was performed with a solution of PVP prepared by a mixture of ethanol and water (17:3 v:v, Montinaro *et al.*, 2015). The simulation is performed with the same conditions as in the experimental setup: applied voltage 30 statV ($\sim$ 10 kV), and the collector placed at 16 cm from the nozzle. The parameters of the PVP solution are taken from the literature: jet initial radius $r_0 = 0.05$ cm (Montinaro *et al.*, 2015), surface tension $\alpha = 21.1\,\mathrm{g\,s^{-2}}$ (Yuya *et al.*, 2010), elastic modulus $G = 5 \cdot 10^4\,\mathrm{g\,cm^{-1}s^{-2}}$ (Morozov and Mikheev, 2012), charge density $\rho_q = 44000\,\mathrm{statC\,cm^{-3}}$, zero-shear viscosity $\mu_0 = 0.2\,\mathrm{g\,cm^{-1}s^{-1}}$ (Yuya *et al.*, 2010; Bühler, 2005), and extensional viscosity $\mu_e \sim 100\mu_0$ (Jian *et al.*, 2006). Reproduced with permission (Lauricella *et al.*, 2015a). Copyright © 2015, Elsevier Ltd.

The $\Theta$ value measured in the simulation is in the range 30°-36°, which is consistent with the experimentally range 29°-37° (Montinaro *et al.*, 2015).

Lagrangian methods are extensively used to span the space of parameters in order to probe the effects of different quantities on the electrospinning process (Thompson *et al.*, 2007; Kowalewski, *et al.*, 2009; Sun *et al.*, 2010; Coluzza *et al.*, 2014; Lauricella *et al.*, 2015a; Lauricella *et al.*, 2016a; Yousefi *et al.*, 2018). For instance, thirteen different parameters were studied varying one at a time in the model in order to determine their effect on final jet cross-sectional radius (Thompson *et al.*, 2007). The growth rate of the bending instability was also investigated by the stability analysis on a similar 3D Lagrangian model (Divvela and Joo, 2017). The jet was perturbed with an initial normal mode disturbance at an amplitude $\delta$ and frequency $\omega_p$ applied on several parameters (position, velocity, jet radius, viscosity, tensile stress and external electric field). For instance, the perturbed $i-$th bead charge $q_{p,i}$ reads $q_{p,i} = q_i + \delta q_0 \exp(\omega_p t)$. At higher voltage and lower viscosity, the non-axisymmetric bending instabilities were found to become dominant.

The versatility of this model allows extra force terms to be incorporated, as well as further algorithmic expedients to probe innovative experimental setups. In the next sub-Section, we review a few of these extensions, together with scenarios for their practical applications.

## C. Advanced Lagrangian models

This Section is focused on more recent advancement in the Lagrangian models for electrified polymer jets. The main improvements are aimed to address essentially two key points: First, refining the model by including force terms which have been neglected as a first approximation. Physical mechanisms considered in the last decade include aerodynamics, focusing electric fields, nanoparticles in the jet, to name but a few. Second, but no less important, improving the jet representation via different descriptive techniques and algorithms. Here, both improvements are reviewed.

## 1. Aerodynamic effects

All Lagrangian models involve a set of EOM which can be modified flexibly to account new force terms and extend applicability. For instance, aerodynamics and air drag effects were not initially considered (Reneker *et al*., 2000), because their contribution to stretching is much smaller than that from electrical forces under the typical conditions of electrospinning experiments. Nonetheless, not only higher accuracy of the models may be desirable, but also there exist variants of the experimental set-up that involve a gas stream surrounding the nozzle, as in electro-blowing (Um *et al*., 2004; Wang *et al*., 2005; Hsiao *et al*., 2012). In these processes, the effects of air-drag and lift cannot be neglected. Hence these terms should be added in the momentum balance, Eq. (18). The dissipative air drag term is a braking force oriented along the tangent axis of the jet, with its magnitude depending on the geometry of the jet, which evolves in time. Based on experimental results (Yarin, 1993), the air drag force acting on $i-$th bead can be assessed by the empirical relation (Ziabicki, 1961; Ziabicki and Kawai, 1991):

$$\mathbf{f}_{drag,i} = -\xi \pi r_i \ell_{ui} \rho_{air} \left( \frac{2r_i}{\nu_{air}} \right)^{-\zeta} \left( \upsilon_i^{\shortparallel} \right)^{1+\eta} \mathbf{t}_{ui} \ , \tag{20}$$

where $\nu_{air}$ and $\rho_{air}$ are the air kinematic viscosity and density, respectively, $\ell_{ui} = |\mathbf{R}_i - \mathbf{R}_{i-1}|$ is the distance module between $i-$th and its upper bead, and the term $\upsilon_i^{\shortparallel} = \left( \mathbf{v}_i - \mathbf{v}_{air} \right) \cdot \boldsymbol{\tau}_{ui}$ denotes the tangent component of the relative jet velocity with respect to the airflow velocity, $\mathbf{v}_{air}$. In Eq. (20), the symbols $\xi$, $\zeta$, and $\eta$ denote three coefficients to be empirically assessed. As plausible values of these coefficients, Ziabicki and Kawai empirically determined them as $\xi = 13/20$, $\zeta = 4/5$, and $\eta = 1/5$ (Ziabicki and Kawai, 1991). These coefficients can be fitted on the drag force data measured in

polymeric free-running filaments by an experimental apparatus equipped with an electronic tensiometer (Ziabicki, A., 1961). Overall, the air drag plays the role of a dissipative term absorbing the air perturbations in a non-linear way with respect to $\upsilon_i^{\shortparallel}$. Collecting several terms of Eq. (20) and assuming the volume conservation so that $r_i = r_0 \sqrt{\ell_{step} / \ell_{ui}}$, it is possible to write the dissipative, friction coefficient, $\gamma_i$, as (Lauricella *et al*., 2016a):

$$\gamma_i = \xi \pi \frac{\rho_{air}}{m_i} \left( \frac{2}{\nu_{air}} \right)^{-\zeta} \ell_{step}^{\frac{1}{2}(1-\zeta)} r_0^{1-\zeta} \ . \tag{21}$$

Inserting Eq. (21) in Eq. (20), the drag force reads:

$$\mathbf{f}_{drag,i} = -m_i \gamma_i \ell_{ui}^{\frac{1}{2}(1+\zeta)} \left( \upsilon_i^{\shortparallel} \right)^{1+\eta} \mathbf{t}_{ui} \ , \tag{22}$$

which dissipates the air fluctuations. The perturbations are in turn due to local interactions of random high-frequency collisions of the gas molecules with the jet, providing a Brownian motion component that contributes to the overall dynamics. The macroscopic force resulting from such fluctuations is taken as a stochastic process with zero mean and diffusion coefficient, $D_\upsilon$, in the velocity space, modelling the total displacement of the jet due to the sum of particle hits over a time window much longer than the inverse of the particle collision frequency. Hence, the random force reads (Lauricella *et al*., 2016a):

$$\mathbf{f}_{rand,i} = \sqrt{2m_i^2 D_\upsilon} \ \boldsymbol{\eta_i}(t) \ , \tag{23}$$

where $D_\upsilon$ is assumed constant and equal for all the beads, and $\boldsymbol{\eta_i}$ is a three-dimensional vector of independent stochastic processes, whereof each component along one of the three unit vectors ($\mathbf{i}$, $\mathbf{j}$, $\mathbf{k}$), is taken as $\eta_i = d\xi_i(t) / dt$. Here $\xi_i(t)$ denotes a Wiener process, namely a stochastic process with stationary independent increments (Durrett, 2019). It is worth to observe that the insertion of the last term in Eq. (18) delivers a non-linear Langevin-like stochastic differential equation in the set of EOM (Lauricella *et al.*, 2015b). As a consequence, the simulation requires a time marching scheme able to integrate the random processes preserving a reasonable order of convergence (Kloeden and Platen, 2013; Tuckerman, 2010). An example of strong convergence scheme with order 3/2 was reported (Lauricella *et al.*, 2015b) for the numerical integration of the stochastic term in Eq. (23). However, a smaller time step $\Delta t \sim 10^{-8}$ s in the integration scheme improves the numerical accuracy, although increasing the computational cost. As an alternative, the energy spectrum (provided as input parameter) of the random process can be exploited to simulate the random velocity displacement as a sum of $m$ harmonic functions (Battocchio et al., 2017). These functions would have frequencies $\omega_j$ with $j \in [1, \ldots, m]$, and amplitude derived from the energy spectrum, assumed constant up to a cut-off frequency and zero at higher frequencies. This approach may be numerically integrated with larger time step, $\Delta t$, by setting a low-frequency cut-off, thus saving significant computational time.

In a 3D framework, the lift force is also acting on the jet dynamics along the normal (curvature) vector. The aerodynamic lift force related to the flow speed can be written for $i-$th bead in the linear approximation (for small bending perturbations) as (Yarin, 1993; Lauricella *et al.*, 2016a):

$$\mathbf{f}_{lift,i} = -\ell_{ui} k_i \rho_{air} \left( \upsilon_i^{\shortparallel} \right)^2 \pi \left( \frac{r_{i+1} + r_i}{2} \right)^2 \mathbf{n}_i \ , \tag{24}$$

where the term $\ell_{ui}$ denotes the magnitude $\ell_{ui} = |\mathbf{R}_i - \mathbf{R}_{i-1}|$. The lift term scales quadratically with $\upsilon_i^{\shortparallel}$, thus playing a significant role for high enough relative velocity of the airflow.

The effects of the three terms given in Eqs. (22), (23) and (24) on the extended EOM were investigated for electrospinning simulations of a PVP solution under three different conditions of air airflow velocity, $\mathbf{v}_{air} \in [0, -10, -20] \, \text{cms}^{-1}$, oriented along the unit vector $\mathbf{k}$ (Lauricella *et al.*, 2016a). The air kinematic viscosity was set to $\nu_{air} = 0.151\,\mathrm{cm^2\,s^{-1}}$ and the air density $\rho_{air} = 1.21 \cdot 10^{-3}\,\mathrm{g\,cm^{-3}}$, while $D_{v,i}$ was taken equal to $\gamma_i$ for all the beads, for the sake of simplicity. In the case $\mathbf{v}_{air} = -20\,\mathrm{cm\,s^{-1}}$, the larger lift force contributes to the bending instabilities (Figure 24). In particular, the synergic action of lift and Coulomb repulsive forces boost bending instabilities at an earlier stage and increase the chaotic behavior of the jet in the subsequent dynamics. This is seen in Figure 24 by the larger statistical dispersion of the shadowed cone (thickness of the instability envelope cone wall) computed as the isosurface with constant value 0.001 of the normalized numerical density field, $\rho_{bead}(x,y,z)$, namely the probability to find a jet bead in a control volume nearby the coordinates $(x,y,z)$. As a consequence, the high-speed gas flow significantly affects the distribution of the deposited nanofibers, providing a broader deposition pattern and a decreased fiber radius due to the larger bending instabilities. In these simulations, the jet was assumed to interact with a uniform airflow field. Nonetheless, computational fluid dynamics software can be used to simulate an arbitrary airflow field. As an example, the velocity flow field, $\mathbf{v}_{air}(x,y,z)$, could be computed by solving the corresponding Navier-Stokes equation (Sun $et$ $al$., 2010). Hence, $\mathbf{v}_{air}$ was coupled with the Lagrangian model in order to investigate the effect of airflows in the spinning process (Sun $et$ $al$., 2010).

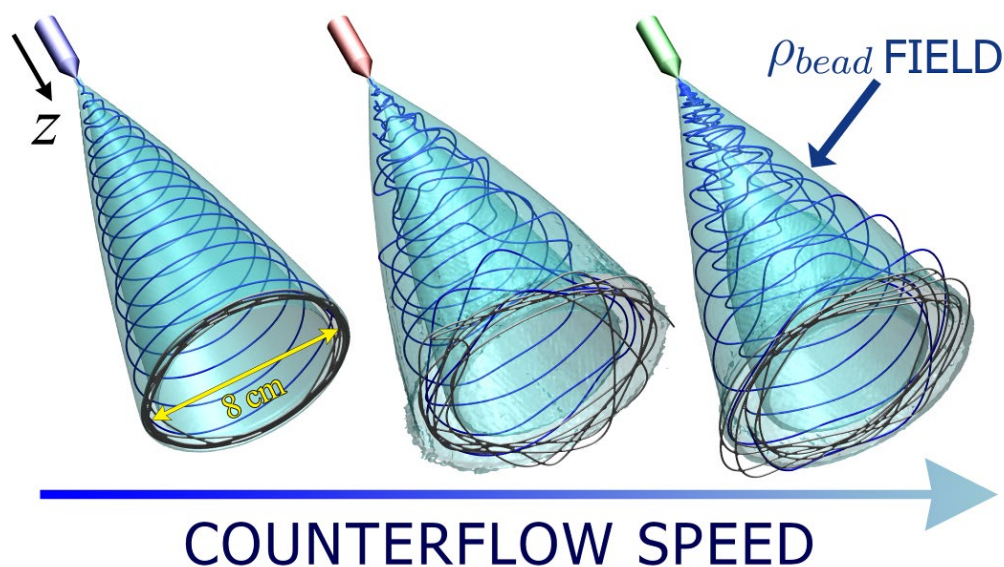

**Figure 24.** Simulation snapshots of the three different cases of an electrospun polymer solution jet with air counterflow. From left to right, the airflow velocity along the unit vector **k** (collinear with the $z$ direction) is $\mathbf{v}_{air} = 0$ cm/s, $\mathbf{v}_{air} = -1000$ cm/s, and $\mathbf{v}_{air} = -2000$ cm/s, respectively. The jet is drawn in dark in the cones, and the nanofibers deposited on the collector are in black at the termination of the cones. The shadowed isosurfaces represent the normalized numerical density field $\rho_{bead}(x,y,z)$ of constant value equal to 0.001. Adapted with permission (Lauricella *et al.*, 2016a). Copyright © 2016, American Chemical Society. Doi: 10.1021/acs.jpca.5b12450. Further permissions related to the material excerpted should be directed to the American Chemical Society.

## 2. Electric and magnetic focusing fields

So far, the external electric field $\mathbf{E}_0 = \left( \Delta\Phi_0 / h \right) \mathbf{k}$ in Eq. (18) was assumed to be oriented along the $\mathbf{k}$ axis over the entire space. Nonetheless, in a real electrospinning setup, the electric field distribution $\mathbf{E}_0(x,y,z)$ depends on both the collector and needle electrode shapes, and this might alter the direction of the jet stretching. Several works investigated the effects of auxiliary electrodes used for the generation of extra electric fields (also named focusing electric fields) that drive the electrospun jet and affect the bending instability (Bellan and Craighead, 2006; Deitzel $et$ $al$., 2001b; Neubert $et$ $al$., 2012). Such effects can be accounted for by resolving the electric field distribution $\mathbf{E}_0(x,y,z)$ over the space. Despite being computationally expensive, finite difference or finite element methods are usually exploited to solve the Poisson equation $\nabla^2 \Phi_0(x,y,z) = \rho_q(x,y,z) / \varepsilon$ and obtain the field $\mathbf{E}_0(x,y,z) = -\nabla\Phi_0(x,y,z)$. For instance, the Lagrangian model was extended to investigate the impact of a conical needle electrode on the radius of electrospun nanofibers (Hamed $et$ $al$., 2018). The Poisson equation could be solved only once at the beginning of the simulation and inserted in Eq. (18) as an input parameter. However, a correct computation of the electric field should also account, in the Poisson equation, for the time-dependent charge distribution on the jet path, paying the high computational costs for the solution of the Poisson problem at each time step. Within this context, since the electric potential difference $\Delta\Phi_0$ should be constant at the tip nozzle, another strategy, shown in Figure 25, exploited the method of images whereas fictitious mirror charges are placed symmetrically to the collector plane, in order to match always the condition $\Phi_0 = 0$ at the collector (Kowalewski *et al.*, 2009). Thus, the total electric field is the superposition of the external static field and of a time-dependent intra-jet field generated by the jet charges, the latter one computed as in Eq. (18) but now as the direct summation running over both real and mirror jet charges.

Furthermore, it is possible to drive both the stretching direction and preferential orientations of the jet at the collector by a time-dependent manipulation of the external electric field, acting as a dynamic focusing field. It was experimentally observed that a time-varying square wave potential is able to periodically deflect the jet between two symmetric, rotatable plate-like electrodes, coaxially placed nearby the needle at a distance of 8.5 cm (Grasl *et al.*, 2013). Rotating electric fields can be also generated by a series of fixed capacitor plates appropriately arranged in space and connected to an alternating power source (Kyselica *et al*., 2016; 2018). For instance, assuming the hexagonal arrangement of the plates and a three-phase power source connected to them (Figure 26), as a first approximation the external electric field can be assumed uniformly distributed in the space, so that the total electric field $\mathbf{E}_0 = \mathbf{E}_0^{\parallel} + \mathbf{E}_0^{\perp}$ is the superposition of a stretching electric field $\mathbf{E}_0^{\parallel} = (E_x, 0, 0)$ parallel to the $\mathbf{k}$ axis and an orthogonal rotating term $\mathbf{E}_0^{\perp} = (0, E_y, E_z)$. In equations:

$$E_y(A, \omega, t) = A\cos(\omega^{\perp} t), \tag{25a}$$

$$E_z(A, \omega, t) = A\sin(\omega^{\perp} t), \tag{25b}$$

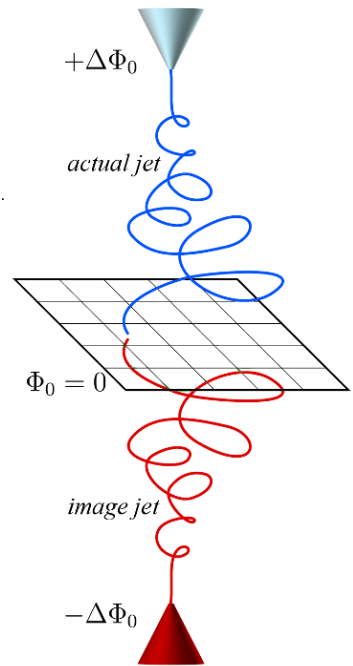

**Figure 25.** Idealized electrostatic setup with fictitious charges of the image jet, which are used to maintain the electric potential condition $\Phi_0 = 0$ on the grounded plane during process modeling.

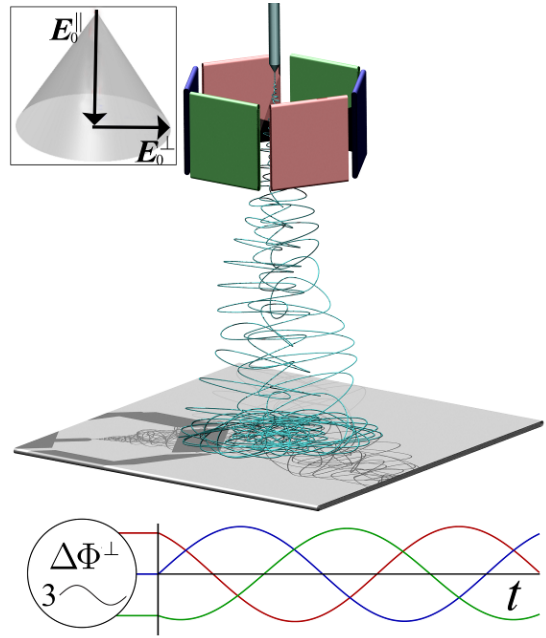

**Figure 26.** Simulation snapshot of the electrospinning process in the presence of an orthogonal rotating electric field at high frequency (~$10^4$ Hz). The jet is stretched by a longitudinal electric field $\mathbf{E}_0^{\parallel}$. The orthogonal electric field ( $\mathbf{E}_0^{\perp}$ ) can be generated by a series of capacitor plates, hexagonally arranged and connected to a three-phase power source, as represented in the bottom part of the figure. Here, the three-phase voltage differences $\Delta\Phi_0$ correspond to the different pairs of capacitor plates. Adapted with permission (Lauricella *et al*., 2017b). Copyright © 2017, AIP Publishing.

where $A(g^{\frac{1}{2}}cm^{-\frac{1}{2}}s^{-1})$ is the magnitude of $\mathbf{E}_0^\perp$ and $\omega(s^{-1})$ is the angular (switching) frequency of the field. By inserting the term $q_i\mathbf{E}_0^\perp(A,\omega,t)$ in Eq. (18), the morphology of electrospun materials could be studied for several pairs of $A$ and $\omega$ values (Lauricella *et al.*, 2017b). In agreement with both experimental observations (Grasl *et al.*, 2013; Kyselica *et al.*, 2016; Kyselica *et al.*, 2018), and further numerical simulations (Kyselica *et al.*, 2019), the jet was found to support rather regular oscillatory patterns, which are controlled by tuning $A$ and $\omega$. This offers the opportunity to deposit oriented nanofibers even on a static collector, which is interesting for the design of new porous materials. Also in the context of magnetic-field assisted electrospinning, Lagrangian models can be easily modified in order to simulate the effect of a stationary, space-distributed magnetic field. In particular, the Lorentz force $\mathbf{f}_L$ for $i-$th bead with velocity $\mathbf{v}_i$ and charge of $q_i$ is given by:

$$\mathbf{f}_{L,i} = q_i\mathbf{v}_i \times \mathbf{B}_0(x,y,z)\,, \tag{26}$$

where $\mathbf{B}_0(x,y,z)$ is the vector field at the coordinates $(x,y,z)$ of $i-$th bead. Hence, the term is added to the momentum Eq. (18) in order to simulate the coupling effects of electric forces and magnetic field. Theoretical works agree well with the experimental data, showing that $\mathbf{B}_0$ could significantly alter the non-axisymmetric (bending/whipping) instabilities of the jet (Xu *et al.*, 2011; Badieyan and Janmaleki, 2015). The jet can be focused, to obtain nanofibers with preferential orientation in given regions of the collector.

## 3. Multiple jets in electrospinning

Multiple jets were introduced in electrospinning to increase the fiber production rate. In addition, the rational design of nozzle patterns, issuing many jets simultaneously, might be important to obtain areas uniformly covered with nanofibers. Thus, numerical investigations were performed to explore the physics of the jet–jet interactions. For instance, the Lagrangian model (sub-Section V.B.3) was generalized for the case of multiple bending jets with mutual Coulombic interactions (Theron *et al*., 2005). In order to compute the Coulomb force acting on $i-$th bead of $k-$th jet, a simple way is to extend in Eq. (18) the direct summation of the Coulomb term as follows:

$$\mathbf{f}_{c,ik} = q_{ik} \sum_{l=1,m} \sum_{\substack{j=1,n(l) \\ jl \neq ik}} \frac{q_{jl}}{|\mathbf{R}_{ik} - \mathbf{R}_{jl}|^3} (\mathbf{R}_{ik} - \mathbf{R}_{jl}), \tag{27}$$

where now $j$ is running over the $n(l)$ beads of the $l-$th jet among the $m$ jets. In this way, the bending instabilities of a jet in a multi-nozzle setup were predicted to be driven not only by self-induced Coulombic forces but also by mutual, inter-jet Coulombic repulsive interactions (Figure 27). Further numerical studies (Angammana and Jayaram, 2011b; Li *et al*., 2015) showed that the stretching ratio in a multi-jet setup is larger than that observed in single-jet electrospinning. In particular, the stretching ratio of a jet increases by decreasing the mutual distance between neighbouring jets.

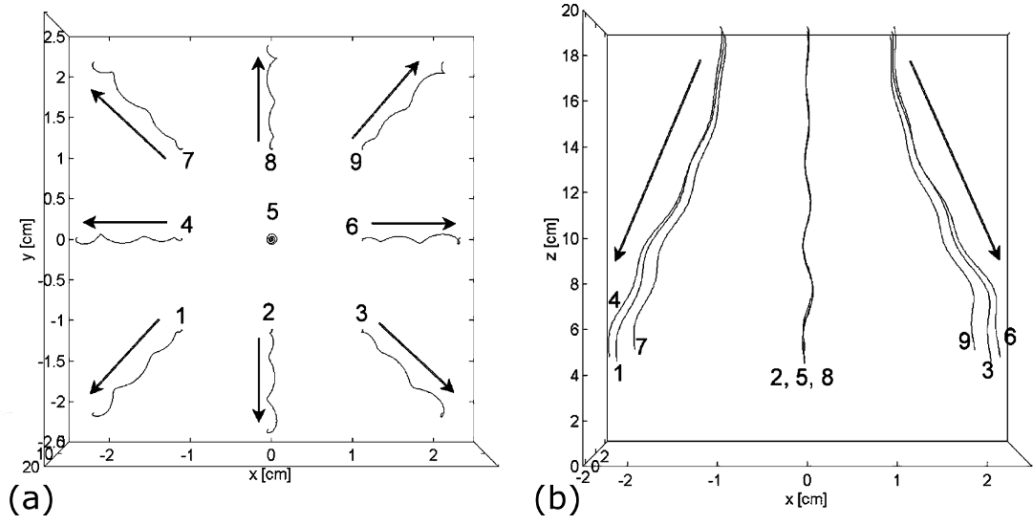

**Figure 27.** The paths of nine jets in a $3 \times 3$ matrix arrangement of the nozzles taken from two different point view: (a) Top view of all the jets of the $3 \times 3$ matrix; (b) Side view of all the jets. Adapted with permission (Theron *et al*., 2005). Copyright © 2005, Elsevier Ltd.

## 4. Dynamic refinement in Lagrangian models

All the Lagrangian models reviewed so far exploit the injection algorithm mentioned in sub-Section V.B.2: all the beads are uniformly inserted at the nozzle, with the mutual distance $\ell_{step}$. Nonetheless, the distances $\ell_{ui}$ and $\ell_{di}$ between any couple of consecutive elements in the bead chain increases along the dynamics, because of the stretching induced by the intense electrical forces. In numbers, assuming the volume conservation, a typical value $r/r_0 = 10^{-2}$ in the jet cross-section measured at the collector provides the increase, $\ell_{ui} = 10^4 \ell_{step}$, in the discretization length of the jet. Thus, the discretization close the collector becomes rather coarse (both $\ell_{ui}$ and $\ell_{di} >> \ell_{step}$) to effectively model the filament, and the information (position, velocity, radius, stress, etc.) describing the jet is scattered downstream. To tackle this issue, an adaptive dynamic refinement procedure was developed, maintaining the lengths $\ell_{ui}$ and $\ell_{di}$, for any $i-$th element, below a prescribed characteristic threshold length, $\ell_{max}$ (Lauricella $et\ al.$, 2016b). Whenever a bead length, $\ell_{ui}$ or $\ell_{di}$, is larger than $\ell_{max}$, the jet description is refined by discretizing uniformly it at the length step value, $\ell_{step}$, given as input. The discretization was performed by cubic spline interpolations (De Boor, 1978) of the main quantities describing the jet beads (positions, jet radius, velocities, stress). To perform the interpolation, the total arc length at time $t$ is assessed as:

$$\ell_{jet}(t) = \sum_{i=1}^{n} \ell_{ui}(t), \tag{28}$$

where the sum runs over the $n$ beads so that all the distances between element pairs $(i+1, i)$ are accounted for. Note that the stretching ratio $\lambda(t)$ at time $t$ is assessed as $\lambda(t) = \ell_{jet}(t)/\ell_{jet}^0$. Hence, the usual curvilinear parameter $s$ is introduced (sub-Sections V.A.4 and V.B.3) and used to define the discretization mesh of the jet. In particular, the discrete set of values $\{s_k\}_{k=1,...,n}$ of the curvilinear parameter $s \in [0, \ell_{jet}^0]$ is estimated for each $k-th$ bead by using the formula:

$$s_k(t) = \frac{1}{\lambda(t)} \sum_{i=1}^{k} \ell_{ui}(t).$$  (29)

Despite $s$ values are still 'frozen' in the jet element, the set $\{s_k(t)\}_{k=1,...,n}$ is now time dependent, since the jet discretization is adapted to satisfy always the condition $\ell_{ui} < \ell_{max}$ during the jet evolution. Hence, the set of $\{s_k(t)\}_{k=1,...,n}$ values represent the mesh used to build the cubic spline. Given a generic quantity $y$, the data $y_k$ are tabulated over the set of values $s_k \lambda$, where $s_k(t)\lambda(t)$ denotes the arc length from the nozzle to the $k-th$ bead of the jet at time $t$. Thus, $y_k = y(s_k \lambda)$ were used to compute the coefficients of the cubic spline, following a well-established algorithm (Press *et al*., 1996). Then, a uniform parametrization was enforced by imposing all the lengths of the elements equal to $\ell_{step}$. The new mesh $s_i^*$ is defined as:

$$s_i^* = \ell_{jet}^0 \frac{i}{n^*}, i = 1, 2, ..., n^*,$$  (30)

where $n^* = \lambda(t)\ell_{jet}^0 / \ell_{step}$ is the number of jet beads in the new representation. Another possible criterion for building the new mesh is to keep all the old beads in their positions and to add only new beads where the bead distance in the pair $(i+1,1)$ is $\ell_{ui} > \ell_{max}$. Indeed, since all the old beads are mantained as knots of the mesh along the dynamic refinement, the errors introduced by the interpolation procedure is mitigated, because it affects only the new inserted bead.

Finally, the new values $y(s_i^* \lambda)$ were computed for any $i-th$ bead by the spline interpolation. The procedure is repeated for positions, jet radius, stress and velocities of the jet beads, in order to provide the quantities in the new mesh $\{s_i^*\}_{i=1,\ldots,n^*}$, (Lauricella *et al.*, 2016b). As a practical application of the mesh refinement procedure, the algorithm was applied to simulate the electrospinning of a polymeric solution containing heavy nanoparticles, which trigger varicosity along jet modifying the path from the nozzle toward the collector (Figure 28, Lauricella *et al.*, 2017a). Here, the mesh refinement showed the capability of representing the fluctuation of the cross-section along the jet, preserving a fine jet representation also close the collector. Clearly, small values of the threshold length, $\ell_{max}$, increase the number of beads in the simulation and the associated computational costs (see also Section IV and strategies reviewed there for saving computational time).

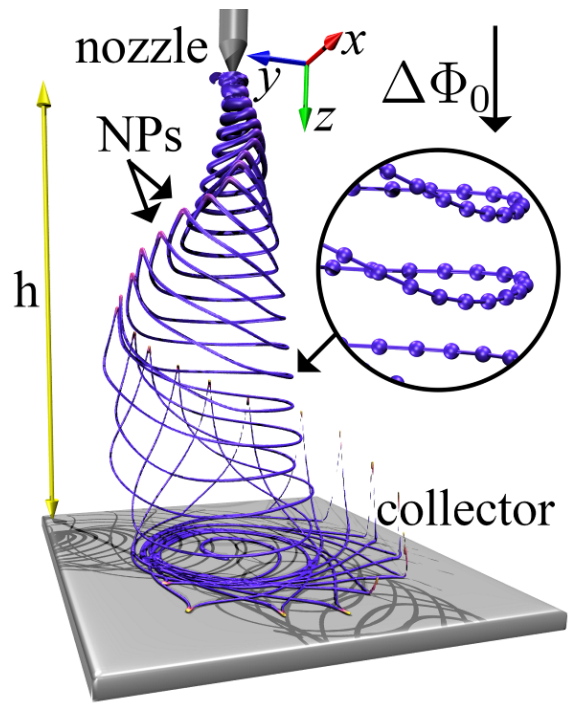

**Figure 28.** Simulation snapshot illustrating an electrified polymer solution jet embedding nanoparticles. $h$: distance between the collector plate and the nozzle. $\Delta\Phi_0$: applied voltage difference between these two elements. Adapted with permission (Lauricella *et al.*, 2017a). Copyright © 2017, EPLA.

## 5. Limits of Lagrangian models

All the Lagrangian models here reviewed assumed the polymeric solution to be fully spinnable. In particular, relevant dimensionless numbers (e.g., capillary and Deborah numbers) should be preliminarily considered in order to justify the assumption of the solution spinnability as described in Section II and sub-Section V.B.2. In other words, the reader should be aware that a Lagrangian model should not be used to foresee the spinnability of a polymer solution numerically because this is already assumed in the model. As a second limit, since the jet is represented as a continuous chain of particles connected pair by pair through viscoelastic springs, Lagrangian models are not effective for investigating failure modes, such as jet breaking and splitting, in fluid polymeric filaments. These deformations could be analyzed for the case of conducting liquids, in terms of a critical value of the linear electric charge density issued by the jets (Zubarev and Zubareva, 2004; 2005). Similarly, Lagrangian models are not able to display varicose instabilities in the jet, such as those due to the capillary pressure.

Further, we point out that the rheological behaviour is highly sensitive to the phenomenological, constitutive law handled in the viscoelastic springs connecting neighbor particles. This fact is not different from what generally remarked in Eulerian descriptions (e.g., numerical solver of Navier-Stokes equations) where a constitutive law is also needed.

Finally, any Lagrangian model exploits proper algorithms to treat the fluid injection at the nozzle (see sub-Section V.B.2). As a consequence, the modelling of Taylor cone is missed. However, this specific limit could be tackled by coupling of Lagrangian models with Eulerian solvers of the EHD equations governing the fluid cone used, for instance, to model the tip streaming of charged droplets (Collins *et al*., 2008; Collins *et al*., 2013).

## D. Polymer network dynamics in electrified jets

This Section reviews efforts describing the polymer network dynamics for highly elongated, electrified polymer solution jets. With strain rate, $\dot{\varepsilon} \geq 10^3$ s$^{-1}$ (Reneker *et al.*, 2000;  Bellan *et al.*, 2007;  Reneker *et al.*, 2007), stretching in these jets can potentially increase the structural order within spun nanofibers, enhance their mechanical properties such as elastic modulus and strength (Rein*, et al.*, 2007; Burman*, et al.*, 2008; 2011; Greenfeld*, et al.*, 2011; Zussman and Arinstein, 2011), and shift the phase transition temperature, e.g., the melting temperature.(Arinstein *et al.*, 2011; Liu *et al.*, 2011) At the same time, the rapid solvent evaporation during electrospinning can lead to increased polymer concentration at the jet boundary (Koombhongse *et al.*, 2001; Guenthner *et al.*, 2006; Dayal and Kyu, 2007; Dayal *et al.*, 2007), forming a solid skin or a heterogeneous and porous structure (Casper*, et al.*, 2004; Dayal *et al.*, 2007; Greenfeld *et al.*, 2011) which may lead to buckling of the fibers (Arinstein *et al.*, 2009). The simultaneous effects of stretching and evaporation (Arinstein and Zussman, 2011) are illustrated in Figure 29. In this respect, the study of electrified polymer solution jets, and specifically of the evolution of the polymer entangled network during electrospinning which is typically followed by stress relaxation (Vasilyev *et al.*, 2017), was clearly aimed at clarifying the microstructure of spun nanofibers, and at improving their mechanical, electrical and optical properties. Modelling of the dynamic evolution of the entangled polymer network in electrospun jets predicted substantial longitudinal stretching and radial contraction of the network, i.e. a transformation from an equilibrium state to an almost fully-stretched state (Greenfeld *et al.*, 2011; Arinstein and Zussman, 2011). This prediction was verified by X-ray phase-contrast imaging of electrospun jets, which revealed a noticeable increase in polymer concentration at the jet center, as well as a concentration crossover within a short distance from the jet initiation point (Greenfeld *et al.*, 2011; Greenfeld *et al.*, 2012).

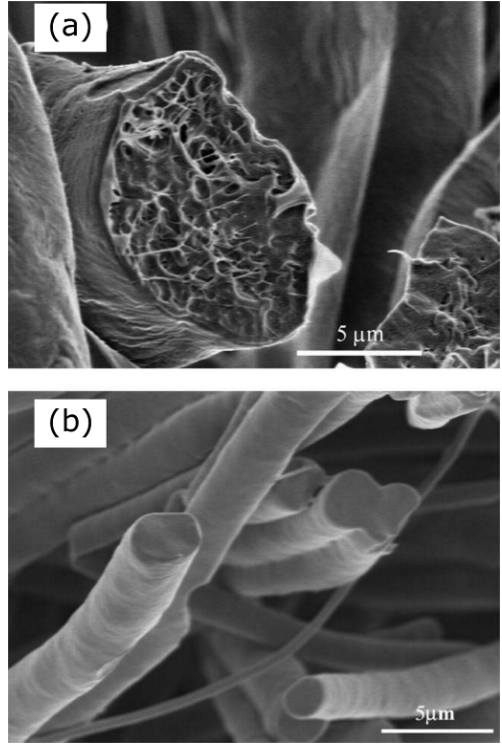

**Figure 29.** Stretching and evaporation involving electrified polymer solution jets. SEM images of nanofibers with 10 wt% PCL with molar mass 80 kDa, dissolved in dichloromethane/DMF (75:25 wt%), electrospun by an electric field of 0.63 kV/cm. (a) High flow rate (20 mL/hr) resulting in heterogeneous fibers. (b) Low flow rate (3 mL/hr) resulting in homogeneous fibers. Reproduced with permission (Arinstein and Zussman, 2011). Copyright © 2011, Wiley Periodicals, Inc.

## 1. Polymer dynamics during electrospinning

The flow of a solution jet clearly consists of both axial and radial velocity components (Figure 30). The analysis of electrically-driven fluid jets evidenced that $\upsilon$, the axial velocity of the jet, reaches an asymptotic regime sufficiently far from the needle, which could be described by a power law of the distance along the jet, $z$:

$$\frac{\upsilon}{\upsilon_0} \approx k_V^2 \left( \frac{z}{r_0} \right)^{2\beta},$$ 

(31)

where the exponent, $\beta$, varies between 1/4 and 1 (Kirichenko *et al.*, 1986; Spivak and Dzenis, 1998; Hohman *et al.*, 2001b; Higuera, 2006; Reznik and Zussman, 2010), $k_V$ is a dimensionless parameter, $r_0$ is the jet initial radius, and $\upsilon_0$ is the initial velocity, $\upsilon_0 = Q / \pi r_0^2$. Assuming volume conservation, the jet local radius, $r_J = r_0 (\upsilon / \upsilon_0)^{-1/2}$ (see Figure 30), and the radial velocity, $\upsilon_r$, can be derived. A rough approximation of $k_V$ in Eq. (31) could be obtained by using a simple scaling approach: the velocity gradient scales as $\nabla \upsilon \approx \upsilon_0 k_V^2 / r_0^2$; the upper bound for $\sigma_q$ (the surface charge density on the jet), assuming static conditions, scales as $\sigma_q / \sigma_e E$, where $\sigma_e$ is the solution electric conductivity and $E$ is the electric field. The electric shear stress is therefore $\tau_t \approx \sigma_q E \approx \sigma_e E^2$, producing a velocity gradient $\nabla \upsilon \approx \tau_t / \mu \sim \sigma_e E^2 \mu^{-1}$. Hence, $k_V \sim \nabla \upsilon^{1/2} \upsilon_0^{-1/2} r_0$, or $k_V \sim r_0^1 \sigma_e^{1/2} \mu^{-1/2} \upsilon_0^{-1/2} E^1$. A more accurate calculation yielded $k_V \cong r_0^{2/3} \upsilon_0^{-2/3} \sigma_e^{1/4} \mu^{-5/12} E^{5/6} \sim 1$ (Reznik and Zussman, 2010).

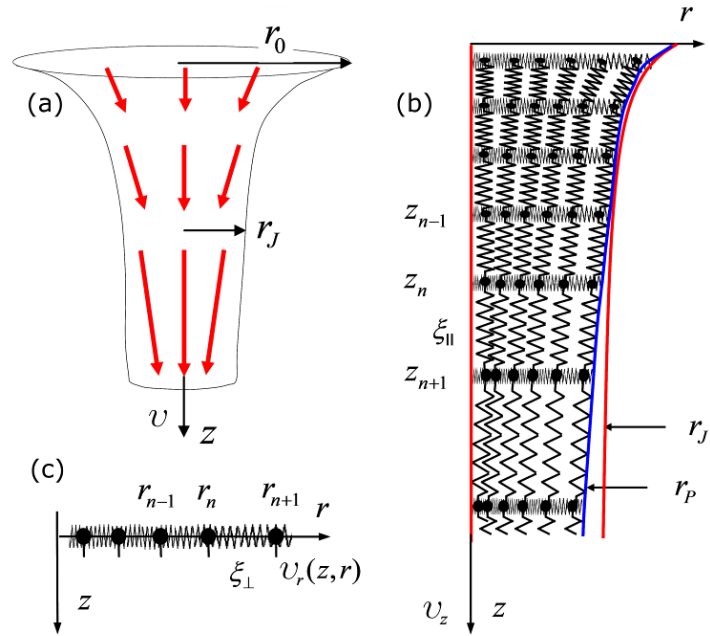

**Figure 30.** (a) Illustration of the flow of an electrified polymer solution jet. (b) Definition of an effective 1D beads-and-springs system in the axial direction and (c) the radial direction (Greenfeld *et al.*, 2011).

Polymer chains dissolved in a sufficiently concentrated solution create an entangled network (Figure 30b), a prerequisite for successful spinnability. A chain section between two subsequent entanglements is a strand, or a subchain, consisting of $N_s$ rigid segments (Kuhn monomers), each of size $b \cong 1$ nm. Given the solution concentration $\phi$ (in terms of the polymer volume fraction), the network mesh size (i.e., average subchain length) is given by $\xi_0 \approx b\phi^{-1} \approx bN_s^{1/2}$ for ideal chains, allowing one to model the network as a 3D beads and springs lattice, where each bead represents the mass $m$ and size $\xi_{eff}$ of a subchain ($\xi_{eff} \propto \xi_0$), and the springs represent the linear entropic elasticity $T/\xi_0^2$ (here, $T$, is the temperature in units of the Boltzmann constant $k_B$) of the subchains connected to each bead. The hydrodynamic force acting on a subchain could be defined, and then the dynamics of the network could be described by the following difference-differential equations (Greenfeld *et al.*, 2011):

$$m\frac{d^2z_n}{dt^2} = \xi_{eff}\,\mu\left[\upsilon(z_n) - \frac{dz_n}{dt}\right] + \frac{T}{\xi_0^2}\left\{[z_{n+1} - z_n - \xi_0] - [z_n - z_{n-1} - \xi_0]\right\} \tag{32a}$$

$$\xi_{eff}\,\mu\upsilon_r(r_n) + \frac{T}{\xi_0^2}\left\{[r_{n+1} - r_n - \xi_\perp] - [r_n - r_{n-1} - \xi_\perp]\right\} = 0. \tag{32b}$$

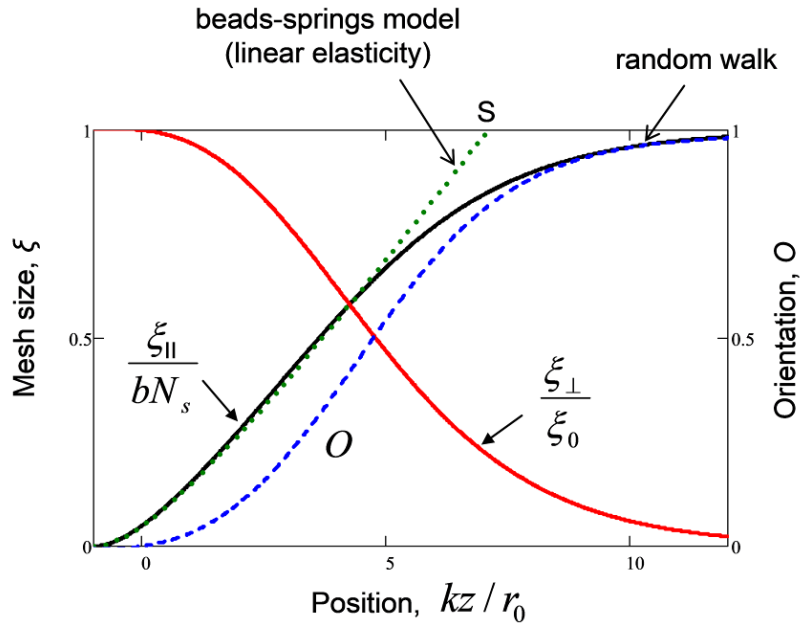

**Figure 31.** Universal plot of the polymer network conformation. Relative axial stretching, $\xi_{II} / bN_s$, radial contraction, $\xi_\perp / \xi_0$, and molecular orientation $O$, *vs.* the normalized axial position, $kz / r_0$. $bN_s$ is the length of a fully extended subchain. The point $S$ on the top of the plot indicates the criterion for 'full' network extension. The results were obtained by random walk simulations and theoretical modeling (Greenfeld *et al.*, 2011).

The corresponding solution, depicted in Figure 31, predicts an affine stretching of the network:

$$\frac{\xi_\parallel(z)}{\xi_0} \approx \frac{\upsilon/\upsilon_0}{1 - \frac{1}{\alpha_\upsilon}d(\upsilon/\upsilon_0)/d(z/r_0)} \approx \frac{\upsilon(z)}{\upsilon_0}, \tag{33}$$

where $\alpha_\upsilon$ is a dimensionless parameter ($\alpha_\upsilon \gg 1$).

The dynamic conformation of subchains can be described by a random walk simulation, where each step represents a single monomer. The probability to step in a specific direction is determined by an effective potential, which arises from the external forces acting at the subchain ends, which propagate along the subchain, and the local hydrodynamic force acting directly on monomers (Figure 32). For a given force vector, $F$, acting on a monomer (normalized as $f=Fb/k_BT$), the probabilities for a random walk step are:

$$P_x^{\pm} = \frac{\exp\left(\pm f_x\right)}{2\sum\limits_x \cosh\left(f_x\right)} \tag{34}$$

in each of the Cartesian directions (Greenfeld $et\ al.$, 2011). The monomer forces in the radial and axial directions are given by:

$$f_z \cong f_0 + \sum_{i=1}^{n} f_i^{hyd} \cong f_0 + \tau_0\left(\upsilon_z - \upsilon_0\right)/b \tag{35a}$$

$$f_r \cong f_0 \approx 3b/\xi_0 \approx 3\phi \tag{35b}$$

where $f_0$ is caused by an effective stretching force acting on subchains in a network at rest, and $\tau_0 \approx \mu_s b^3/k_BT$ is the monomer relaxation time given the solvent viscosity $\mu_s$. The hydrodynamic force in the radial direction is small compared to the force in the axial direction, since $f_r f_z \cong \upsilon_r/\upsilon \cong r/z << 1$, and was therefore neglected. Since the dominant force acts at the subchain ends, the stepping probabilities were assumed to remain uniform along the subchain, and the axial stretching and radial contraction could be written using Eqs. (34) and (35) (Greenfeld $et\ al.$, 2011):

$$\frac{\xi_{\parallel}}{\xi_{max}} \cong \frac{\sinh\left(f_z\right)}{Q} \tag{36a}$$

$$\frac{\xi_{\perp}}{\xi_{max}} \cong \frac{\sinh\left(f_r\right)}{Q} \tag{36b}$$

$$Q = \cosh\left(f_z\right) + 2\cosh\left(f_0\right) \tag{36c}$$

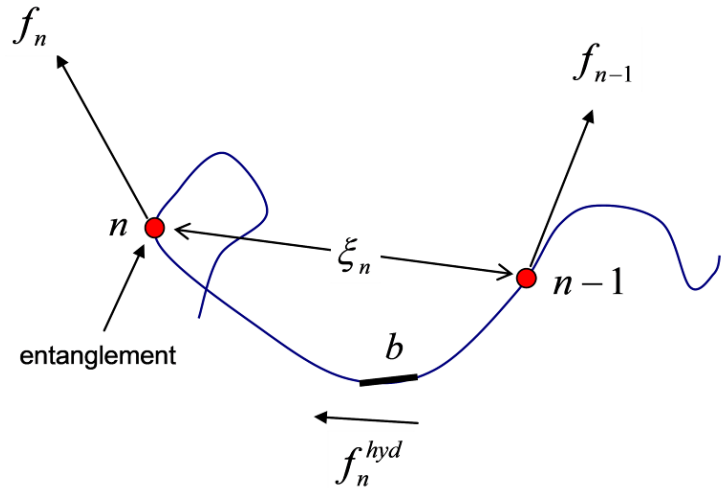

**Figure 32.** Forces acting on a subchain: external forces ($f_n$, $f_{n-1}$) act at chain ends, and a hydrodynamic force ($f_n^{hyd}$) acts on each monomer.

Shortly after the jet start, but before the network approaches full stretching ($f < 1$), the relative longitudinal elongation of a subchain can be approximated by:

$$\frac{\xi_\parallel}{\xi_0} \approx \frac{\xi_0 \upsilon_0 \tau_0}{3b^2}\left(\frac{\upsilon}{\upsilon_0}\right) \approx \frac{\upsilon}{\upsilon_0}, \tag{37}$$

as obtained by the dynamic model, Eq. (33). The condition for affine stretching is satisfied with a prefactor, $\xi_0 \upsilon_0 \tau_0 / (3b^2) = 1$. The simulation allowed the analysis to be expanded to large chain elongations with non-linear elasticity, showing that subchains approach full extension not far (<1 mm) from the jet start (Figure 31). The criterion established for such full extension (marked by point $S$ in Figure 31), using Eq. (33) or (37), is when the jet velocity rises above its initial value by a factor equal to the inverse of the polymer volume fraction:

$$\frac{\upsilon_S}{\upsilon_0} \approx \frac{\xi_{\parallel,S}}{\xi_0} \approx \frac{bN_s}{bN_s^{1/2}} \approx N_s^{1/2} \approx \phi^{-1}, \tag{38}$$

which occurs at a jet radius reduction ratio of $r_0 / r_S \approx N_s^{1/4} \approx \phi^{-1/2} \sim 2 \div 10$. One should note that the relative velocity and radius at the stretching crossover point are found to depend only on the solution concentration, and to be completely independent of the electrospinning materials and conditions (e.g., molar mass and electric field). The transformation of subchains from a coil-like equilibrium state into a stretched state was found to occur as a continuous crossover, and no phase transition was observed, in contrast to the well-known coil stretch transition in unentangled chains (de Gennes, 1974; de Gennes, 1979). The dominant local force on a subchain is the elastic force arising from the action of the linked subchains, whereas the local hydrodynamic forces, whose accumulation along the network gives rise to the global elastic stretching, are negligible. Theoretically, since a vertical sequence of subchains in a network is analogous to a very long chain, a network stretch transition is possible if the jet strain rate is very low; however, under such conditions, the flow will be dominated by viscosity and network relaxation rather than elasticity. The strong increase in the longitudinal mesh size, $\xi_{\parallel}$, results in a decrease in the radial mesh size, $\xi_{\perp}$, due to redistribution of the random walk stepping probabilities (Figure 31). This results in a lateral contraction of the network toward the jet center, which is proportional to the decrease in the subchains radial mesh size. An approximation for the decrease of the polymer network radius with respect to the jet radius is given by:

$$r_P(z) \cong \frac{\xi_{\perp}(z)}{\xi_0} r(z). \tag{39}$$

As depicted in Figure 33, this shows the dominant effect of axial stretching on the radial contraction. This result allowed a significant increase of the polymer concentration at the jet center to be predicted. The concentration (in terms of the polymer volume fraction) can be calculated by $\phi_p = b^3 N_s / \left( \xi_{\parallel} \xi_{\perp}^2 \right)$ or by using Eqs. (36) and (39) and the relationship $\phi \approx b / \xi_0$, where $\phi$ is the solution initial concentration:

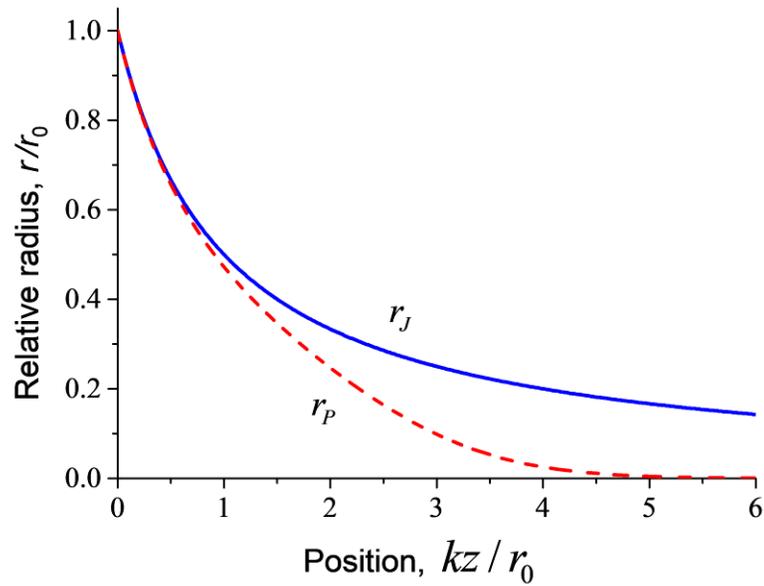

**Figure 33.** Universal plot of the simulated polymer network radius, $r_P / r_0$, vs. the normalized axial position, $kz / r_0$, compared to the jet radius ($r_J / r_0$). Adapted with permission (Greenfeld *et al.*, 2011). Copyright © 2011, American Physical Society.

$$\phi_p = \phi \left( \frac{r}{r_p} \right)^2 = \phi \left( \frac{\xi_0}{\xi_\perp} \right)^2 \cong \frac{\phi}{9} \left[ \cosh \left( 3\phi \frac{\upsilon}{\upsilon_0} \right) + 2 \right]^2. \tag{40}$$

Here, the right term assumes affine stretching by means of the vertical force, $f_z \cong 3\phi \upsilon / \upsilon_0$. When full stretching occurs, the polymer at the jet core is fully compacted ($\phi_p \cong 1$), and the corresponding jet radius can be approximated by $r_j / r_0 \approx \phi^{1/2}$, the same result as in Eq. (38). These predictions were validated by X-ray absorption measurements on electrospun jets (Greenfeld *et al.*, 2011; Greenfeld *et al.*, 2012) as better explained below. The validity of the network modeling is restricted to the initial stage of the jet (first few millimeters), where elastic elongation is still possible, and therefore the model does not describe the final state of the polymer matrix in electrospun nanofibers. In fact, additional processes, such as rapid solvent evaporation and polymer entanglement loss (Greenfeld and Zussman, 2013) which can result in chain relaxation, are not accounted for in this model. Nevertheless, the results strongly indicate non-equilibrium, ordered nanostructures that could remain in the nanofibers after solidification, set a new internal scale, and affect the nanofiber elasticity through confinement (Arinstein *et al.*, 2007).

## 2. Experiments: X-ray imaging of electrospun jets

The theoretically-predicted longitudinal stretching and lateral contraction of the polymer network, as well as the additional effects of rapid evaporation, were investigated experimentally by fast X-ray, phase-contrast, high-resolution imaging of the first 10 mm of electrospun jets (Greenfeld *et al.*, 2011; Greenfeld *et al.*, 2012), using solutions of PEO (Figure 34). The power-law jet geometry, assumed in Eq. (31), was validated by detailed measurements of the jet profile under a wide range of electrospinning conditions, demonstrating that the jet diameter narrows faster under higher electric fields, lower flow rates, and lower polymer concentrations (Figure 35).

The polymer concentration mapping along and across the jet makes use of the different X-ray mass absorption coefficients of the polymer and solvent, $\varsigma_p$ and $\varsigma_s$, respectively. The absorption coefficient of the polymer solution is given by $\text{Abs}(r,z) = \varsigma_p\, m_p\,(r,z) + \varsigma_s\, m_s\,(r,z)$ (Roe, 2000), where $m_p$ and $m_s$ denote the mass concentrations of the polymer and solvent, respectively, the $(r,z)$ coordinates are the radial and axial position in the jet, respectively, and $(m_s\,/\rho_s) + (m_p/\rho_p) = 1$, where $\rho_s$ and $\rho_p$ indicate the densities of the solvent and of the polymer. Thus, the change in the local polymer concentration, $\Delta\phi_p(r,z)$, with respect to the initial concentration ($\phi$), is linearly dependent on the change in the local absorption coefficient $\Delta$Abs:

$$\Delta\phi_p(r,z) = \phi_p(r,z) - \phi_p = \Delta\text{Abs}(r,z)\,\rho_p / (\varsigma_p\,\rho_p - \varsigma_s\,\rho_s). \tag{41}$$

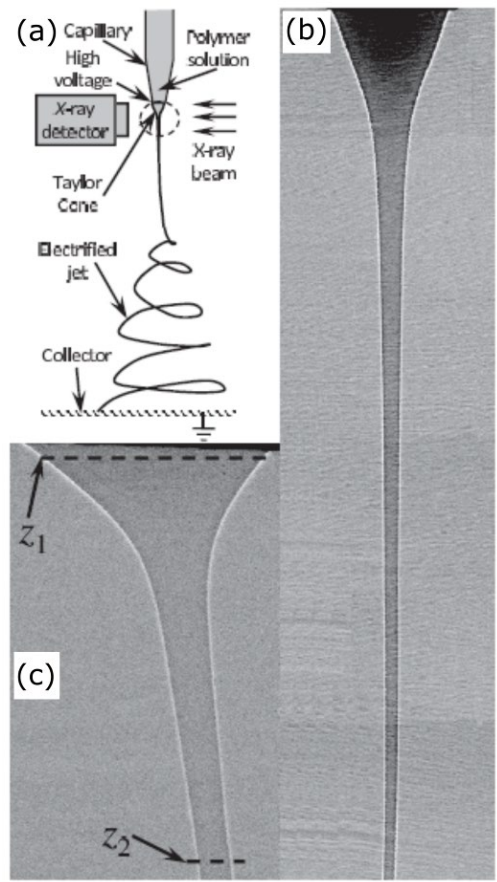

**Figure 34.** (a) Schematics of electrospinning with in-process X-ray imaging. The imaged region is circled. Electric field gap: 6.5 cm. (b) Straight jet region (5 mm length), consisting of a sequence of 10 images. Solution: 5 wt% PEO (600 kDa) in water, electric field: 0.6 kV/cm, flow rate: 3.2 mL/hr. (c) Solution: 3 wt% PEO in water, electric field: 1.6 kV/cm, flow rate: 2 mL/hr. The lines at $z_1$ and $z_2$ (0.02 mm and 0.5 mm, respectively) highlight the cross-sections of regions used for absorption measurements across the electrospun jet. Reproduced with permission (Greenfeld *et al.*, 2011). Copyright © 2011, American Physical Society.

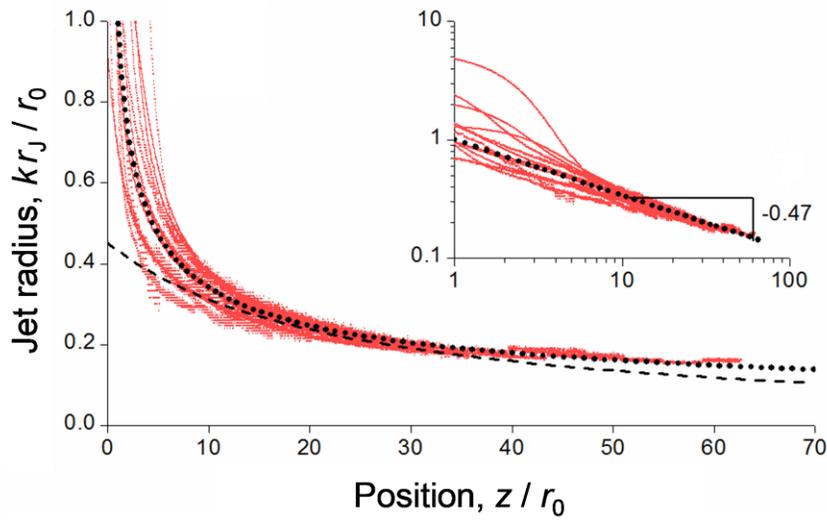

**Figure 35.** Normalized jet radius, $kr_J / r_0$, at position $z / r_0$ along a jet, for combined data from various electrospinning tests, where each experimental set was multiplied by the constant $k$ pertaining to that test. Dotted line: power fit with the expression $r_J / r_0 = (z / r_0)^{-0.47}$, where the exponent is that measured in the inset. Dashed line: hyperbolic fit with the expression $r_J / r_0 = (z / r_0 + p)^{-1}$, with $p = 23.1$. Inset: data shown in bi-log scale, highlighting the power fit exponent. The measured dimensionless parameter $k$ compares well with the theoretical prediction. Adapted with permission (Greenfeld *et al.*, 2012). Copyright © 2012, American Chemical Society.

$\Delta$Abs$(r,z)$ was calculated by comparing the measured X-ray transmission (Tr$_{exp}$) to a simulated transmission for a 'still' jet (Tr$_{sim}$), at a given beam travel distance, $d(r,z)$, through the jet:

$$\Delta\text{Abs}(r,z) \cong -\frac{1}{d(r,z)}\ln\left[\frac{\text{Tr}_{exp}(r,z)}{\text{Tr}_{sim}(r,z)}\right]\left[\frac{\text{Abs}_0}{\text{Abs}_{sim}(r,z)}\right], \tag{42}$$

where the correction factor, Abs$_0$/Abs$_{sim}$, filters out the effects of scattering in a homogenous jet, and is used as an approximation for the heterogeneous electrified polymer solution jet. Concentrations were found to rapidly increase below a critical jet radius of ~25 μm (equivalent to radius reduction ratio of 0.2, Figure 36), a possible evidence for full network extension and for rapid evaporation that occurs much earlier than theoretical predictions found in previous literature. When depicted versus the jet radius, the concentration curves collapse into groups of common initial solution concentration. The concentration crossover occurs at a lower radius for the lower solution concentration, as predicted in Eq. (38). The theoretical prediction [Eq. (40)], depicted in the inset of Figure 36, was found to conform well to the experimental results, hence favoring the stretching premise over evaporation. These results confirmed that the stretching crossover position depends predominantly on the solution initial concentration. Note that in Figure 36 the concentration increase is slightly slower in the experiment compared to the theoretical curve, possibly accounting for the stretching and stress relaxation.

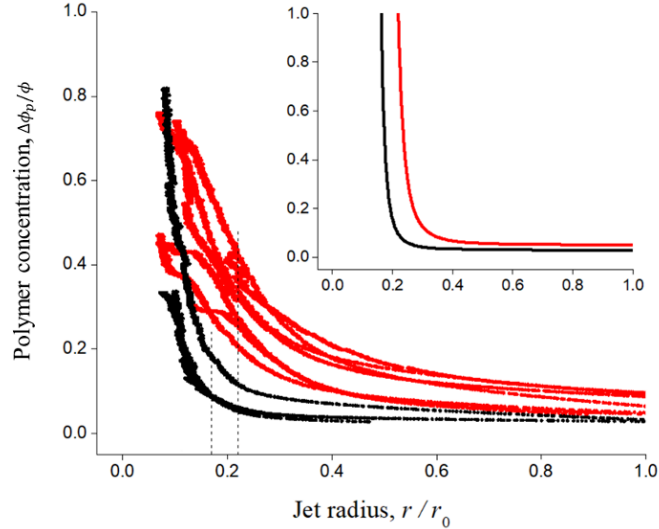

**Figure 36.** Polymer concentration *vs.* relative jet radius, $r_J / r_0$, derived from X-ray absorption measurements at jet center, for experiments with solution jets made of PEO 3 wt% (black lines) and PEO 5 wt% [red lines (grey in print)]. The predicted crossover radius from Eq. (38) is highlighted for both solution concentrations. Adapted with permission (Greenfeld *et al.*, 2012). Copyright © 2012, American Chemical Society. Inset: corresponding theoretical prediction.

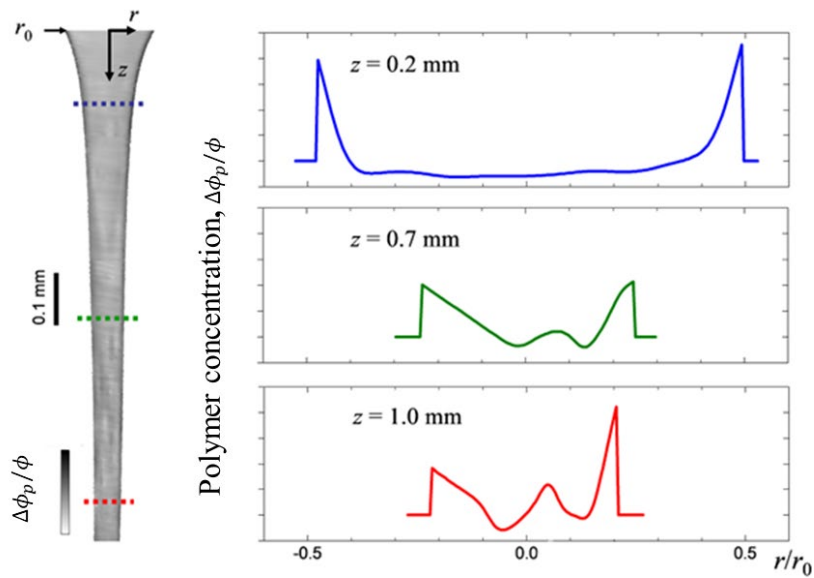

**Figure 37.** Relative polymer concentration change across the jet *vs.* relative radial distance from the jet center, $r/r_0$, for several axial positions ($z$) along the jet. Data are from X-ray absorption measurements of PEO 5 wt%, electric field: 2.8 kV/cm, flow rate: 1.9 mL/h. Adapted with permission (Greenfeld *et al.*, 2012). Copyright © 2012, American Chemical Society.

The variation of concentration across the jet also revealed high concentrations at the jet boundary due to evaporation, as well as a concentration rise at the jet center within ~1 mm from the jet start (Figure 37), in agreement with the model and simulation. Evaporation becomes dominant when stretching is weaker, e.g., at lower electric field and/or higher flow rate, inhibiting the concentration peaks measured at the jet center. Such tuning of parameters evidenced the balance between the effects of evaporation and stretching, which determines the non-equilibrium conformation of the polymer network during electrospinning, and explains the diversity of macrostructures and properties found in solid nanofibers.

## VI. MODELLING OF POLYMER SOLUTION BLOWING

This Section reviews the works dealing with detailed modeling of solution-blowing, which allows one to predict the evolution of polymer jets toward nanofibers, as well as such laydown properties as thickness, porosity and permeability, i.e. the 3D micro-structure, fiber-size distribution and polymer mass distribution resulting from the process. Two monographs (Yarin, 1993; Yarin *et al.*, 2014), triggered a group of inter-related works devoted to modeling of meltblowing (Sinha-Ray *et al.*, 2010; 2011; 2013; Yarin *et al.*, 2010; Ghosal *et al.*, 2016a) and solution blowing (Sinha-Ray *et al.*, 2015, Ghosal *et al.*, 2016b). The effects of the governing process parameters on variation of laydown properties were predicted, primarily the influence of the velocity of moving collectors. For instance, it was shown and explained how an increase in the velocity of the collector screen leads to an increase in the porosity and permeability of the nonwoven laydown. The modeling was based on the system of quasi-1D equations of the dynamics of polymer solution jets moving, evaporating and solidifying when driven by a surrounding air jet, as discussed below. The governing equations were solved numerically. Multiple polymer jets were considered simultaneously when they are deposited on a moving screen and forming a nonwoven laydown of nanofibers. A scheme that visualizes the process in the context of the model reported in the present Section is shown in Figure 38a, where two nozzles, at which the polymer solution and compressed air are delivered, are supported by a nosepiece. The polymer jet/air jet interaction, and the various forces acting on the jet, together with the solvent evaporation, are illustrated in Figure 38b. Similarly to electrified jets, the initial part (~ 1 mm long) of the solution blown jets is too thick to bend since its bending stiffness is high. Here, the straight parts of polymer jets are stretched by the surrounding high-speed air jets, leading to a strong elongational flow and corresponding cross-sectional reduction. Similarly to electrospinning, when the jet becomes thin, its bending stiffness ($\sim r^4$) is strongly reduced and a vigorous bending, which in the case of solution blowing is driven by aerodynamic forces, begins. Both the straight and the bending regimes were described in the framework of quasi-1D equations, with the following continuity and momentum balance equations (Yarin, 1993; Yarin *et al*., 2014; Sinha-Ray *et al*., 2015; Ghosal *et al*., 2016b):

$$\frac{\partial \lambda f}{\partial t} + \frac{\partial f W}{\partial s} = -D_a b \pi \lambda \, , \tag{43a}$$

$$\frac{\partial \lambda f \mathbf{v}}{\partial t} + \frac{\partial f W \mathbf{v}}{\partial s} = \frac{1}{\rho}\frac{\partial P \boldsymbol{\tau}}{\partial s} + \lambda f \mathbf{g} + \frac{\lambda}{\rho}\mathbf{q_{TOT}} \, , \tag{43b}$$

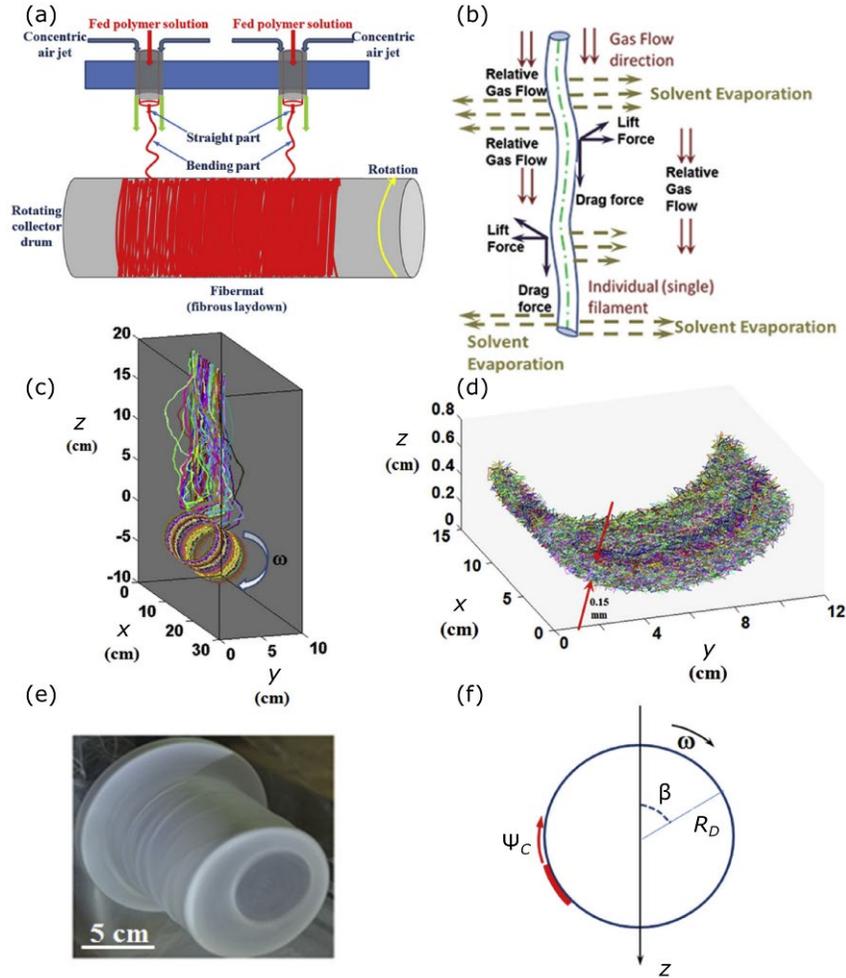

**Figure 38.** (a) Scheme of polymer solution blowing, with multiple jets located along a nosepiece, and a rotating collector. (b) Different forces acting on the polymer solution jet. (c) Snapshot of 60, numerically-simulated polymer solution jets. (d) Section of a deposited solution-blown laydown. (e) A drum used in the experiments. (f) Schematics of the drum cross-section with corresponding coordinate system. Adapted with permission (Ghosal *et al*., 2016b). Copyright © 2016, Elsevier Ltd..

In Eqs. (43), most of the symbols have the same meaning as in Eqs. (12). In particular, $\lambda$ is the stretching factor, $f = \pi\, r^2$ is the cross-sectional area of the jet, $s$ is an arbitrary coordinate reckoned along the jet axis, that might be understood, as usual, as a Lagrangian coordinate marking material elements along the jet axis, $D_a$ is the vapor diffusion coefficient in air, $\mathbf{v}$ is the absolute velocity of polymer solution in the jet, $W$ is the liquid velocity along the jet relative to a cross-section with a certain value of $s$. $P$ denotes the magnitude of the longitudinal internal viscoelastic force in the jet cross-section, $\boldsymbol{\tau}$ the unit tangent vector of the jet axis, $\mathbf{g}$ the gravity acceleration, and $\mathbf{q_{TOT}}$ the overall force acting on a jet element. In the present case, this force is the aerodynamic force applied by the surrounding gas on a unit jet length, rather than the electric force (Sinha-Ray *et al*., 2010a; 2011; 2013; 2015; Yarin *et al*., 2010; Ghosal *et al*., 2016a; 2016b). The term on the right-hand side in the continuity equation (43a) describes the solvent evaporation. The factor, $b$, in this term reads (Yarin *et al*., 2014):

$$b = 0.495\, \mathrm{Re}_a^{1/3}\, Sc^{1/2}\left[ C_{s,eq}(T) - C_{s,\infty} \right], \tag{44}$$

where $\mathrm{Re}_a$ is the local Reynolds number of a jet element based on its relative velocity to the surrounding air, $\mathrm{Sc}$ is the Schmidt number ($\mathrm{Sc} = \nu_{air}\, /\, D_a$, where $\nu_{air}$ is the kinematic viscosity of air), and $C_s$ is solvent concentration. The subscript *eq* corresponds to the equilibrium vapor pressure over the polymer solution surface determined by temperature ($T$), whereas the subscript $\infty$ corresponds to the vapor content far away from the jet surface in the surrounding air. Therefore, the solvent evaporation rate is dependent on $T$, through the equilibrium solvent concentration, $C_{s,eq}(T)$. This dependence can be derived from the Antoine equation (Reid *et al*., 1987), or similar equations (Seaver *et al*., 1989; Alduchov and Eskridge, 1996) that are available.

The projections of the momentum balance equation onto the accompanying trihedron of the jet axis, namely, the unit tangent vector $\boldsymbol{\tau}$ , the unit principal normal vector $\mathbf{n}$ , and the unit binormal vector $\mathbf{b}$ , are kindred to a hyperbolic wave equation. Accordingly, they could be solved numerically using the implicit numerical scheme of the generalized Crank-Nicolson type, with the central difference spatial discretization at three time levels (Sinha-Ray *et al*., 2010a; 2011; 2013; 2015; Yarin *et al*., 2010; Ghosal *et al*., 2016a; 2016b). The implementation of the initial and boundary conditions, and the post-processing procedure which allows one to reconstruct the 3D architecture of the laydown of nanofibers were discussed in detail (Sinha-Ray *et al*., 2015; Ghosal *et al*., 2016b), using the touch-down times of the individual jet elements, their locations on the collecting screen and the cross-sectional radii of as-deposited filaments (Ghosal *et al*., 2016a; 2016b).

Solution blowing is an isothermal process, with the temperature $T$ in Eq. (44) being room temperature. This temperature is typically above the $\theta$-temperature at which solvent-polymer and polymer-polymer interactions equal each other, and thus the solvents are initially good, which means that polymer molecules preferentially possess extended, elongated configurations. However, during solution blowing the solvent concentration in the polymer jet decreases due to evaporation, and, accordingly, the polymer concentration, $C_p$, increases. In addition, the local polymer concentration varies due to stretching as $C_p = C_{p,0} \lambda_0 f_0 / \lambda f$ , where, as usual, the subscript 0 denotes the values at the initial cross-section of the jet bending part. The rheological parameters of the viscoelastic polymer solution, namely its viscosity and relaxation time also vary along the jet as (Yarin *et al*., 2014):

$$\mu = \mu_0 \times 10^{J(C_p^\infty - C_{p,0}^\infty)} , \tag{45a}$$

$$\theta = \theta_0 C_{p,0} / C_p, \tag{45b}$$

where $\mu_0$ and $\theta_0$ are the initial values of the viscosity and the elastic relaxation time (namely, again, those at the initial cross-section of the jet bending part), and $J$ and $m$ are material-dependent constants. This strongly nonlinear $C_p$-dependence of the zero-shear viscosity and of the relaxation time practically arrests the deformation of the polymer solution at some moment, which corresponds to polymer precipitation at a high enough concentration where polymer-polymer self-interactions prevail. Also, the longitudinal force, $P$ (Eq. 43b), is a function of ($\tau_{\tau\tau} - \tau_{nn}$), where $\tau_{\tau\tau}$ and $\tau_{nn}$ are the longitudinal and normal deviatoric stresses in the jet cross-section, respectively. Since $\tau_{\tau\tau} >> \tau_{nn}$, $P$ is practically a function of $\tau_{\tau\tau}$ only. The deviatoric stresses could be calculated using an appropriate rheological constitutive equation, for example the UCM model (Yarin, 1993; Yarin *et al*., 2014). Hence, the rheological behavior of polymer solutions could be described using phenomenological constitutive equations that do not directly utilize any physical information related to macromolecular chains and their conformations. However, a link between UCM and micromechanical models of polymer solutions and jets was established (Yarin, 1993), showing that in strongly stretched polymer jets the higher values of the longitudinal deviatoric stress $\tau_{\tau\tau}$ correspond to the macromolecular chain stretching and orientation in the axial stretching direction. Jet stretching in flight was also studied in detail in the framework of meltblowing (Yarin *et al*., 2014).

In both experiments and numerical simulations of solution blowing (Ghosal *et al*., 2016b), solidified polymer jets (i.e., nanofibers) were collected on a rotating drum (Figure 38a,c-f). Here, the direction of the blowing is defined as *z*. In particular, Figure 38c shows this geometry along with a snapshot of 60 numerically simulated polymer solution jets in flight, which are wound on the collecting drum. Figure 38d shows a cut portion of the numerically-simulated laydown on the drum. In order to model the process, the drum is assumed rotating with the angular velocity, $\omega = d\beta / dt$, where $\beta$ is the angular coordinate around the drum axis, which is parallel to the nosepiece (Figure 38f). The drum has cross-sectional radius $R_D$. The angular coordinate of a material element after its touch-down at the drum (or at the preceding fiber laydown on the drum) is so found as $\beta = \beta_{touch} + \omega(t - t_{touch})$, where the touch-down happens at the angle angle $\beta_{touch}$ and at time $t_{touch} < t$. It should be emphasized that this expression allows the angle $\beta$ to grow beyond $\beta_{touch} + 2\pi$, which means that the deposited fibers have made a full rotation with the drum and are being covered by a newly deposited fiber layer. The corresponding circumferential coordinate of a material element of a polymer filament on the drum axis denotes as $\psi_C$ and is found as:

$$\psi_C = \psi_{C,touch} + \omega(t - t_{touch})R_D ,\tag{46}$$

where $\psi_{C,touch} = \beta_{touch} \times R_D$. Therefore, the coordinate of a material element of a polymer filament on the drum, along a direction normal to $z$ and aligned parallel with the nosepiece (Figure 38a) does not change after the touch-down. On the other hand, its $z$-coordinate varies as (Figure 38f):

$z = z_{touch} + R_D \cos(\pi - \beta) = z_{touch} - R_D \cos\beta$, where $z_{touch}$ corresponds to the touch-down position. This leads to:

$$z = z_{touch} - R_D \cos\left[\beta_{touch} + \omega(t - t_{touch})\right].\tag{47}$$

The previously deposited and new layer of fibers (generated after a full rotation of the drum) could have the same value of the coordinate $z$ as per Eq. (47), albeit they are distinguished by their coordinates $\psi_C$, as per Eq. (46). In other words, $\psi_C$ corresponds to a longitudinal coordinate along an unrolled laydown. These equations allow one to pose the boundary conditions at the end of a free polymer jet already deposited on the rotating cylindrical collector (the drum), similarly to the boundary conditions used on planar collector screens (Sinha-Ray *et al.*, 2010a; 2011; 2013; 2015; Yarin *et al.*, 2010; Ghosal *et al*., 2016a; 2016b). These boundary conditions affect backward the oncoming part of the polymer jet/filament through the corresponding viscoelastic force acting along the jet/filament as predicted by the model (Ghosal *et al*., 2016b).

Results from simulations favorably compare with experimental data. For instance, the predicted mean turbulent velocity field in the axisymmetric air jet surrounding polymer solution jets was calculated (Sinha-Ray *et al*., 2015). The simulated bending jet domain, for a single solution-blown jet, was found to occupy a cylinder of about 0.38 cm in diameter, whereas the experimentally observed jet was located inside a cylinder of 0.33 cm in diameter as highlighted by vertical lines in Figure 39 (Sinha-Ray *et al*., 2015). The predicted cross-sectional fiber diameter distribution is shown in Figure 40.

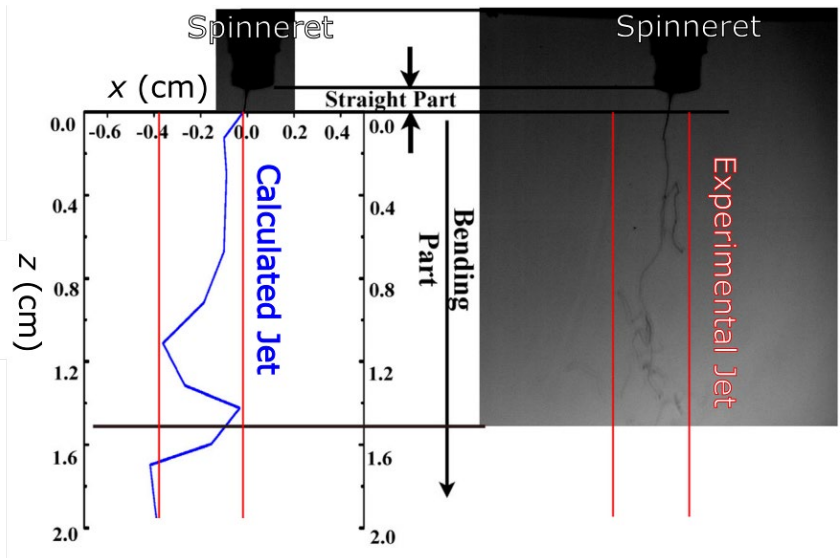

**Figure 39.** Snapshot of jet configuration at the beginning of the bending part, for blown polymer solutions. The bending jet domain that is found experimentally is highlighted by two vertical straight lines. The predicted snapshot of the jet axis at the beginning of the bending part wiggles in between these lines, showing good agreement between model and experiment. The experimental data was acquired with a Phantom V210 fast camera. Adapted with permission (Sinha-Ray *et al*., 2015). Copyright © 2014, Elsevier Ltd..

Two additional and interesting properties of solution-blown nonwovens are the volumetric porosity and permeability. In modelling works, the volumetric porosity $p_{vol}$ was defined following its basic definition, namely, based on the predicted volume, $V_E$, of the laydown envelope and the volume, $V_F$, of the polymer fibers encompassed by this envelope. Accordingly, $p_{vol} = [1 - (V_F / V_E)] \times 100\%$. Volumetric porosity values under different conditions were found by post-processing the simulated nonwoven laydowns formed by 60 polymer solution jets for 1 min, and the calculated volumetric porosity was found to increase upon increasing the angular speed of the rotating drum collector (Ghosal *et al.*, found to increase upon increasing the angular speed of the rotating drum collector (Ghosal *et al.*, 2016b), which is reminiscent of the case of meltblowing onto a moving surface. A similar behavior was found for permeability (Ghosal *et al.*, 2016b). Finally, the comparison of experimentally measured and numerically-simulated laydown landscapes was also favorable, with similar found morphological profiles and height variations (Figure 41).

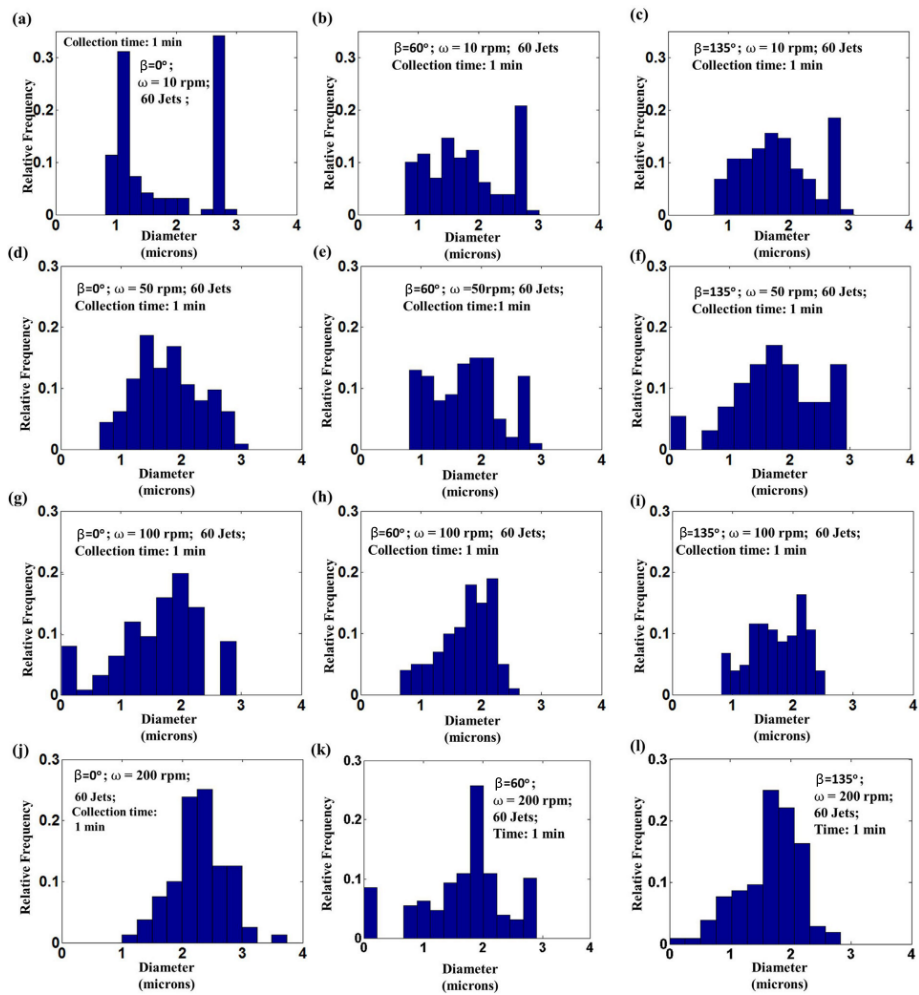

**Figure 40.** Predicted diameter distributions of solution-blown laydowns from 60 jets (parameters shown in Figure 38f), for various angular velocities (ω) of the rotating drum. (a-c) Diameter distributions for β=0º, 60º and 135º, respectively, for ω=10 rpm. (d-f) Same angles, with ω=50 rpm. (g-i) Same angles, with ω=100 rpm. (j-l) Same angles, with ω=200 rpm. Adapted with permission (Ghosal *et al.*, 2016b). Copyright © 2016, Elsevier Ltd..

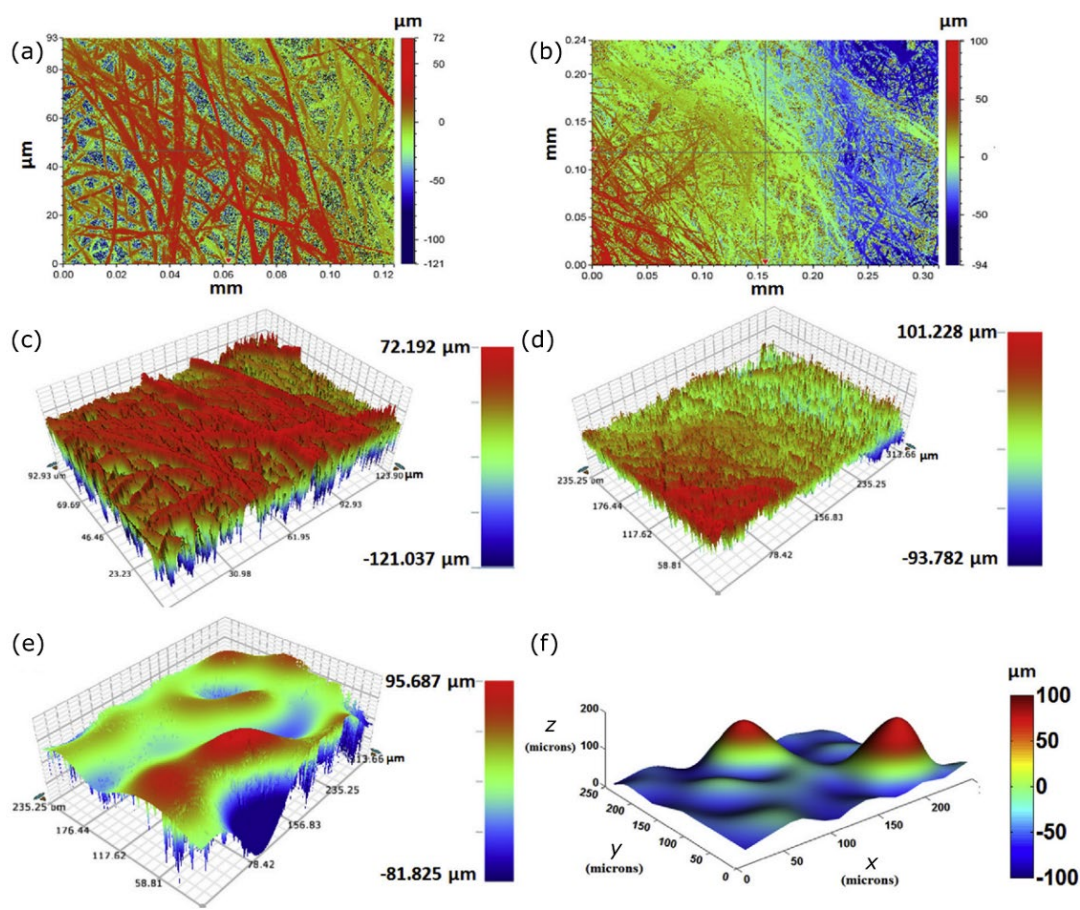

**Figure 41.** Layout produced by solution blowing of polymer nanofibers. Height profiles of the laydown measured by optical profilometry in different sample points (a-e), and simulated average profile (f) for a deposition region (*x,y*) of comparable size (order of $10^4$ μm$^2$). Polymer solution concentration: 15 wt%. Angular drum speed: 240 rpm. *z*-numerical data (f) are obtained starting from a large simulated laydown area (20×10 cm$^2$), which is then subdivided into several smaller areas of 250×250 μm$^2$, and averaging the simulated mean elevation over those smaller areas. Adapted with permission (Ghosal *et al.*, 2016b). Copyright © 2016, Elsevier Ltd..

## VII. PERSPECTIVE AND CONCLUSIONS

Here, we provide a summary of currently open challenges, as well as of possible future developments in the vibrant field of modeling of electrified polymer jets, solution blowing, and of materials obtained by these processes.

*From fluid dynamics to material properties.* Modeling of electrospinning and solution blowing is still to be generalized, for instance, to be able to predict the degree of crystallinity in the as-spun nanofibers. This would be a step toward the solution of a formidable problem, namely, the prediction of the ultimate mechanical properties of both individual nanofibers and of materials assembled by them. In fact, fiber-forming methods and laydown post-processing encompass such processes as spunbonding and hydroentanglement, respectively. Modeling of these processes has already begun to develop along the same lines as that of electrospinning and solution blowing, i.e., based on quasi-1D Eqs. (12) (Li *et al.*, 2019a; 2019b). Only the very first steps in this direction were done, however, and significant efforts would be required in future to predict the mechanical properties of the obtained materials. Electrospinning and solution blowing of core-shell nanofibers, such as those used in self-healing vascular, nano-textured composite materials (Yarin *et al.*, 2019), also poses multiple modelling questions. Novel electromechanical devices based on electrospun and solution-blown nanofibers (An *et al.*, 2018a; 2018b; Kang *et al.*, 2019) involved multiple issues that require detailed modeling of material properties, and of how they descend from the jet properties and dynamics. That would enable development of software capable of control on such nano-textured devices.

Other challenges are given by our capability to understand the internal structure of obtained nanofibers, which is also strictly related to the jet behavior. The variation of polymer concentration across the jet was found to go through high concentrations at the jet boundary due to evaporation as well as a concentration rise at the jet center (Greenfeld *et al.*, 2011; Greenfeld *et al.*, 2012). The latter phenomenon is attributed to polymer stretching that causes lateral contraction of the polymer network toward the jet center, and is in good agreement with theoretical models. Moreover, it was shown that evaporation is dominant when stretching is weaker (e.g., at lower electric field and/or higher flow rate), canceling the concentration peaks measured at the jet center. The balance between the effects of evaporation and stretching determines the polymer network non-equilibrium conformation during electrospinning, and can help in clarifying the reasons for the diverse structures and properties found in solid nanofibers. In particular, the size-dependent mechanical, thermomechanical, and thermodynamic properties of as-spun nanofibers, such as the rise of the elastic modulus at small diameters, and the shift of the glass transition temperature, are attributed to the internal molecular and supramolecular structure of the polymer matrix in nanofibers. The current implementation of the method is, however, limited to the initial section of the jet. Further investigation downstream could provide evidence for disentanglement of the polymer chains and reveal non-uniform flow regime due to rapid evaporation, possibly with streamlines toward the jet boundary. Such studies may be important for several applications of electrospinning, such as drug delivery and nano-composites, and aerogels based on short-fibers.

*Beyond Lagrangian models for a multiscale description.* The last advances in both mesoscopic and atomistic simulations lead to a new, multiscale paradigm. New high-tech experimental setups of electrospinning processes prompt a quantitative understanding of different complex phenomena that span a broad range of scales in both space and time. Several physical quantities may be described using different representations already at the same length scale. For instance, as pointed out in the sub-Section V.C.3, the simulation of electrospinning in the presence of a space distributed electric field $\mathbf{E}(x, y, z)$ can be performed by coupling the Eulerian representation of the vector field to the Lagrangian jet beads. Moreover, the electrospinning process is a non-equilibrium transport problem, driven by strong force terms which span across many length and time scales. A typical example is the repulsive Coulomb force. For each $i$–th bead, the Coulomb term, $\mathbf{f}_{c,i}$, is assessed as the summation of bead pairs $(i, j)$ as the inverse of the distance $|\mathbf{R}_j - \mathbf{R}_i|$ [see Coulomb term in Eq. (18)], distributed over several orders of magnitude in length ($|\mathbf{R}_j - \mathbf{R}_i|$ in the range from 0.01 up to 10 cm). Denoted $r_{pcut}$ a primary cutoff for the Coulomb pair interactions, the Coulomb force can be split in two terms of different time scales: a contribution from all the neighbour beads with $|\mathbf{R}_j - \mathbf{R}_i| \le r_{pcut}$, which changes quickly in time, and a second, slower contribution due to the remaining far-field beads. Hence, it is possible to assess the force $\mathbf{f}_{c,i}$ by the so-called multistep method (Allen *et al*., 2017), where the contribution of the neighbor beads is computed at every time step (usually $\Delta t \sim 10^{-8}$ s), while that from the remaining beads is updated at a larger time scale (typically every $100\Delta t$).

On the other hand, other force terms cannot be easily described by a multiscale approach. For instance, the tensile stress force is related to the behavior of the polymeric matrix pushed towards non-equilibrium states when subject to the external, intense electric field. Hence, the mechanical response and fracture phenomena would require an atomistic description in order to be fully resolved (without using a rheological constitutive law). In the last decade, several works have investigated the mechanical response of stretched polymeric matrices at the atomistic level by Lagrangian methods (usually by molecular dynamics simulations) where particle-like points represent the atomic positions (Buell *et al*., 2009; Park and Joo, 2014; Miao *et al*., 2015; 2017; Lolla *et al*., 2016). However, the largest available system size in molecular dynamics simulations is about $\sim 10^{-6}$ cm, while the typical time step for the integration of the EOM is $\Delta t \sim 10^{-15}$ s. On the other hand, the typical length and time scales in the Lagrangian models (Section V.B) are about $\sim 10$ cm and $\Delta t \sim 10^{-8}$ s, respectively. This range of several orders of magnitude in both space and time is challenging to cover by a multiscale approach, which is, therefore, not a viable route in this specific context. Nonetheless, it is possible to exploit alternative strategies based on mesoscale physics in order to represent other phenomena, which are not described in the Lagrangian models. As an example, in the bead models the fluid is assumed to be spinnable (sub-Section V.B.2), and the capillary breakup phenomena close to the nozzle are entirely missed in the description. Nonetheless, it is possible to bridge the gap by a mesoscale solver of the Navier-Stokes equation of the fluid close to the nozzle. Then, for instance, the information could be transferred to the Lagrangian bead representation within a multiscale scheme.

The mesoscale approaches are grounded into the intermediate level of the description of matter, namely kinetic theory: the main versions are Boltzmann's kinetic theory and Langevin stochastic particle dynamics. A versatile simulation technique for solving Navier-Stokes equation is the lattice Boltzmann method (LBM), which exploits the Boltzmann's kinetic theory (Succi, 2018; Krüger *et al*., 2017). Here, the fluid is represented as pseudo particles propagating and colliding over a discrete lattice domain in space. Briefly, each pseudo particle represents a statistical probability, $f(x, y)$, of observing a fluid parcel in a lattice point moving over a discrete set of allowed directions, so that the degrees of freedom of the system decrease enormously if compared to the correspondent atomistic representation. Although LBM suffers the typical problems of grid-based methods (Section IV and sub-Section V.B.1), it can be implemented with outstanding computational efficiency, especially on parallel computers (Feichtinger *et al*., 2015; Succi *et al*., 2019). In the context of electrospinning modelling, the quasi-straight jet path (sub-Section V.B.2) can be effectively described by the LBM (Lauricella *et al*., 2018). Figure 42 displays a comparison between the hyperbolic profiles of a stretched jet, as observed in simulation and experiment. In particular, the profile in the simulation appears to be in qualitative agreement with the characteristic shape of the jet, experimentally observed close to the injecting nozzle by the Rafailovich and Zussman groups (Greenfeld *et al.*, 2011) and in consistency with previous theoretical results on the jet conical shape (Feng, 2002; 2003). By its very mesoscopic nature, LBM is conceptually at a vantage point for multiscale/level coupling, both upwards, towards Lagrangian bead representations and downwards, towards atomistic models. In perspective, a possible route is to devise new algorithms that allow for reproducing electrospinning setups of increasing complexity, by coupling different descriptions from the bottom-up or, vice versa, top-down overview.

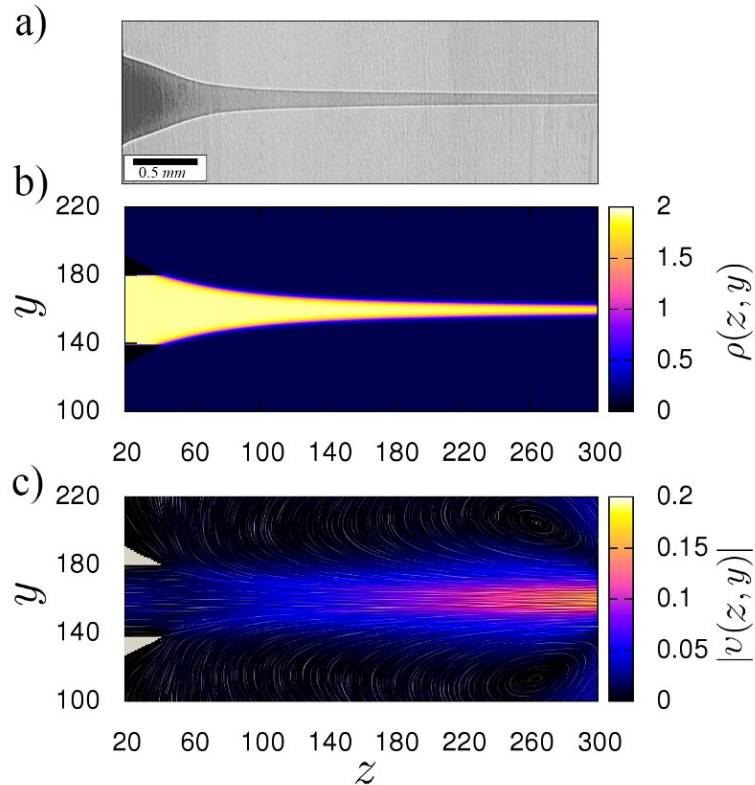

**Figure 42.** (a) The quasi-straight section of a jet in an electrospinning experiment with a solution of 5 wt% PEO in water. Adapted with permission (Greenfeld *et al.*, 2011). Copyright © 2011, American Physical Society. (b) A snapshot of the fluid density, $\rho(z, y)$, in the stationary regime after the jet has touched the right side of the simulation box. (c) The corresponding velocity field magnitude $|\upsilon(z, y)|$, and the line integral convolution representation (Forssell *et al.*, 1995) of the velocity field. Reproduced with permission (Lauricella *et al.*, 2018). Copyright © 2018, American Physical Society.

**Acknowledgements**

E.Z. acknowledges the support of the Winograd Chair of Fluid Mechanics and Heat Transfer at Technion. D.P. and E.Z. acknowledge the European Research Council (ERC) for supporting, under the European Union's Horizon 2020 research and innovation programme, the ERC Consolidator Grant "*x*PRINT" (grant agreement no. 682157). M.L. and D.P. acknowledge the project "3D-Phys" (PRIN 2017PHRM8X) from MIUR. D.P. also acknowledges the support from the project PRA_2018_34 ("ANISE") from the University of Pisa, and the European Research Council (ERC) for previous support, under the European Union's Seventh Framework Programme (FP/2007-2013), to the ERC Starting Grant "NANO-JETS" (grant Agreement no. 306357). M.L. and S.S. acknowledge the European Research Council (ERC) for supporting, under the European Union's Horizon 2020 Framework Programme (No. FP/2014-2020)/ERC Grant Agreement No. 739964 (COPMAT).